\documentclass[12pt]{article}
\pdfoutput=1
\usepackage{jheppub}
\usepackage{amsmath}
\usepackage{wrapfig}

\newcommand{\CN}{{\cal N}}
\newcommand{\CB}{{\cal B}}

\newcommand{\C}{{\mathbb C}}

\title{Vertex Algebras at the Corner}
\abstract{We introduce a class of Vertex Operator Algebras which arise at junctions of supersymmetric interfaces in ${\cal N}=4$ Super Yang Mills gauge theory.
These vertex algebras satisfy non-trivial duality relations inherited from S-duality of the four-dimensional gauge theory. 
The gauge theory construction equips the vertex algebras with collections of modules labelled by supersymmetric interface line defects. 
We discuss in detail the simplest class of algebras $Y_{L,M,N}$, which generalizes $W_N$ algebras. We uncover tantalizing relations between $Y_{L,M,N}$,
the topological vertex and the $W_{1+\infty}$ algebra. }

\author[1]{Davide Gaiotto,}
\author[1]{Miroslav Rap\v{c}\'{a}k}
\affiliation[1]{Perimeter Institute for Theoretical Physics, Waterloo, Ontario, Canada N2L 2Y5}

\begin{document}
\maketitle

\section{Introduction}
The objective of this paper is to introduce a new class of Vertex Operator Algebras $Y_{L,M,N}[\Psi]$ labelled by three integers and a continuous coupling $\Psi$, 
which generalize the standard W-algebras $W_N$ of type $\mathfrak{sl}(N)$. These algebras are important building blocks of a general class 
of VOA's which can be defined in terms of junctions of boundary conditions and interfaces in the GL-twisted ${\cal N}=4$ Super Yang Mills gauge theory
\cite{Kapustin:aa}. 

The concrete definition of $Y_{L,M,N}[\Psi]$ is somewhat laborious: it involves a BRST reduction of 
the combination of WZW algebras for super-groups $U(N|L)_{\Psi} \times U(M|L)_{1-\Psi}$ 
and a certain collection of $bc$ and $\beta \gamma$ systems.\footnote{As pointed out by Mikhail Bershtein, algebras with similar structure were defined 
in \cite{1512.08779,Litvinov:2016mgi} in terms of a kernel of screening charges. It would be interesting to explore the relation further.} We will introduce it and motivate it in the main text
of the paper.\footnote{Here and throughout the main text we will follow a somewhat unusual notation for the level of unitary (super) WZW
algebras, such that the current algebra $U(N|L)_{\Psi}$ contains a standard $SU(N|L)_{\Psi +L-N}$ WZW current sub-algebra at level $\Psi +L-N$. 
In particular, the critical level corresponds to $\Psi=0$.
We will review our definitions in Appendix \ref{app:conventions} and scatter frequent reminders about this notational choice throughout the text. }

\subsection{Dualities}
Our definition will be manifestly symmetric under the reflection $\Psi \leftrightarrow 1-\Psi$ accompanied by the exchange $N \leftrightarrow M$. 
Our main conjecture is that our definition is also symmetric under an  ``S-duality'' transformation $\Psi \leftrightarrow \Psi^{-1}$ accompanied by the exchange $L \leftrightarrow M$.
The two transformations combine into an $S_3$ triality symmetry which acts by permuting the three integral labels $L$,$M$,$N$ while acting on the coupling $\Psi$ by appropriate $PSL(2,Z)$ duality transformations. In particular, we have cyclic rotations:
\begin{equation} \label{eq:cyclic}
Y_{L,M,N}[\Psi] = Y_{N,L,M}[\frac{1}{1-\Psi}]= Y_{M,N,L}[1-\frac{1}{\Psi}]
\end{equation}

An alternative, instructive way to describe the $S_3$ symmetry is to introduce three parameters $\epsilon_i$ which satisfy 
\begin{equation}
\epsilon_1 + \epsilon_2 + \epsilon_3 = 0 \qquad\qquad  \Psi = - \frac{\epsilon_2}{\epsilon_1}
\end{equation}

Then the $S_3$ symmetry acts on 
\begin{equation}
Y^{\epsilon_1, \epsilon_2, \epsilon_3}_{N_1,N_2,N_3} \equiv Y_{N_1,N_2,N_3}[- \frac{\epsilon_2}{\epsilon_1}]
\end{equation}
by a simultaneous permutation of the $\epsilon_i$ and $N_i$ labels. 

We can illustrate this type of relations for $W_N \equiv Y_{0,0,N}$.
The $Y_{0,0,N}[\Psi]$ VOA is defined as the regular quantum Drinfeld-Sokolov reduction of $U(N)_{\Psi}$ and
thus coincides with the standard W-algebra $W_N$ with parameter $b^2 = -\Psi$ combined with a free $U(1)$ 
current.\footnote{Recall our choice of notation in Appendix \ref{app:conventions} which defines $U(N)_{\Psi}$ in terms of $SU(N)_{\Psi-N}$
and $U(1)$ WZW currents.} The $W_N$ algebra has a symmetry $b \to b^{-1}$ known as Feigin-Frenkel duality,
demonstrating immediately the expected S-duality relation between $Y_{0,0,N}[\Psi]$ and $Y_{0,0,N}[\Psi^{-1}]$.

On the other hand, our definition of $Y_{N,0,0}[1-\Psi^{-1}]$ involves a BRST reduction of a product of elementary VOAs
\begin{equation}
U(N)_{-\frac{1}{1-\Psi}} \times U(N)_{\frac{\Psi}{1-\Psi}} \times \mathrm{Ff}^{U(N)} \times \mathrm{bc}^{\mathfrak{u}(N)},
\end{equation}
where $\mathrm{Ff}^{U(N)}$ denotes the VOA 
of $N$ complex free fermions transforming in a fundamental representation of $U(N)$ and 
$\mathrm{bc}^{\mathfrak{u}(N)}$ a $bc$ ghost system valued in the $\mathfrak{u}(N)$ Lie algebra. \footnote{In terms of the $\epsilon_i$, the WZW levels become $U(N)_{\frac{\epsilon_1}{\epsilon_3}} \times U(N)_{\frac{\epsilon_2}{\epsilon_3}}$.}

The BRST complex is essentially a symmetric description of a coset construction, 
leading to a third realization of $W_N$ as 
\begin{equation}
W_N = \frac{SU(N)_{\frac{\Psi}{1-\Psi}-N} \times SU(N)_{1}}{SU(N)_{\frac{1}{1-\Psi}-N}}
\end{equation}
which is the analytic continuation of the well-known coset definition of $W_N$ minimal models. 
See e.g. \cite{Gaberdiel:2012kq} for a review and further references on this ``triality'' enjoyed by $W_N$ algebras. 

One of the most important features of the $W_N$ W-algebra is the existence of two distinct collections of degenerate modules 
labelled by weights of $SU(N)$ and permuted by the Feigin-Frenkel duality. These degenerate modules 
have very special fusion and braiding properties. 

An extension of our main conjecture is the claim that $Y_{L,M,N}[\Psi]$ will have three collections 
$D_\nu$, $H_\lambda$, $W_\mu$ of degenerate modules which are permuted by the $S_3$ triality symmetry and 
are labelled respectively by weights of $U(L|M)$, $U(M|N)$, $U(N|L)$. 
These modules will also have special fusion and braiding properties.

\subsection{Gauge theory construction}
Our conjecture is motivated by a four-dimensional gauge theory construction, involving local operators sitting at a $Y$-shaped junction of three interfaces between GL-twisted ${\cal N}=4$ Super Yang Mills theories with gauge groups $U(L)$, $U(M)$, $U(N)$. The conjectural $S_3$ triality symmetry follows from a conjectural  invariance of this system under permutations of the interfaces combined with $PSL(2,Z)$ S-duality transformations. Degenerate representations 
for the vertex algebra arise at the endpoints of topological line defects running along either of the three interfaces.

The full derivation of the VOA from the gauge theory setup involves a certain extension of the beautiful results of \cite{Witten:2010aa,Mikhaylov:2014aa} 
relating Chern-Simons theory and GL-twisted $\CN=4$ SYM. It extends and generalizes the results of \cite{Nekrasov:2010aa} which give a gauge-theory construction of $W_N$ conformal blocks where S-duality implies Feigin-Frenkel duality and degenerate representations arise from boundary Wilson and 't Hooft loops.  

The action of S-duality on the gauge theory setup involves both a small generalization of 
the known action of S-duality on half-BPS interfaces discussed in \cite{Gaiotto:2008aa,Gaiotto:2008ab,Gaiotto:2008ac} and a novel statement about the 
S-duality co-variance of the junctions we employ. We motivate such statement by a string theory brane web construction, 
involving a junction between an NS5 brane, a D5 brane and a $(1,1)$ fivebrane 
together with $L$,$M$ and $N$ D3 branes filling in the three angular wedges between the fivebranes. 

In this paper we will only sketch the relation to the gauge theory and brane constructions, mostly in order to produce 
instructive pictures. Instead, we will bring evidence for our conjecture from the VOA side, 
matching central charges, the structure of degenerate modules, etc. It would be interesting to 
fill in the gaps in our analysis and give a rigorous gauge theory derivation of our proposal.

For various values of parameters the VOA $Y_{L,M,N}[\Psi]$ coincides with known and well-studied examples of W-algebras. 
Our conjecture unifies a large collection of known dualities relating different constructions of these W-algebras and 
makes a variety of predictions about their representation theory. 

\subsection{Melting crystals from characters}
In the process of computing the characters of the vacuum module and of degenerate modules, we stumbled on 
a beautiful combinatorial conjecture: the characters are counting functions of 3d partitions, possibly with 
semi-infinite ends of shapes $\lambda$, $\mu$, $\nu$, restricted to lie in the difference between the standard 
positive octant and the positive octant with origin at $z=L$, $x=M$, $y=N$. If we send $L$, $M$ or $N$ to infinity, 
the characters are thus related to the topological vertex \cite{Okounkov:uq}.

Recall that $W_N$, through the AGT correspondence \cite{Alday:2009fj,Wyllard:2009qv}, plays a role in localization calculations
in $\CN=2$ gauge theory. Mathematically, this appears as an action of $W_N$ on the equivariant cohomology of $U(N)$ instanton moduli spaces 
\cite{Maulik:2012rm,Schiffmann:2012gf,Braverman:2014ys}. The $\epsilon_1$ and $\epsilon_2$ parameters appear as equivariant 
parameters on $\C^2$. Physically, one expects the $W_N$ generators to appear as BPS local operators 
in five-dimensional maximally supersymmetric $U(N)$ gauge theory. The cohomology of instanton moduli spaces may also be interpreted 
in terms of BPS bound states of N $D4$ branes and a generic number of $D0$ branes. 

The relation between the basic fivebrane junction and the toric diagram of $\C^3$, combined with some judicious string dualities, 
suggests that $Y_{L,M,N}^{\epsilon_1,\epsilon_2,\epsilon_3}$ may act on the equivariant cohomology of some generalization of instanton moduli spaces, 
involving three stacks of $D4$ branes wrapping three orthogonal $\C^2$ in $\C^3$ bound to any number of $D0$ branes. 
Such moduli spaces (and further generalizations to $\C^4$) have been introduced recently in \cite{Nekrasov:2016dn}.

More general VOAs discussed at the end of this paper may be associated to moduli spaces of D0 branes bound to D4 branes 
wrapping cycles in general toric Calabi-Yau three-folds. The characters for these general VOA's are conjecturally assembled 
from the characters of $Y_{L,M,N}$ in a manner akin to the composition of topological vertices. 

\subsection{Relation to $W_{1+\infty}$}
The computation of the vacuum character of $Y_{L,M,N}[\Psi]$ strongly suggests that all these 
VOA can be interpreted as truncations of a $W_{1+\infty}$ algebra, such as the two-parameter family of algebras 
introduced in \cite{Gaberdiel:2012aa}. The $W_{1+\infty}$ algebras have families of fully degenerate modules which are analogue to the degenerate modules we encounter,
with characters associated to the topological vertex \cite{Prochazka:2015aa}. 

These algebras admit truncations along certain families of lines in the parameter space, where the vacuum module acquire a null vector \cite{Prochazka:2014aa}. 
It is tempting to speculate that $Y_{L,M,N}[\Psi]$ coincide with such truncations. 

\subsection{Orthogonal and symplectic groups}
The addition of $O3$ orientifold plane modifies our construction and leads to vertex algebras $Y^\pm_{L,M,N}[\Psi]$ and $\tilde Y^\pm_{L,M,N}[\Psi]$ associated to  $OSp$-type supergroups. 
These include as a special case the $N=1$ super-Virasoro algebra and many other known W-algebras. We conjecture that they enjoy similar properties as the $Y_{L,M,N}[\Psi]$.

\subsection{Structure of the paper}

The paper will be structured as follows. Sections \ref{sec:branes}, \ref{sec:cs}, \ref{sec:cstwo} contain a quick review of some useful facts (some well-known, some conjectural) about interfaces and junctions in four-dimensional gauge theory, their Chern-Simons interpretation and the relation to VOAs. A reader which is only interested in the definition of our VOAs can safely skip these sections. Section \ref{sec:definition} contains the actual definition of the VOAs. 
Section \ref{sec:modules} discusses the three sets of degenerate modules exchanged by triality. Section \ref{sec:abelian}
presents in detail examples which arise from $U(1)$ gauge theory. Section \ref{sec:utwo} presents several examples which involve $U(2)$ gauge theory.
Section \ref{sec:central} contains a computation of central charges and anomalous dimensions of degenerate modules 
for general $L$,$M$,$N$. It also contains a computation of characters for vacuum modules and degenerate modules. 
Section \ref{sec:orthosymplectic} discusses the ortho-syplectic generalization of our VOAs. Section \ref{sec:junctions}
discusses the possible definition of a general class of VOAs associated to more complicated fivebrane junctions. 

\section{A quick review of $(p,q)$-fivebrane interfaces and their junctions. } \label{sec:branes}

Brane constructions in Type IIB string theory imply the existence of a family of half-BPS interfaces 
$\CB_{(p,q)}$ for 4d ${\cal N}=4$ SYM with unitary gauge groups, 
parameterized by two integers $(p,q)$ defined up to an overall sign. 
The main property of these interfaces is that they are covariant under the action of 
$PSL(2,Z)$ S-duality transformations, which act in the obvious way on the integers $(p,q)$. 
Concretely, these interfaces arise as the field theory limit of 
a setup involving two sets of D3 branes ending on a single $(p,q)$-fivebrane from opposite sides \cite{Gaiotto:2008ac,Gaiotto:2008ab,Gaiotto:2008aa}. 

Most of the $\CB_{(p,q)}$ interfaces do not admit a straightforward, weakly coupled definition. Rather, they involve some intricate 3d SCFT coupled to 
the $U(N)$ and $U(M)$ gauge theories on the two sides of the interface. The exceptions are $\CB_{(0,1)}$ and $\CB_{(1,q)}$ interfaces.

The $\CB_{(0,1)}$ interface, also denoted as a D5 interface, has a definition which depends on the relative value of $N$ and $M$: 
\begin{itemize}
\item If $N=M$, a D5 interface breaks the $U(N)_L \times U(N)_R$ gauge symmetry of the bulk theories to a diagonal $U(N)$. 
A set of 3d hypermultiplets transforming in a fundamental representation of $U(N)$ is coupled to the $U(N)$ gauge fields. 
Concretely, the 4d fields on the two sides of the interface 
are identified at the interface, up to some discontinuities involving bilinears of the 3d fields. 
\item If $N>M$, a D5 interface breaks the $U(M)_L \times U(N)_R$ gauge symmetry of the bulk theories to a block-diagonal $U(M)$. Concretely, 
$U(N)_R$ is broken to a block-diagonal $U(N-M)_R \times U(M)_R$ and $U(N-M)_R \times U(M)_L \times U(M)_R$ is further broken to the diagonal $U(M)$. 
The breaking of $U(N-M)_R$ involves a Nahm pole boundary condition of rank $N-M$. No further matter fields are needed at the interface.
\item If $M>N$, a D5 interface breaks the $U(M)_L \times U(N)_R$ gauge symmetry of the bulk theories to a $U(N)$, including a Nahm pole 
of rank $M-N$. 
 \end{itemize}
 
The $\CB_{(1,0)}$ interface, also denoted as an NS5 interface, has a uniform definition for all $N$ and $M$ \cite{Gaiotto:2008aa}:
the gauge groups are unbroken at the interface and coupled to 3d hypermultiplets transforming in a bi-fundamental 
representation of $U(M)\times U(N)$. The $\CB_{(1,q)}$ interface is obtained from a $\CB_{(1,0)}$ interface by adding $q$ units of Chern-Simons coupling 
on one side of the interface, $-q$ on the other side. 

A well known property of $(p,q)$-fivebranes is that they can form quarter-BPS webs \cite{Aharony:ab,Aharony:aa}, configurations with five-dimensional super-Poincare
invariance involving fivebrane segments and half-lines drawn on a common plane, with slope determined by the phase of their central charge. 
For graphical purposes, the slope can be taken to be $p/q$, though the actual slope depends on the IIB string coupling $\tau$ and is the phase of $p \tau + q$. 
The simplest example of brane web is the junction of three semi-infinite branes of type $(1,0)$, $(0,1)$ and $(1,1)$. It has an $S_3$ triality 
symmetry acting simultaneously on the branes and the IIB string coupling. 

Five-brane webs are compatible with the addition of extra D3 branes filling in faces of the web. These configurations preserve four super-charges, 
organized in a $(0,4)$ 2d supersymmetry algebra. One may thus consider a setup with $L$,$M$,$N$ D3 branes respectively filling the faces of the 
junction in between the $(1,1)$ and $(1,0)$ fivebranes, the $(1,0)$ and $(0,1)$ fivebranes and the $(0,1)$ and $(1,1)$ fivebranes.  

\begin{figure}[h]
  \centering
      \includegraphics[width=0.8\textwidth]{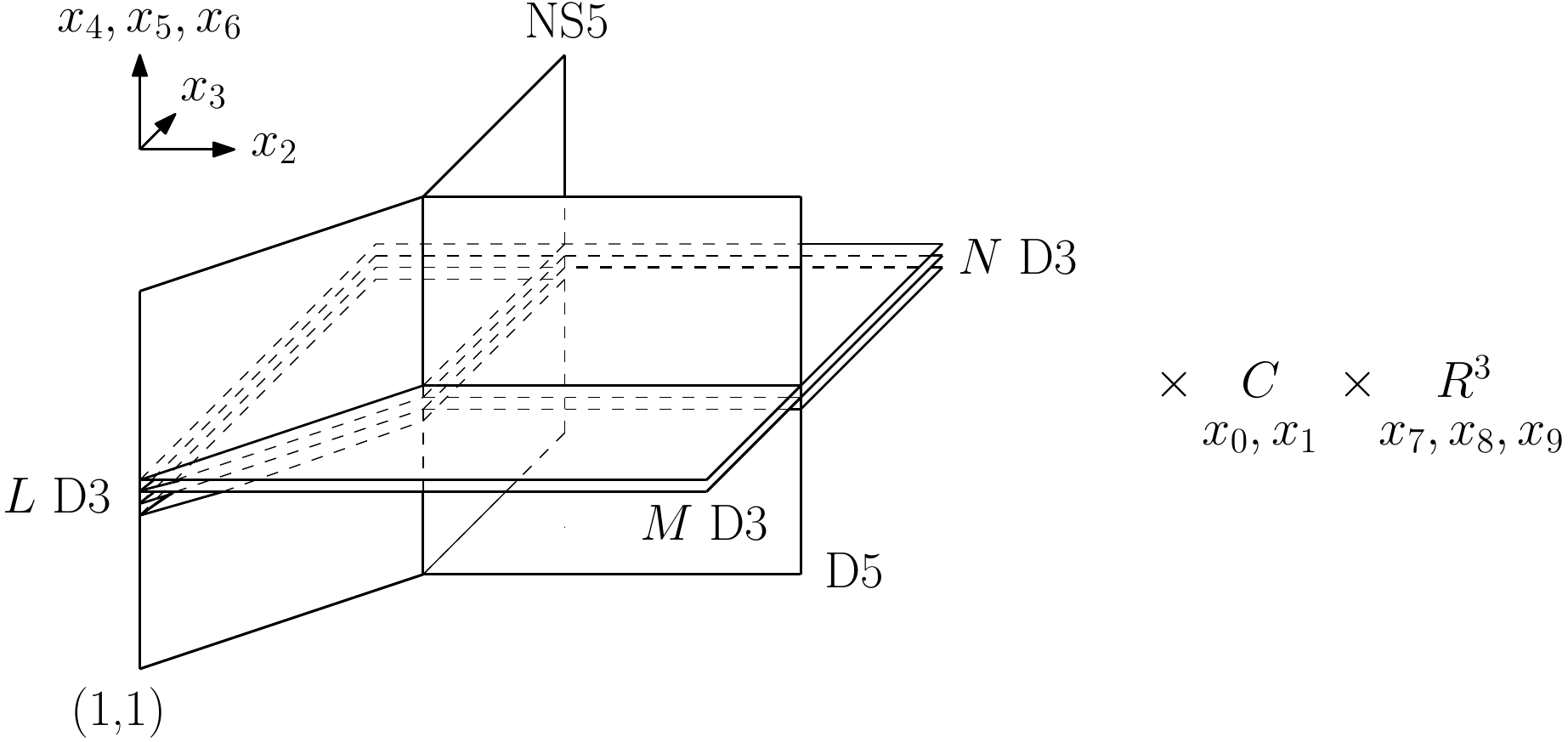}
\caption{The brane system engineering our Y-junction for four-dimensional ${\cal N}=4$ SYM. The three fivebranes extend along 
the $01456$ directions together with a ray in the $23$ plane. The stacks of D3 branes extend along the $01$ directions and fill wedges 
in the $23$ plane. Notice the $SO(3)_{456}\times SO(3)_{789}$ isometry of the system, which becomes the R-symmetry of a 
$(0,4)$ 2d super-symmetry algebra.}
\label{fig:1}
\end{figure}

The resulting configuration is invariant under $S_3$ triality transformations, the combination of permutations of $L$,$M$,$N$ and duality transformations.
The field theory limit of such a configuration, from the point of view of the D3 brane worldvolume, is that of a junction between $\CB_{(1,0)}$, 
$\CB_{(0,1)}$ and $\CB_{(1,1)}$ interfaces between $U(L)$, $U(M)$ and $U(N)$ ${\cal N}=4$ SYM defined on three wedges in the plane 
of the junction. The junction will be invariant under the triality transformations. Notice that this statement will only hold 
if we identify correctly the field theory description of the junction. 

\begin{figure}[h]
  \centering
      \includegraphics[width=0.8\textwidth]{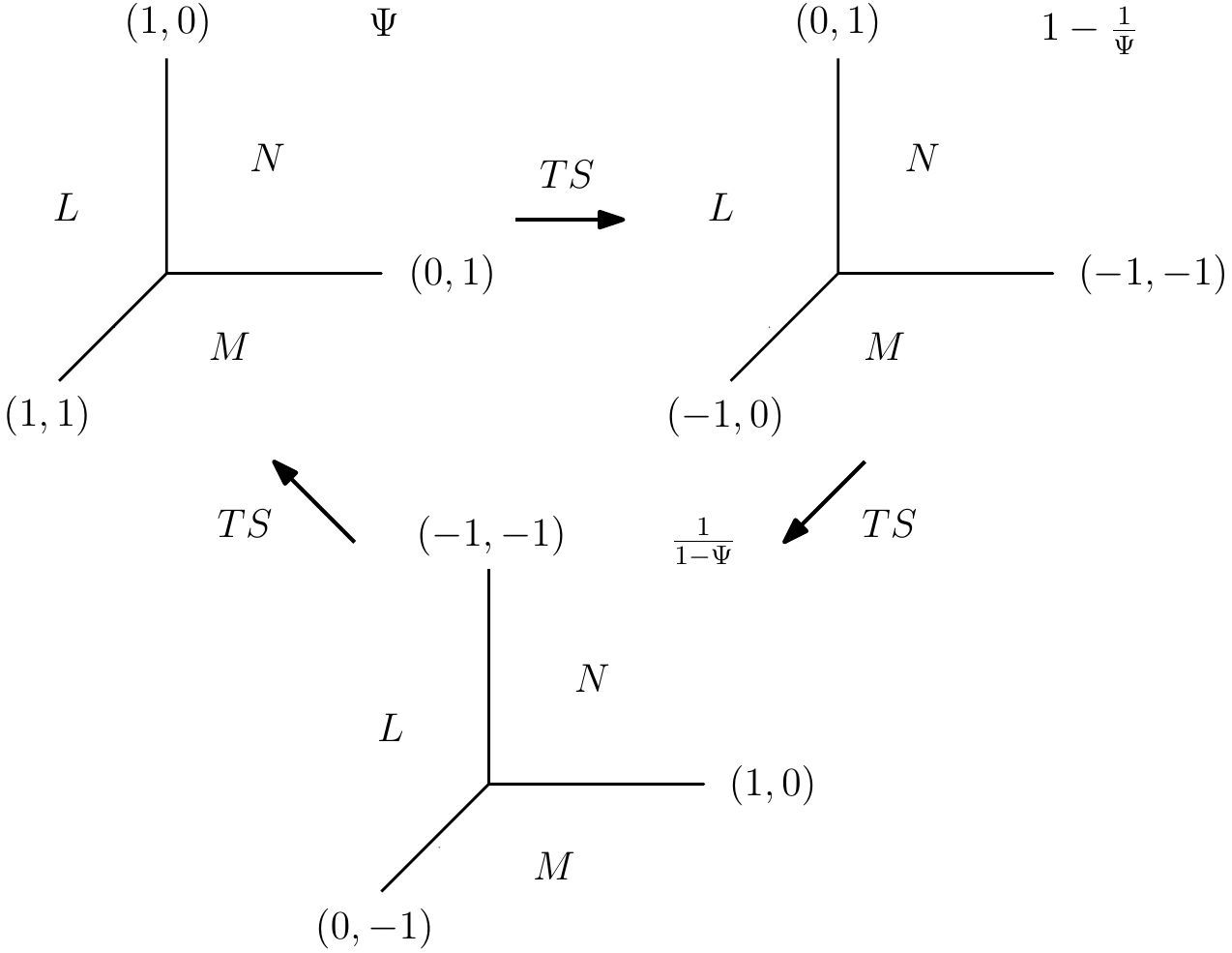}
\caption{The dualities which motivate the identification (\ref{eq:cyclic}) of the VOA 
$Y_{L,M,N}[\Psi]$, $Y_{N,L,M}[\frac{1}{1-\Psi}]$ and $Y_{M,N,L}[1-\frac{1}{\Psi}]$.}
\label{fig:2c}
\end{figure}

We will next conjecture the field theory description of the junction. Our conjecture will be motivated by some matching of 2d anomalies and 
consistency with the GL-twisted description in the next section. 

A field theory description in a given duality frame is naturally given in a very weak coupling limit. In that limit, it is natural to take the 
$(1,q)$ fivebranes to be essentially vertical in the plane, and the D5 brane to be horizontal. Thus we will describe a $T$-shaped junction, 
with an $U(L)$ gauge theory on the negative $x$ half-plane, $U(N)$ on the top right quadrant and $U(M)$ in the bottom right quadrant. 

\begin{figure}[h]
  \centering
      \includegraphics[width=0.7\textwidth]{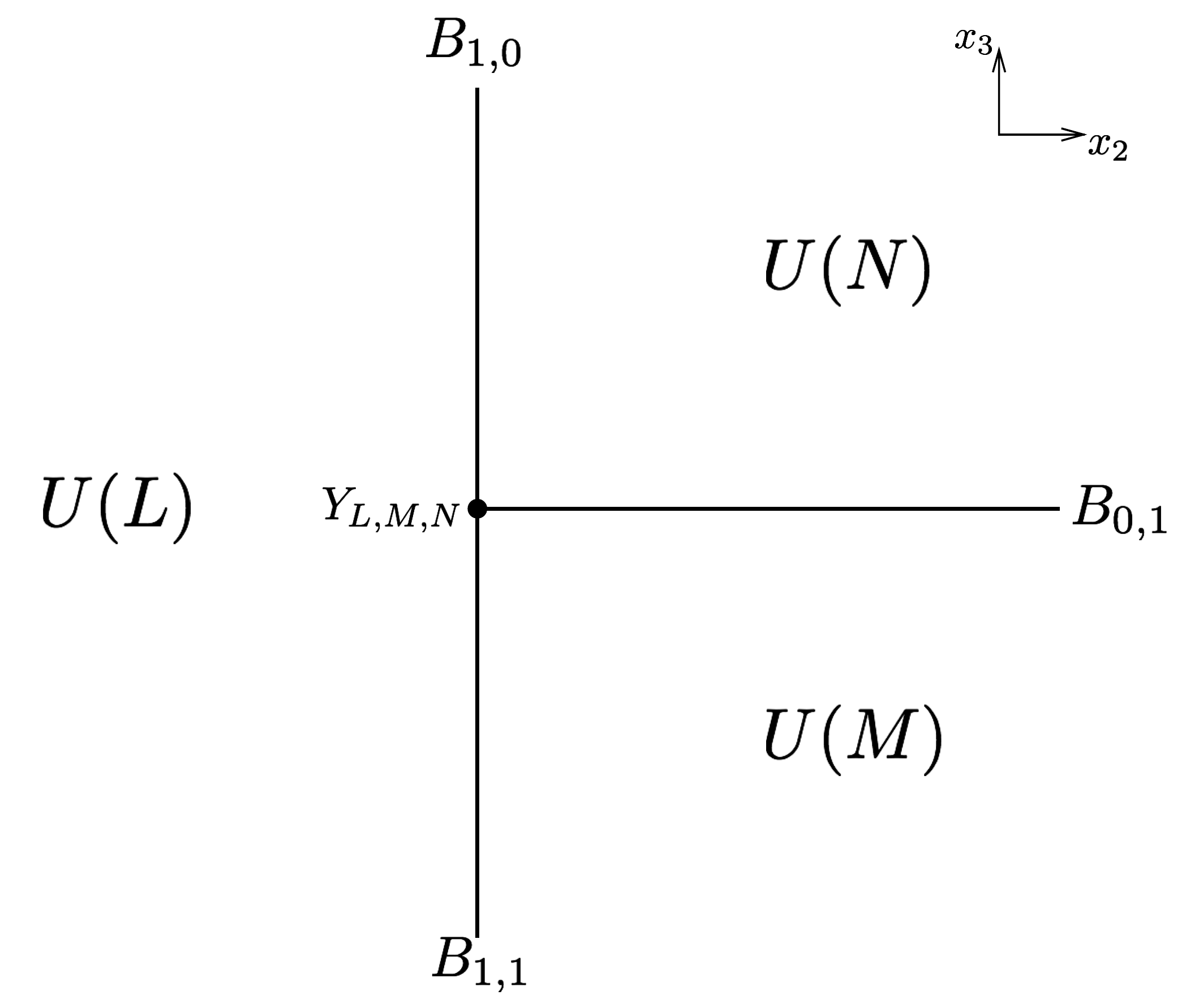}
\caption{The gauge theory image of a Y-junction on the $23$ plane. We denote the specific junction as 
$Y_{L,M,N}$. The $Y_{L,M,N}[\Psi]$ VOA will arise as a deformation of the algebra of BPS local operators at the junction.}
\label{fig:3}
\end{figure}

\subsection{The $L=0$ junctions}

\subsubsection{The $N=M$ case}
At first, we can set $N=M$ and $L=0$. That means we have a $U(N)$ gauge theory defined on the $x_2>0$ half-space with
Neumann boundary conditions at $x_2=0$. The boundary conditions are deformed by an unit of 
Chern-Simons boundary coupling on the $x_3<0$ half of the boundary. We also have an interface at $x_3=0$, 
where the $U(N)$ gauge theory is coupled to a set of $N$ 3d hypermultiplets transforming in a fundamental representation of the gauge group. 

The interface meets the boundary at $x_2=x_3=0$. 
The hypermultiplets must have some boundary conditions at the origin of the plane, preserving $(0,4)$ supersymmetry,
such as these described in Appendix \ref{app:hypers}. 
There is a known example of such a boundary condition, involving Neumann boundary conditions for all the scalar fields. 
We expect it to appear in the field theory limit of the junction setup. The choice of Neumann b.c. is natural for the following reasons: 
the relative motion of the D3 branes on the two sides of the D3 interface involves the 3d hypermultiplets acquiring a vev. The junction 
allows for such a relative motion to be fully unrestricted and thus the 3d hypermultiplets boundary conditions should be of Neumann type. 

The $(0,4)$ boundary conditions for the hypermultiplets have an important feature: they set to zero 
the left-moving half of the hypermultiplet fermions at the boundary. Such a chiral boundary condition 
has a 2d gauge anomaly which is cancelled by anomaly inflow from the boundary $U(N)$ Chern-Simons 
coupling along the negative imaginary axis. This anomaly will reappear in a similar role in the next section. 

\subsubsection{The $N\neq M$ cases}
Next, we can consider $N = M+1$ and $L=0$. Now we do not have 3d matter along the positive real axis, 
but the gauge group drops from $U(M+1)$ to $U(M)$ across the boundary. The four-dimensional  
gauginoes which belong to the $U(M+1)$ Lie algebra but not to the $U(M)$ subalgebra live on the upper right quadrant 
of the junction plane with non-trivial boundary conditions on the two sides. They may in principle contribute a 2d $U(M)$ gauge anomaly 
at the corner. It is a bit tricky to compute it, but we will recover it from a vertex algebra computation in Section 
\ref{sec:cs}. Again, we expect it to cancel the anomaly inflow from the boundary $U(M)$ Chern-Simons coupling along the negative imaginary axis. 

Similar considerations for general $N \neq M$ and $L=0$, though the positive real axis now supports a partial Nahm pole boundary condition 
along with the reduction from $U(N)$ to $U(M)$ or vice-versa. Again, we will describe the corresponding anomalies and their cancellations in Section 
\ref{sec:cs}.

\subsection{The $L>0$ junctions}
\subsubsection{The $N=M$ case}
Next, we can set $N=M$ but take general $L$. That means we have an $U(N)$ gauge theory defined on the $x_2>0$ half-space 
and an $U(L)$ gauge theory defined on the $x_2<0$ half-space. Both boundary conditions are deformed by an unit of Chern-Simons boundary coupling on the $x_3<0$ half of the boundary, with opposite signs for the two gauge groups. At the common boundary at $x_2=0$, the gauge fields are 
coupled to 3d $L \times N$ bi-fundamental hypermultiplets. We also have an interface at $x_3=0$, $x_2>0$, 
where the $U(N)$ gauge theory is coupled to a set of $N$ 3d hypermultiplets transforming in a fundamental representation of the gauge group. 

The interfaces meet at $x_2=x_3=0$. The fundamental hypermultiplets should be given a boundary condition at the origin which preserve 
$(0,4)$ symmetry. The boundary condition may involve the bi-fundamental hypermultiplets restricted to the origin and, potentially, 
extra 2d degrees of freedom defined at the junction only. 

We can attempt to define the boundary condition starting from the basic $(0,4)$ Neumann b.c. for 
the fundamental hypers and adding extra couplings at the origin. These couplings will not play a direct role for us
but help us conjecture the correct choice of auxiliary 2d degrees of freedom needed at the corner in order to reproduce the field theory 
limit of the brane setup. 

Indeed, the values at $x_2=x_3=0$ of the bi-fundamental and fundamental hypers behave as $(0,4)$ hypermultiplets and $(0,4)$ twisted hypermultiplets
respectively. There is a known way to couple these types of fields in a $(0,4)$-invariant way, but it requires 
the addition of an extra set of $(0,4)$ fields: Fermi multiplets transforming in the fundamental representation of 
$U(L)$ which can enter in a cubic fermionic superpotential with the hypermultiplets and twisted hypermultiplets \cite{Tong:2014aa}.
 
This coupling is known to occur in similar situations involving multiple D-branes ending on an NS5 brane \cite{Gomis:2016aa}.
The Fermi multiplets should arise from $D3-D5$ strings and the coupling from a disk amplitude involving 
$D3-D5$, $D3-D3'$ and $D3'-D5$ strings in the presence of an NS5 brane. 

The $U(L)$ fundamental Fermi multiplets also play another role: they consist of 2d left-moving fermions, whose anomaly 
compensates the inflow from the boundary $U(L)$ Chern-Simons coupling along the negative imaginary axis.

\subsubsection{The $N\neq M$ cases}

Next, we can consider $N = M+1$ and general $L$. Now the number of hypermultiplets along the imaginary axis 
drops from $L \times N$ to $L \times M$ across the origin of the junction's plane. We can glue together 
$M \times L$ of them according to the embedding of $U(M)$ in $U(N)$ along the real axis, 
but we need a boundary condition for the remaining $L$ hypermultiplets. 

Neumann boundary conditions for these $L$ hypermultiplets would contribute an anomaly of the wrong side to 
cancel the inflow from the boundary $U(L)$ Chern-Simons coupling along the negative imaginary axis. 
The opposite choice of boundary conditions, i.e. Dirichlet b.c. for the scalar fields, imposes the opposite boundary condition 
on the hypermultiplet's fermions and seems a suitable choice. We will thus not need to add extra 
2d Fermi multiplets at the corner.\footnote{Notice that one can obtain such boundary conditions 
starting from Neumann boundary conditions and coupling them to $(0,4)$ 2d Fermi multiplets,
which get eaten up in the process. It would be nice to follow in detail in the field theory the process of separating a $D3$ brane segment 
from the $N=M$ system and flowing to the $N=M+1$ system, by giving a vev to the fundamental hypermultiplets which induces 
a bilinear coupling of the 2d Fermi multiplets}

Similar considerations for general $N \neq M$ and general $L$, though the positive real axis now supports a partial Nahm pole boundary condition 
along with the reduction from $U(N)$ to $U(M)$ or vice-versa. The boundary conditions at the corner for the $|N-M| \times L$ 
hypermultiplets which do not continue across the corner will be affected by the Nahm pole. We will refrain from discussing them in detail here and 
focus on the GL-twisted version in the next section.

\section{From junctions in four dimensions to interfaces in analytically continued Chern-Simons theory}  \label{sec:cs}

The analysis of \cite{Witten:2010aa} gives a prescription for how to embed 
calculations in (analytically continued) Chern-Simons theory into GL-twisted 
four-dimensional $\CN=4$ Super-Yang-Mills theory. 

Concretely, a Chern-Simons calculation on a three-manifold $M_3$ maps to a four-dimensional gauge theory 
calculation on $M_3 \times {\mathbb R}^+$ with a specific boundary condition which deforms the standard supersymmetric Neumann boundary conditions. 
The (analytically continued) Chern-Simons level is related to the coupling $\Psi$ of the GL-twisted $\CN=4$ SYM as \cite{Witten:2011aa}
\begin{equation}
k+ h = \Psi
\end{equation}

It is natural to wonder about the possible implications in Chern-Simons theory of the S-duality group of the four-dimensional 
gauge theory \cite{Dimofte:2011aa}. In order to do so, we need to overcome a simple problem: supersymmetric Neumann boundary conditions
are {\it not} invariant under S-duality.  For example, they are mapped to a regular Nahm pole boundary condition 
by the $S$ element of $PSL(2,Z)$. 

Assuming that the deformed Neumann boundary conditions transform in a manner analogous to the undeformed ones,   
that means the $S$ transformation will map the Chern-Simons setup to a different setup involving a deformed Nahm pole boundary condition. 
This was a basic step in the gauge-theory description of categorified knot invariants in \cite{Witten:2011aa}.

In general, we expect the $\CB_{(p,q)}$ boundary conditions to admit deformations 
$\tilde \CB_{(p,q)}$ compatible with the GL twist, such that $\tilde \CB_{(1,0)}$ coincides with deformed Neumann boundary conditions
and $PSL(2,Z)$ duality transformations act in the obvious way on the integers $(p,q)$. 
In Appendix \ref{app:bound} we discuss briefly the deformations $\tilde \CB_{(1,0)}$ and $\tilde \CB_{(0,1)}$.

Some elements of $PSL(2,Z)$ do leave Neumann b.c. invariant: the Nahm pole boundary conditions are invariant under $T$ and 
thus Neumann b.c. are invariant under $S T^n S$. This invariance ``explains'' why the partition function of Chern-Simons theory is a function of 
$q \equiv e^{\frac{2 \pi i}{k + h}} = e^{\frac{2 \pi i}{\Psi}}$ : they are invariant under $\Psi^{-1} \to \Psi^{-1} + n$.\footnote{This statement has to be 
slightly modified for gauge groups which are not their own Langlands dual, so that the duality group is reduced to a subgroup of $PSL(2,Z)$.} 

In order to broaden the set of interesting S-duality transformations and obtain further duality relations, 
one may consider configurations involving multiple boundary conditions. That is the basic idea we pursue in this paper. 

\subsection{Corner configurations}
The general formalism of \cite{Witten:2010aa} relates a variety of analytically continued path integrals in $d$ dimensions 
and topological field theory calculations in $d+1$ dimensions, possibly including local observables or defects. 
Intuitively, observables which are functions of the $d$-dimensional fields 
will map to the same functions applied to the boundary values of $(d+1)$-dimensional fields, but modifications 
of the $d$-dimensional path integral may propagate to modifications of the $(d+1)$-dimensional bulk. Extra degrees of freedom 
added in the $d$-dimensional setup may remain at the boundary of the $(d+1)$-dimensional bulk or analytically continued to 
extra degrees of freedom in the bulk. 

A simple, rather trivial example of this flexibility is the observation that one can split off a well-defined multiple of the Chern-Simons action before 
analytic continuation, giving rise to a bulk theory with coupling $\Psi + q$ with a $\tilde \CB_{(1,q)}$ boundary condition.

A more important example is analytically continued Chern-Simons theory defined on a manifold with boundary, $M_3 = C \times {\mathbb R}^+$,
with some boundary condition $B_{3d}$. This setup will map to a calculation involving four-dimensional gauge theory on a corner geometry 
$C \times {\mathbb R}^+ \times {\mathbb R}^+$. One of the two sides of the corner will have deformed Neumann boundary condition $\tilde \CB_{(1,0)}$.
The other side will have some boundary condition $B_{4d}$ which can be derived from the boundary condition $B_{3d}$ in a systematic fashion. 
At the corner, the two boundary conditions will be intertwined by some interface which is also derived from the boundary condition $B_{3d}$.

The simplest possibility is to consider holomorphic Dirichlet boundary conditions $D^{3d}$ in Chern-Simons theory,  
given by $A_{\bar z} = 0$ at the boundary. It is well known that these boundary conditions support 
WZW currents $J = A_z |_{\partial}$ of level $\Psi - h$, given by the holomorphic part of the connection restricted to the boundary.
These boundary conditions will lift to a deformation of Dirichlet boundary conditions in SYM. 

A slightly more refined possibility is to consider a generalization of holomorphic Dirichlet boundary conditions $D_\rho^{3d}$ which is labelled by an $\mathfrak{su}(2)$ 
embedding in the gauge group \cite{Verlinde:1989ua,Bilal:1991cf}. These boundary conditions require the boundary gauge field to be a generalized oper
of type $\rho$.
They are expected to support the Vertex Operator Algebras $W_\rho[G_{\Psi - h}]$
obtained from $G_{\Psi - h}$ WZW by a Quantum Drinfeld Sokolov reduction. In particular, for the regular $\mathfrak{su}(2)$ embedding one obtains the standard W-algebras. 
These boundary conditions will lift to a deformation of the regular Nahm pole boundary conditions in SYM. We describe the deformation briefly in Appendix 
\ref{app:bound}. 

The regular Nahm pole boundary condition in SYM is precisely $\CB_{(0,1)}$.  That means the Chern-Simons setup leading to the standard W-algebras 
lifts to a corner geometry in SYM with $\tilde \CB_{(1,0)}$ on one edge and a boundary condition we expect to coincide with $\tilde \CB_{(0,1)}$
on the other edge. This is supported by the analysis of \cite{Nekrasov:2010aa}, which reduced the problem on a compact Riemann surface $C$ 
and found conformal blocks for the corresponding W-algebras. 

In particular, the symmetry of the standard W-algebras under the Feigin-Frenkel duality, which maps $\Psi \to \Psi^{-1}$, 
strongly suggests that the junction at the corner should be S-duality invariant. We expect that for a $U(N)$ gauge group the junction 
will take the form of a deformation of the corner configuration in the previous section, for $L = M = 0$.

\subsection{From junctions in four dimensions to interfaces in three-dimensions}
At this point it is natural to seek configurations in Chern-Simons theory which could be uplifted to 
a deformation of the junctions in the previous section for general $L$, $M$, $N$, 
involving $\tilde \CB_{(1,0)}$, $\tilde \CB_{(0,1)}$ and $\tilde \CB_{(1,1)}$ interfaces. 

We take the same coupling $\Psi$ uniformly in the whole plane of the $Y_{L,M,N}$ junction and the T-shaped configuration 
of Figure \ref{fig:3}: the construction of \cite{Witten:2010aa} applied along the $x_2$ direction 
maps the four-dimensional gauge theory with $\tilde \CB_{(1,0)}$ boundary conditions at $x_3>0$ 
to a Chern-Simons theory with $k+h=\Psi$ and the four-dimensional gauge theory with $\tilde \CB_{(1,1)}$ boundary conditions at $x_3<0$ 
to a Chern-Simons theory with $k+h = \Psi-1$. The interface at $x_3=0$ together with the junction 
will encode some two-dimensional interface between the two Chern-Simons theories, as described in the following. 

\subsubsection{The $L=0$ and $N>M$ cases}
At first, we can take $L=0$ and $N>M$. In order to re-produce the (deformation of the) bulk Nahm pole, we can consider the following
interface between $U(N)$ and $U(M)$ Chern-Simons theories at levels $\Psi - N$ and $\Psi - M -1$. First, we take the boundary condition $D_{N-M,1,\cdots,1}^{3d}$
for the former CS theory, defined by the same $\mathfrak{su}(2)$ embedding in $U(N)$ as the Nahm pole we need to realize, which decomposes the fundamental 
of $U(N)$ into a dimension $N-M$ irrep together with $M$ copies of the trivial representation. This boundary condition 
preserves an $U(M)$ subgroup of the $U(N)$ gauge group, which we couple to the $U(M)$ gauge fields on the other side of the interface. 

Classically, the $U(N)$ connection at the interface decomposes into blocks
\begin{equation}
A^{U(N)}|_\partial = \begin{pmatrix} *_{(N-M) \times (N-M)} & *_{(N-M) \times M} \cr *_{M \times (N-M)} & A^{U(M)}|_\partial \end{pmatrix}
\end{equation}
with one block identified with the $U(M)$ connection and the other blocks subject to the oper boundary condition. 

In order for this interface to make sense quantum mechanically, the anomaly of the $U(M)$ WZW currents in the VOA $W_{N-M,1,\cdots,1}[U(N)_{\Psi}]$ 
must be cancelled by anomaly inflow from the expected level $\Psi - M -1$ of the $U(M)$ Chern-Simons theory.\footnote{We remind the reader again that the VOA we denote as $U(N)_{\Psi}$ 
has an $SU(N)_{\Psi-N}$ current subalgebra.} We will demonstrate this fact for general $N-M$ 
later on with a detailed  Quantum Drinfeld Sokolov reduction. Essentially, the naive level $\Psi-N$ is shifted to $\Psi - M -1$
by boundary ghost contributions. For $N=M+1$ it is almost obvious: the $SU(M)$ currents in $U(N)_{\Psi}$ currents have anomaly 
$\Psi-N = \Psi - M - 1$, just as expected. See Appendix \ref{app:conventions} for further details. 

\subsubsection{The $L=0$ and $N=M$ case}
Next, we can take $L=0$ and $N=M$. Recall that the bulk setup involves fundamental hypermultiplets 
extended along the $\tilde \CB_{(0,1)}$ interface. We show in Appendix \ref{app:hypers} that the topological twist of these 3d degrees of freedom 
implements an analytically continued two-dimensional path integral for a theory of free {\it chiral symplectic bosons}.
This is another name for a $\beta \gamma$ system here the dimension of both $\beta$ and $\gamma$ 
are $1/2$, so that they can be treated on the same footing. Each hypermultiplet provides a single copy of the symplectic bosons VOA.
See Appendix \ref{app:conventions} for details on the symplectic boson VOA.  

Thus we will consider a simple interface 
between $U(N)_{\Psi - N}$ and $U(N)_{\Psi - N-1}$ Chern-Simons theories: we identify the 
gauge fields across the interface, but couple them to the theory $\mathrm{Sb}^{U(N)}$ of $N$ chiral symplectic bosons
transforming in a fundamental representation of $U(N)$. This VOA includes $U(N)$ WZW currents $\beta^a \gamma_b$ 
whose anomalies precisely compensate the shift of CS levels. We refer the reader to Appendix \ref{app:conventions} for details. 
This is just another manifestation of the corner anomaly cancellation discussed in the previous 
section. 

\subsubsection{The $L>0$ cases}
Next, we can consider general $L$. Now we will have $\tilde \CB_{(1,0)}$ and $\tilde \CB_{(1,1)}$ interfaces between $U(L)$ and $U(N)$
gauge theories. According to \cite{Mikhaylov:2014aa}, a $\tilde \CB_{(1,0)}$ interface between $U(L)$ and $U(N)$ GL-twisted gauge theories 
will map to a $U(N|L)$ Chern-Simons theory at level $\Psi - N + L$. We can thus proceed as before and consider 
interfaces between $U(N|L)$ and $U(M|L)$ Chern-Simons theories at levels $\Psi - N + L$ and $\Psi - M + L-1$. 

If $N \neq M$, the interface should be a super-group generalization $D_{N-M,1,\cdots,1|1,\cdots,1}^{3d}$ 
of the Nahm-pole-like boundary condition, preserving an $U(M|L)$ subgroup of the gauge group which can be coupled to the 
Chern-Simons gauge fields on the other side of the interface. The oper-like boundary conditions have an obvious generalization to 
supergroups, with $\mathfrak{su}(2)$ embedding into the bosonic subgroup. It would be interesting to determine the corresponding 
boundary condition on the bi-fundamental hypermultiplets present on the $\tilde \CB_{(1,0)}$ and $\tilde \CB_{(1,1)}$ interfaces.

If $N=M$, we need to generalize the symplectic boson VOA to something which admits an 
action of $U(N|L)$ with appropriate anomalies. The obvious choice is to add at the interface both $N$ copies of the chiral symplectic bosons VOA 
and $L$ chiral complex fermions. The fermions do not need to be uplifted to 3d fields and can 
instead be identified in four-dimensions with the $(0,4)$ Fermi multiplets at the origin of the junction. 

The symplectic bosons and fermions combine into a fundamental representation of $U(N|L)$ and define together a VOA $\mathrm{Sb}^{U(N|L)}$
which includes the required $U(N|L)$ WZW currents. See Appendix \ref{app:conventions} for details.
\footnote{Notice that the coupling of the $\mathrm{Sb}^{U(N|L)}$ VOA to the 3d Chern Simons theory induces a discontinuity of $A_z$ across the interface
proportional to the WZW currents in the VOA. In particular, the discontinuity of the odd currents in 
$U(N|L)$ is proportional to products of a 2d symplectic boson and a 2d fermion. This must correspond to the effect of the 
junction coupling between the $(0,4)$ Fermi multiplets and the restrictions of the fundamental and bi-fundamental 
hypermultiplets to the junction. }

\section{From Chern-Simons theory to VOA's} \label{sec:cstwo}
In the gauge theory constructions of Section \ref{sec:cs} we have encountered a variety of 
boundary conditions and interfaces for (analytically continued) Chern-Simons theory. In this section 
we discuss the chiral VOA of local operators located at these boundaries or interfaces. 

The best known example, of course, is the relation between Chern-Simons theory and WZW models \cite{Witten:1988hf}:
a Chern-Simons theory with gauge group $G$ at level $k$ defined on a half-space with appropriate orientation 
and an anti-chiral Dirichlet boundary condition $A_{\bar z} = 0$ supports at the boundary a 
chiral WZW VOA $G_k$ based on the Lie algebra of $G$, with currents $J$ of level $k$ which are proportional the restriction of $A_z$ to the boundary.\footnote{The proportionality factor is $k$.}

Dirichlet boundary conditions are associated to a full reduction of the gauge group at the boundary: 
gauge transformations must go to the identity at the boundary and constant gauge transformations at the boundary become a global symmetry 
of the boundary local operators. For our purposes, we need to consider a more general situation, 
where the gauge group is only partially reduced and may be coupled at the boundary to extra two-dimensional 
degrees of freedom. 

First, we should ask if Neumann boundary conditions could be possible, so that the gauge group is fully preserved at the boundary. 
In the absence of extra 2d matter fields, this is not possible, because of the boundary gauge anomaly inflowing from the bulk 
Chern-Simons term. We would like to claim that Neumann boundary conditions are possible if extra 2d matter fields are added, 
say a 2d chiral CFT $T^{2d}$ equipped with chiral, $G$-valued WZW currents $J^{2d}$ of level $-k - 2 h$. 

Indeed, we can produce Neumann boundary conditions by coupling auxiliary two-dimensional chiral gauge fields to the combination of $T^{2d}$ and 
standard Dirichlet boundary conditions. The level of $T^{2d}$ is chosen in such a way to cancel the naive bulk anomaly inflow when combined with the ghost contribution to the boundary anomaly. The effect if coupling two-dimensional gauge fields to VOA is well understood from the study of coset 
conformal field theory \cite{Karabali:1989dk,Hwang:1993nc}.

The VOA of boundary local operators should be built from the combination of $G_k$, $T^{2d}$ and a $bc$ ghost system $\mathrm{bc}^{\mathfrak{g}}$ valued in the Lie algebra of $G$,
taking the cohomology of the BRST charge
\begin{equation}
Q_{\mathrm{BRST}} = \oint dz \mathrm{Tr} \left[ \frac12 :b(z) [c(z),c(z)]: + c(z) (J(z) + J(z)^{2d}) \right] + Q^{2d}_{\mathrm{BRST}}
\end{equation}
which implements quantum-mechanically the expected boundary conditions $J(z) + J(z)^{2d} =0$. We included $Q^{2d}_{\mathrm{BRST}}$ to account for the possibility that $T^{2d}$ itself was defined in a BV formalism. We will denote such procedure as a $\mathfrak{g}$-BRST reduction. 

The relation to coset constructions is related to the observation that the BRST cohomology 
includes the sub-algebra of local operators in $T^{2d}$ which are local with the WZW currents $J^{2d}$.
In other words, the boundary VOA includes the current algebra {\it coset}
\begin{equation}
\frac{T^{2d}}{G_{-k-2h}}
\end{equation}
which generalizes the idea that Neumann boundary conditions support local gauge-invariant operators in $T^{2d}$.
One can envision the $\mathrm{bc}^{\mathfrak{g}}$ ghosts as cancelling out  both the $G_k$ and the $G_{-k-2h}$
currents, in a sort of Koszul quartet or Chevalley complex, leaving behind precisely the coset. 
\footnote{For example, the final central charge is the expected 
\begin{equation}
c^{2d} + c^{G_k} + c^{\mathrm{bc}^{\mathfrak{g}}}  = c^{2d} + \frac{d_G k}{k+h} - 2 d_G = c^{2d} - \frac{d_G (-k-2 h)}{-k-2 h+h} = c^{2d} - c^{G_{-k-2h}}
\end{equation}
as expected from the coset. Because
\begin{equation}
[Q_{\mathrm{BRST}}, \frac{1}{k+h} \mathrm{Tr} b(J - J(z)^{2d}) = T^{G_k} +T^{G_{-h-k}} - \mathrm{Tr} b \partial c
\end{equation}
the total stress tensor is indeed equivalent to the coset stress tensor
\begin{equation}
T^{T^{2d}} + T^{G_k} - \mathrm{Tr} b \partial c = T^{T^{2d}} - T^{G_{-h-k}}
\end{equation} }
The BRST complex above can be thought of as a sort of differential graded or derived version of a coset, 
perhaps better suited than the usual complex to non-unitary VOAs. 

Interfaces can be included in this discussion by a simple folding trick. The change in orientation maps $k \to - 2 h - k$. 
Thus we can consider a Neumann-type interface between $G_k$ and $G'_{k'}$ Chern-Simons theories coupled 
to a 2d chiral CFT $T^{2d}$ equipped with chiral, $G\times G'$-valued WZW currents of levels $-k - 2 h$ and $k'$.

The interface VOA will be the $\mathfrak{g} \oplus \mathfrak{g}'$-BRST reduction 
of the combination of $G_k$, $G'_{- k' - 2h'}$, $T^{2d}$ and $\mathrm{bc}^{\mathfrak{g} \oplus \mathfrak{g}'}$.
This implements a coset 
\begin{equation}
\frac{T^{2d}}{G_{-k-2h}\otimes G'_{k'}}
\end{equation}

The construction above has an obvious generalization to mixed boundary conditions, where the gauge group is reduced to a subgroup $H$
at the boundary and coupled with extra degrees of freedom $T^{2d}$ equipped with chiral, $H$-valued WZW currents $J^{2d}$ of level $-k_H - 2 h_H$.
The boundary VOA will consist of the $\mathfrak{h}$-BRST reduction of the combination of $G_k$, $T^{2d}$ and $\mathrm{bc}^{\mathfrak{h}}$.

The simplest example of this construction is a trivial interface between $G_k$ and $G_k$ Chern-Simons theories. 
The interface breaks the $G \times G$ gauge groups to the diagonal combination, gluing together the 
gauge fields on the two sides. The VOA of local operators should be the BRST cohomology of 
$G_k \times \hat G_{-k-2 h}$ combined with one set $\mathrm{bc}^{\mathfrak{g}}$ of $bc$ ghosts valued in the Lie algebra of $G$.
This BRST cohomology is trivial: the trivial interface in Chern-Simons theory supports no local operators except for the identity. 

A more interesting example is an interface where the $G_k$ Chern-Simons theory is coupled 
to some 2d degrees of freedom $T^{2d}$ equipped with chiral, $G$-valued WZW currents $J^{2d}$ of level $k'$. 
Notice that the levels on the two sides of the interface should be $k$ and $k+k'$. 

Then the interface VOA will be given by the BRST cohomology of $G_k \times T^{2d} \times G_{-k-k'-2 h}$
combined with one set of $b\,c$ ghosts valued in the Lie algebra of $G$. This can be interpreted as either of two conjecturally equivalent cosets 
\begin{equation}
\frac{G_k \times T^{2d}}{G_{k+k'}} \stackrel{?}{=} \frac{G_{-k-k'-2 h} \times T^{2d}}{G_{-k- 2 h}}
\end{equation}
An example of this was discussed in \cite{Costello:2016aa} with $T^{2d}$ taken to be a set of chiral fermions transforming 
in the fundamental representation of $U(N)$, resulting in the coset 
\begin{equation}
\frac{SU(N)_k \times SU(N)_1}{SU(N)_{k+1}}
\end{equation}
which is a well-known realization of a $W_N$ VOA. That construction was a source of inspiration for this project. 

A second important topic we need to discuss is the Quantum Drinfeld-Sokolov reduction $W_\rho[G_k]$ of $G_k$,
the VOA which appear at ``oper-like'' boundary conditions for a $G_k$ Chern-Simons theory, labelled by an 
$\mathfrak{su}(2)$ embedding $\rho$.

As a starting point, we may recall the construction for $SU(2)$ gauge group and the regular $\mathfrak{su}(2)$ embedding \cite{Verlinde:1989ua}. 
The classical boundary condition takes the schematic form 
\begin{equation}
A_{\bar z} = \begin{pmatrix} \frac12 a^{K}_{\bar z} & 0 \cr * &  -\frac12 a ^{K}_{\bar z}\end{pmatrix} \qquad \qquad A_z = \begin{pmatrix} * & 1 \cr * &  *\end{pmatrix} 
\end{equation}
where the $*$ denotes elements which are not fixed by the boundary condition and $\frac12 a^{K}_{\bar z}$ is the connection on the canonical bundle. 

Gauge-transformations can be used to locally gauge-fix the holomorphic connection to 
\begin{equation}
A_z = \begin{pmatrix} 0 & 1 \cr t(z) &  0\end{pmatrix} 
\end{equation}
with $t(z)$ behaving as a classical stress tensor. 

Quantum mechanically, one proceeds as follows \cite{ALEKSEEV1989719,Polyakov:1987zb,Bershadsky:1989mf}. 
The stress tensor of the usual $SU(2)_k$ WZW currents is shifted by $\partial J^3$, in such a way that 
$J^+(z)$ acquires conformal dimension $0$ and $J^-(z)$ acquires conformal dimension $2$. Furthermore, a single pair of $b \,c$ ghosts is added, allowing us to 
define a BRST charge 
\begin{equation}
Q_{\mathrm{BRST}} = \oint dz c(z) (J^+(z) -1) 
\end{equation}
enforcing the $J^+(z) = 1$ constraint. The total stress tensor 
\begin{equation}
T = T_{SU(2)_k} - \partial J^3 - b \partial c
\end{equation}
is in the BRST cohomology and generates it. It has central charge 
\begin{equation}
\frac{3 k}{k+2} - 6 k - 2 = 13 - \frac{6}{k+2} - 6(k+2) = 1 + 6 (b + b^{-1})^2
\end{equation}
with $b^2 = - (k+2)$. 

The construction generalizes as follows \cite{Bershadsky:1989mf,Feigin:1990pn,Bais:1990bs}. 
Take the $t^3$ element in the $\mathfrak{su}(2)$ embedding $\rho$. 
The Lie algebra $\mathfrak{g}$ decomposes into eigenspaces of $t^3$ as 
\begin{equation}
\mathfrak{g} = \oplus_i \mathfrak{g}_{i/2}
\end{equation}

The raising generator $t^+$ of $\rho$ is an element in $\mathfrak{g}_1$. Naively, we want to set to zero all currents of positive degree 
under $t^3$ except for the one along $t^+$, which should be set to $1$. We cannot quite do so because 
if we set to zero all currents in $\mathfrak{g}_{1/2}$ we will also set to zero their commutator, including the current along the $t^+$ 
direction. The commutator together with the projection to $t^+$ gives a symplectic form on $\mathfrak{g}_{1/2}$ 
and we are instructed to only set to zero some Lagrangian subspace $\mathfrak{g}^+_{1/2}$ in $\mathfrak{g}_{1/2}$.

Then $W_\rho[G_k]$ is defined as the BRST cohomology of a complex which is almost the same as 
the one we would use to gauge the triangular sub-group 
\begin{equation}
\mathfrak{n} = \mathfrak{g}^+_{1/2} \oplus \bigoplus_{i>\frac12} \mathfrak{g}_{i/2}
\end{equation} 
In particular, we add to $G_k$ a set of $b\,c$ ghosts valued respectively in $\mathfrak{n}$ and $\mathfrak{n}^*$ . 

The main difference is that we will shift the stress tensor by the $t^3$ component of $\partial J$ and 
by a similar ghost contribution $[b,t^3]\cdot c$ in such a way that 
currents and $b$-ghosts in $\mathfrak{g}_{i/2}$ have conformal dimension $1 - i/2$. 
This allows us to add the crucial extra term setting the $t^+$ component of $J$ to $1$: 
\begin{equation}
Q^{qDS}_{\mathrm{BRST}} = \oint dz \mathrm{Tr} \left[ \frac12 :b(z) [c(z),c(z)]: + c(z) J(z) \right] - t^+ \cdot c(z)
\end{equation}

In general, if the $\mathfrak{su}(2)$ embedding $\rho$ commutes with some subgroup $H$ of $G$, 
the WZW currents in $H$ can be corrected by ghost contributions to give $H$ WZW currents 
in $W_\rho[G_k]$. The ghost contributions will shift the level away from the value inherited from $G_k$.  

The oper-like boundary conditions can be further modified by gauging subgroups of $H$ coupled to appropriate 
2d degrees of freedom and/or promoted to interfaces by identifying the $H$ subgroup of the $G$ connection with 
an $H$ connection on the other side of the interface. This will lead to further $\mathfrak{h}$-BRST cosets involving 
$W_\rho[G_k]$ as an ingredient. 

\section{From Chern-Simons interfaces to $Y_{L,M,N}[\Psi]$} \label{sec:definition}
We now have all ingredients we need in order to provide a definition of $Y_{L,M,N}[\Psi]$.

We can start from the case $N=M$ and $L=0$. Recall that we have a $U(N)$ Chern Simons theory 
with an interface supporting a two-dimensional theory $\mathrm{Sb}^{U(N)}$ consisting of $N$ pairs of 
symplectic bosons transforming in a fundamental representation of $U(N)$. See Appendix \ref{app:conventions} for details of the
corresponding VOA.

The $SU(N)$ level of the CS theory is $\Psi - N$ on one side of the interface, $\Psi - N - 1$ on the other side. According to the 
prescription in Section \ref{sec:cstwo}, the interface VOA $Y_{0,N,N}[\Psi]$ is the $\mathfrak{u}(N)$-BRST reduction of the product 
\begin{equation}
U(N)_{\Psi} \times \mathrm{Sb}^{U(N)} \times U(N)_{- \Psi +1} \times \mathrm{bc}^{\mathfrak{u}(N)}
\end{equation}
Recall our conventions that $U(N)_\Psi$ contains $SU(N)_{\Psi-N}$ WZW currents. 
The level of the $SU(N)$ currents in $\mathrm{Sb}^{U(N)}$ is $-1$. The anomalies of the $U(N)$ WZW currents in $\mathrm{Sb}^{U(N)}$
are precisely such that they can be added to $U(N)_\Psi$ currents to give $U(N)_{\Psi-1}$ currents. We refer the reader to 
Appendix \ref{app:conventions} for details. 

Thus the VOA $Y_{0,N,N}[\Psi]$ can be identified with either of the two cosets
\begin{equation}
Y_{0,N,N}[\Psi] = \frac{U(N)_{\Psi} \times \mathrm{Sb}^{U(N)}}{U(N)_{\Psi -1}}= \frac{\mathrm{Sb}^{U(N)} \times U(N)_{- \Psi +1} }{U(N)_{-\Psi}}
\end{equation}
Notice that the BRST definition is symmetric under $\Psi \leftrightarrow 1-\Psi$. 

Next, we can consider the case $N=M+1$ and $L=0$. Recall that we have an interface between 
a $U(N)$ Chern Simons theory and a $U(M)$ Chern-Simons theory defined simply by reducing the $U(N)$ 
gauge symmetry to $U(M)$ at the boundary and identifying with the gauge symmetry on the other side. 

The $SU(N)$ level of the CS theory is $\Psi - N$ on one side of the interface, $\Psi - M - 1=\Psi-N$ on the other side. 
According to the prescription in Section \ref{sec:cstwo}, the interface VOA $Y_{0,N-1,N}[\Psi]$ is the $\mathfrak{u}(N-1)$-BRST reduction of the product 
\begin{equation}
U(N)_{\Psi} \times U(N-1)_{- \Psi +1} \times \mathrm{bc}^{\mathfrak{u}(N-1)}
\end{equation}
Notice that the $U(N)_\Psi$ currents precisely contain a block-diagonal $U(N-1)_{\Psi-1}$ WZW subalgebra. See Appendix \ref{app:conventions} for details. 

Thus the VOA $Y_{0,N-1,N}[\Psi]$ can be identified with  the coset
\begin{equation}
Y_{0,N-1,N}[\Psi] = \frac{U(N)_{\Psi} }{U(N-1)_{\Psi - 1}}
\end{equation}

Similarly, for $N=M-1$ and $L=0$ we would define $Y_{0,N+1,N}[\Psi]$ as the $\mathfrak{u}(N)$-BRST reduction of the product
\begin{equation}
U(N)_{\Psi} \times U(N+1)_{- \Psi +1} \times \mathrm{bc}^{\mathfrak{u}(N)}
\end{equation}
Equivalently, the coset
\begin{equation}
Y_{0,N+1,N}[\Psi] = \frac{U(N+1)_{-\Psi +1} }{U(N)_{-\Psi}}
\end{equation}

For general $N>M$ and $L=0$, we need to reduce the $U(N)$ gauge symmetry at the interface 
by the oper-like boundary condition involving the $\mathfrak{su}(2)$ embedding 
which decomposes the fundamental representation of $U(N)$ into an $(N-M)$-dimensional 
irrep and $M$ copies of the trivial irrep. The residual $U(M)$ symmetry can be identified with 
the gauge symmetry on the other side. 

First, we need to make sure that the Chern-Simons levels work out. The block-diagonal $U(M)_{\Psi-N+M}$ subalgebra of $U(N)_\Psi$ 
should be combined with the ghost contributions to give BRST-closed total $U(M)$ currents. 
It is easy to see that the triangular subalgebra $\mathfrak{n}$ includes $N-M-1$ copies of the fundamental representation of $U(M)$ and
several copies of the trivial representation. 
Each set of ghosts transforming in the fundamental representation of $U(M)$ 
will shift by $1$ unit the level of the $U(M)$ WZW currents. Thus $W_{N-M, 1, \cdots, 1}[U(N)_{\Psi}]$
has an $U(M)_{\Psi-1}$ subalgebra. See Appendix \ref{app:conventions} for details.

We are ready to define $Y_{0,M,N}[\Psi]$ as the $\mathfrak{u}(M)$ BRST reduction of the product
\begin{equation} 
W_{N-M, 1, \cdots, 1}[U(N)_{\Psi}] \times U(M)_{-\Psi +1} \times \mathrm{bc}^{\mathfrak{u}(M)}
\end{equation}
i.e. the coset 
\begin{equation} 
Y_{0,M,N}[\Psi] = \frac{W_{N-M, 1, \cdots, 1}[U(N)_{\Psi}]}{U(M)_{\Psi -1}}
\end{equation}
We can combine the quantum DS reduction and the $\mathfrak{u}(M)$-BRST coset into a
single deformed $\left(\mathfrak{n} \oplus \mathfrak{u}(M)\right)$-BRST quotient of the product 
\begin{equation}
U(N)_{\Psi} \times U(M)_{-\Psi +1}\times \mathrm{bc}^{\mathfrak{n}\oplus \mathfrak{u}(M)}
\end{equation}

Similarly, for $N<M$ and $L=0$ we would get a BRST coset of the form 
\begin{equation} 
Y_{0,M,N}[\Psi] = \frac{W_{M-N, 1, \cdots, 1}[U(M)_{-\Psi+1}]}{U(N)_{-\Psi}}
\end{equation}
Notice that these definitions are trivially symmetric under $N \leftrightarrow M$ together with $\Psi \leftrightarrow 1-\Psi$.

For general $L$ we need to upgrade all the constructions described above to super-groups. 
The quantum DS reduction for WZW VOA's based on super-Lie algebras 
works in the same way as for standard Lie algebras, except that fermionic ghosts are replaced by bosonic ghosts in the odd components of $\mathfrak{n}$ \cite{2003CMaPh.241..307K}.

According to the prescription in Section \ref{sec:cstwo}, the interface VOA $Y_{L,N,N}[\Psi]$ is the $\mathfrak{u}(N|L)$-BRST reduction of the product 
\begin{equation}
U(N|L)_{\Psi} \times \mathrm{Sb}^{U(N|L)} \times U(N|L)_{- \Psi +1} \times \mathrm{bc}^{\mathfrak{u}(N|L)}
\end{equation}
The $\mathrm{Sb}^{U(N|L)}$ consists of $N$ sets of symplectic bosons and $L$ complex fermions. 
The $U(N)$ WZW currents $J^{\mathrm{Sb}^{U(N|L)}}$ are precisely such that they can be added to $U(N|L)_\Psi$ currents to give $U(N|L)_{\Psi-1}$ currents. We refer the reader to Appendix \ref{app:conventions} for details. 

Thus the VOA $Y_{L,N,N}[\Psi]$ can be identified with either of the two cosets
\begin{equation}
Y_{L,N,N}[\Psi] = \frac{U(N|L)_{\Psi} \times \mathrm{Sb}^{U(N|L)}}{U(N|L)_{\Psi -1}}= \frac{\mathrm{Sb}^{U(N|L)} \times U(N|L)_{- \Psi +1} }{U(N|L)_{-\Psi}}
\end{equation}
Notice that the BRST definition is symmetric under $\Psi \leftrightarrow 1-\Psi$. 

For $N>M$ we will use a quantum DS reduction of the super-groups. We will define $Y_{L,M,N}[\Psi]$ as the $\mathfrak{u}(M|L)$-BRST reduction of the product
\begin{equation} W_{N-M, 1, \cdots, 1|1,\cdots,1}[U(N|L)_{\Psi}] \times U(M|L)_{-\Psi +1} \times \mathrm{bc}^{\mathfrak{u}(M|L)}
\end{equation}
i.e. the coset 
\begin{equation} 
Y_{L,M,N}[\Psi] = \frac{W_{N-M, 1, \cdots, 1|1,\cdots,1}[U(N|L)_{\Psi}]}{U(M|L)_{\Psi -1}}
\end{equation}
We can combine the quantum DS reduction and the $\mathfrak{u}(M|L)$-BRST coset into a
single deformed $\left(\mathfrak{n} \oplus \mathfrak{u}(M|L)\right)$-BRST quotient of the product 
\begin{equation}
U(N|L)_{\Psi} \times U(M|L)_{-\Psi +1}\times \mathrm{bc}^{\mathfrak{n}\oplus \mathfrak{u}(M|L)}
\end{equation}

Similarly, for $N<M$ we would get a BRST coset of the form 
\begin{equation} 
Y_{L,M,N}[\Psi] = \frac{W_{M-N, 1, \cdots, 1|1,\cdots,1}[U(M|L)_{-\Psi+1}]}{U(N|L)_{-\Psi}}
\end{equation}
Notice that these definitions are trivially symmetric under $N \leftrightarrow M$ together with $\Psi \leftrightarrow 1-\Psi$.

\section{Modules} \label{sec:modules}
\subsection{Degenerate modules}
The standard $W_N$ algebras have maximally degenerate modules $M_{\lambda, \mu}$ labelled by a pair of dominant weights of $\mathfrak{su}(N)$. The Feigin-Frenkel duality $b \to b^{-1}$ exchanges the role of the two weights. 

These modules are expected to arise in the gauge theory construction from local operators at 
the corner which are attached to a boundary Wilson line of weight $\lambda$ along the NS5 boundary 
and a boundary 't Hooft line of weight $\mu$ along the D5 boundary. These two line defects are correspondingly 
exchanged by S-duality. 

If we denote $W_\lambda = M_{\lambda,0}$ and $H_\mu = M_{0,\mu}$, then the following facts hold true:
\begin{itemize}
\item The $W_\lambda$ have the same fusion rules 
\begin{equation}
W_\lambda \times W_{\lambda'} \sim \sum_{\lambda''} c^{\lambda''}_{\lambda, \lambda'} W_{\lambda''}
\end{equation}
as finite-dimensional $SU(N)$ irreps. They have non-trivial braiding and fusion matrices which are closely related to these of $SU(N)_{\Psi - N}$. 
Conformal blocks with $W_\lambda$ insertions satisfy BPZ differential equations.
\item The $H_\mu$ also have the same fusion rules 
\begin{equation}
H_\mu \times H_{\mu'} \sim \sum_{\mu''} c^{\mu''}_{\mu, \mu'} H_{\mu''}
\end{equation}
as finite-dimensional $SU(N)$ irreps. They have non-trivial braiding and fusion matrices which are closely related to these of $SU(N)_{\Psi^{-1} - N}$. Conformal blocks with $H_\mu$ insertions satisfy BPZ differential equations
\item The $W_\lambda$ and $H_\mu$ vertex operators are almost mutually local. They are local if we restricts the weights to those of GL-dual groups. The fuse in a single channel $M_{\lambda, \mu}$. 
\end{itemize}

We expect analogous statements for maximally degenerate modules of $Y_{L,M,N}[\Psi]$, involving local operators sitting at the end to three boundary lines, one for each 
component of the gauge theory junction. These modules should thus carry three labels, permuted by the $S^3$ triality symmetry, corresponding to the possible labels 
of BPS line defects living on the $\tilde \CB_{(p,q)}$ boundary conditions. 
It is known \cite{Mikhaylov:2014aa} that such line defects include analogues of Wilson lines, labelled by data akin to dominant weights
of $U(N|L)$, $U(L|M)$, $U(M|N)$ respectively. 

In particular, we expect the following to be true: if we denote $W_\lambda$, $H_\mu$ and $D_\sigma$ the modules associated to either type of boundary lines
\begin{itemize}
\item The $W_\lambda$ should have the same fusion rules as finite-dimensional $U(N|L)$ irreps, with appropriate non-trivial braiding and fusion matrices and BPZ-like differential equations. 
\item The $H_\mu$ should have the same fusion rules as finite-dimensional $U(M|N)$ irreps, with appropriate non-trivial braiding and fusion matrices and BPZ-like differential equations.
\item The $D_\sigma$ should have the same fusion rules as finite-dimensional $U(L|M)$ irreps, with appropriate non-trivial braiding and fusion matrices and BPZ-like differential equations.
\item The $W_\lambda$, $H_\mu$ and $D_\sigma$ vertex operators should be mutually local and fuse together into a single channel $M_{\lambda, \mu,\nu}$
\end{itemize}

\begin{figure}[h]
  \centering
      \includegraphics[width=0.25\textwidth]{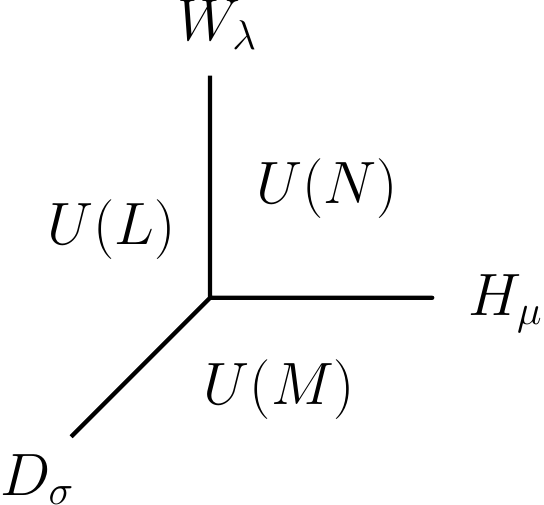}
\caption{Modules $W_{\lambda},H_{\mu},D_{\sigma}$ associated to the three classes of boundary lines.}
\label{fig:2b}
\end{figure}

In the coset constructions for $Y_{L,M,N}[\Psi]$, the data for $U(N|L)$ and $U(L|M)$ representations 
appears rather naturally, as one may implement the BRST reduction starting from Weyl modules 
of the current algebras built from irreducible representations of the zeromode algebra, up to subtleties in relating weights and representations for 
supergroups. 

The data of $U(M|N)$ is much harder to uncover, though in principle it can be done with the help of the gauge theory 
description in \cite{Mikhaylov:2014aa}. In general, the line defect along the D5 interface 
will map to some disorder local operator at the interface between Chern-Simons theories. 

We will analyze some basic examples through the rest of the paper and then come back to the general story in Section 
\ref{sec:central}. 

\subsection{Other modules}
Another natural enrichment of the four-dimensional gauge theory setup is to include surface defects 
which fill the whole wedge between two interfaces. Gukov-Witten surface defects are labelled by a 
Levi subgroup of the gauge group and have non-trivial couplings. In the GL-twisted theory the couplings are
essentially valued on products of elliptic curves with modular parameter $\Psi$, up to some discrete identifications. 

Upon reduction to 3d, these GW surface defects are known to implement the insertion of analytically continued 
versions of Wilson loops in the Chern-Simons theory away from integral weights. 

At the intersection with the junction, the surface defects will produce a variety of modules for 
the $Y_{L,M,N}[\Psi]$ algebras. We leave a general analysis to future work. 

\section{Abelian examples} \label{sec:abelian}

In this section we discuss the VOA associated to junctions in $U(1)$ gauge theory. 
The building blocks of the corresponding vertex algebras will be $U(1)$ and $U(1|1)$ current algebras together with symplectic bosons. 

\subsection{$U(1)_\Psi$}

The simplest example is a $U(1)$ gauge theory defined on the upper right quadrant of the plane, with deformed Neumann and Dirichlet boundary conditions at the two sides. 

\begin{wrapfigure}{l}{0.2\textwidth}
\vspace{-20pt}
  \begin{center}
      \includegraphics[width=0.175\textwidth]{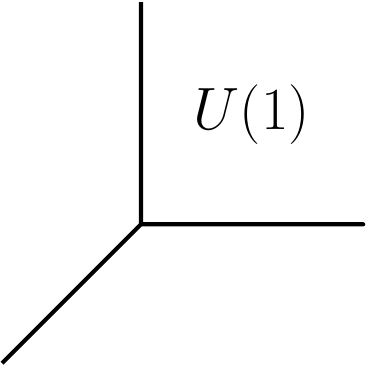}
\end{center}
\end{wrapfigure}

This four-dimensional setup can be related first to three-dimensional analytically continued $U(1)$ Chern-Simons theory at level $\Psi$
with standard boundary conditions. 

In turns, these boundary conditions support a $U(1)_\Psi$ current algebra, with OPE
\begin{equation}
J_\Psi(z) J_\Psi(w) \sim \frac{\Psi}{(z-w)^2}
\end{equation}
and Sugawara stress tensor
\begin{equation}
T_{U(1)_\Psi} = \frac{1}{2 \Psi} J_\Psi J_\Psi
\end{equation}
of central charge $1$. 

Thus we define 
\begin{equation}
Y_{0,0,1}[\Psi] \equiv U(1)_\Psi
\end{equation}
Notice that the level of a $U(1)$ current is a mere formality. The actual effect of the bulk CS coupling is to determine 
which boundary vertex operators arise at the end of a bulk Wilson loop of charge $n$: they will be vertex 
operators of charge $n$ under $J_\Psi$. We will come back to that momentarily. 

There are two other inequivalent definition of the junction vertex algebra which must give us the same answer as $Y_{0,0,1}[\Psi]$:
$Y_{0,0,1}[\Psi^{-1}]$ and $Y_{1,0,0}[\frac{1}{1-\Psi}]$. The other three configurations $Y_{1,0,0}[\frac{\Psi}{\Psi-1}]$,$Y_{0,1,0}[1-\Psi]$,$Y_{0,1,0}[1-\Psi^{-1}]$
do not produce a new definition. 

The second definition, $Y_{0,0,1}[\Psi^{-1}]$ gives obviously a $U(1)_{\Psi^{-1}}$ current algebra. We can identify it with $Y_{0,0,1}[\Psi]$ by the trivial rescaling 
\begin{equation}
J_\Psi = \Psi J_{\Psi^{-1}}
\end{equation}

\begin{wrapfigure}{l}{0.2\textwidth}
\vspace{-20pt}
  \begin{center}
      \includegraphics[width=0.175\textwidth]{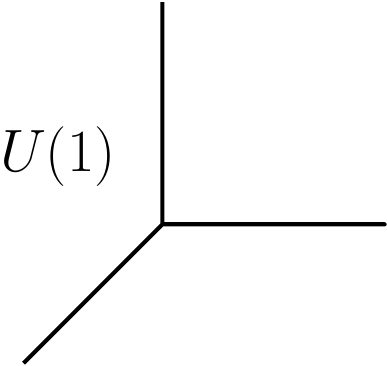}
\end{center}
\end{wrapfigure}

The third definition $Y_{1,0,0}[\frac{1}{1-\Psi}]$ is more intricate. Recall that the three-dimensional setup involves a $U(1)$ Chern-Simons theory 
coupled to a single complex free fermion at a two-dimensional interface. According to our prescription, 
the resulting VOA is the $\mathfrak{u}(1)$-BRST coset of
\begin{equation} 
U(1)_{-\frac{1}{1-\Psi}} \times U(1)_{-\frac{\Psi}{\Psi-1}} \times \mathrm{Ff}^{U(1)} \times \mathrm{bc},
\end{equation}
with $\mathrm{Ff}^{U(1)}$ being the VOA of a single complex free fermion with generators $(\psi, \chi \equiv \psi^\dagger)$.
 
The level are such that total $U(1)$ current
\begin{equation}
J_{\mathrm{tot}} = J_{-\frac{1}{1-\Psi}} + J_{-\frac{\Psi}{\Psi-1}} + J_{\psi \chi}
\end{equation}
has level $0$. The BRST charge is 
\begin{equation}
Q_{BRST} = \oint c J_{\mathrm{tot}}
\end{equation}
In a more conventional language, we would express $Y_{1,0,0}[\frac{1}{1-\Psi}] \equiv Y_{1,0,0}[\frac{\Psi}{\Psi-1}]$
as either of the two cosets
\begin{equation}
Y_{1,0,0}[\frac{1}{1-\Psi}] \equiv \frac{U(1)_{-\frac{1}{1-\Psi}} \times \mathrm{Ff}^{U(1)}}{U(1)_{-\frac{\Psi}{1-\Psi}}} = \frac{U(1)_{\frac{\Psi}{1-\Psi}} \times \mathrm{Ff}^{U(1)}}{U(1)_{\frac{1}{\Psi-1}}}
\end{equation}

The linear combination 
\begin{equation}
J_\Psi = \Psi J_{-\frac{1}{1-\Psi}} + J_{-\frac{\Psi}{\Psi-1}} 
\end{equation}
is BRST close, as it has trivial OPE with the total $U(1)$ current. It has level $\Psi$. We expect it to generate the BRST cohomology
and coincide with the current which appears in the $Y_{0,0,1}[\Psi]$ definition. 

For completeness, we can compute the vacuum character in the different descriptions. 
The character for a single $U(1)$ current is 
\begin{equation}
\chi_{Y_{0,0,1}}(q)\equiv \chi_{U(1)}(q) = \frac{1}{\prod_{n>0} (1-q^n)}
\end{equation}
The character for the $\mathfrak{u}(1)$-BRST coset can be computed as a Witten index, though one has to 
deal separately with the $c$ ghost zeromode and implement ``by hand'' the projection 
on the global gauge singlets by a contour integral:
\begin{equation}
\chi_{Y_{1,0,0}}(q) \equiv \oint \frac{dz}{2 \pi i z} \chi_{U(1)}^2(q) \chi'_{\mathrm{bc}}(q) \chi_{\mathrm{Ff}^{U(1)}}(q;z)
\end{equation}
The ghosts (excluding the $c$ zeromode) cancel precisely the $U(1)$ currents contributions. 
Because of the well-known relation 
\begin{equation}
 \chi_{\mathrm{Ff}^{U(1)}}(q;z) \equiv \prod_{n\geq 0} (1 - z q^{n+\frac12})(1 - z^{-1} q^{n+\frac12}) = \frac{1}{\prod_{n>0} (1-q^n)}\sum_n (-z)^n q^{\frac{n^2}{2}}
\end{equation}
then we recover the desired 
\begin{equation}
\chi_{Y_{1,0,0}}(q)\equiv \oint \frac{dz}{2 \pi i z}\frac{\sum_n (-z)^n q^{\frac{n^2}{2}}}{\prod_{n>0} (1-q^n)} = \frac{1}{\prod_{n>0} (1-q^n)}
\end{equation}

\subsubsection{``Degenerate'' modules}

Going back to $Y_{0,0,1}[\Psi]$, the local operators which sit at the end of a boundary Wilson line in the Neumann boundary 
correspond to endpoints of charge $n$ Wilson lines in the 3d CS theory and thus to charge $n$ ``electric'' vertex operators for $J_\Psi$:
\begin{equation}
W_n(z) = V^{\Psi}_{n} \equiv e^{i \frac{n}{\Psi} \phi(z)}  
\end{equation}
of conformal dimension $\Delta_{W_n} = \frac{1}{2 \Psi} n^2$. Here $\phi$ is the bosonization of the current $J_\Psi =- i \partial \phi$.

The dyonic operators are absent in this example. On the other hand, the gauge theory description of Abelian boundary 't Hooft operators 
is simple enough that we can attempt to identify the corresponding three-dimensional configuration and then define directly in $Y_{0,0,1}[\Psi]$
the corresponding junction local operators $H_m$.

We expect the boundary 't Hooft lines on the Dirichlet boundary to map to boundary local operators in the Chern-Simons theory
defined by a Hecke modification of the boundary condition on $A_{\bar z}=0$. In turn, these should map to ``magnetic'' vertex operators 
\begin{equation}
H_m(z) = V^{\Psi}_{m \Psi} \equiv e^{i m \phi(z)} 
\end{equation}
of conformal dimension $\Delta_{H_m} = \frac{\Psi}{2} m^2$ and charge $m \Psi$ under $J_\Psi$. 

The identification is motivated by the observation that these operators induce the correct classical singularity 
in the boundary value of the connection $A_z|_\partial = \Psi^{-1} J_\Psi$ and that they induce 
zeroes or poles of order $n m$ in the expectation values of vertex operators of electric charge $n$. 

This answer is perfectly compatible with the S-dual description $Y_{0,0,1}[\Psi^{-1}]$: under the identification $J_\Psi = \Psi J_{\Psi^{-1}}$ we see that
$H_m$ are electric vertex operators of charge $m$ for $J_{\Psi^{-1}}$ and $W_n$ are magnetic vertex operators of charge $n$ for $J_{\Psi^{-1}}$.

It is a bit more interesting to look at the realization of these vertex operators in $Y_{1,0,0}[\frac{1}{1-\Psi}]$. 
Both sets of boundary line defects in four dimensions map to Wilson lines in the Chern Simons theories on either sides of the interface. 
Because the $U(1)$ gauge symmetry is unbroken at the interface, a Wilson line of charge $n$ ending on the interface from either side will 
need to end on a local operator of charge $n$ in the free fermion interface theory. 

In the BRST construction, that is the BRST close combination of 
a charge $n$ vertex operator for either $U(1)$ theory and and the simplest charge $-n$ operator $O_{-n}$ built from the fermions: 
\begin{align}
W_n &\equiv V_n^{-\frac{\Psi}{\Psi-1}} O_{-n} \cr
H_m &\equiv V_m^{-\frac{1}{1-\Psi}} O_{-m}
\end{align}
The dimensions of these BRST close representatives are indeed 
\begin{equation}
\Delta_{W_n} = \frac{n^2}{2} (\Psi^{-1} -1) + \frac{n^2}{2} \qquad \qquad \Delta_{H_m} = \frac{m^2}{2} (\Psi -1) + \frac{m^2}{2} 
\end{equation} 
as expected. 

Notice that $W_n$ fuse as $W_n \times W_{n'} \sim W_{n+n'}$ and have OPE singularities controlled by $n n' \Psi^{-1}$. 
Similarly, $H_m$ fuse as $H_m \times H_{m'} \sim H_{m+m'}$ and have OPE singularities controlled by $m m' \Psi$. 
On the other hand, $W_n$ and $H_m$ are mutually local and fuse to 
\begin{equation}
M_{m,n}(z) = e^{i \left(m+ \frac{n}{\Psi} \right)\phi(z)}  
\end{equation}
of conformal dimension $\Delta_{n,m} = \frac{\Psi}{2} m^2+\frac{1}{2\Psi} n^2 + n m$.

Notice as well that these vertex operators define perfectly normal modules for the $U(1)_\Psi$ VOA, with no null vectors. 
The moniker ``degenerate'' here only indicates that they play an analogous role as the degenerate modules 
for the $W_N$ algebra. 

\subsubsection{General modules}
General vertex operators for $U(1)_\Psi$ arise from 4d gauge theory configurations involving a Gukow-Witten 
surface defect, which descends to a monodromy defect in the 3d Chern-Simons theory and then to the generic vertex operator 
\begin{equation}
S_p(z) = e^{i \frac{p}{\Psi} \phi(z)} 
\end{equation}
with the momentum $p$ being the complex combination of the GW defect parameters which survives the GL twist. This has dimension $\Delta_{S_p} = \frac{p^2}{2\Psi}$. 

Notice that the $p$ parameter is not periodic: although the parameters of the GW defect are valued in a torus of modular parameter $\Psi$, 
$p$ encodes an extra choice of boundary conditions on the two sides of the corner. We can change these boundary conditions by 
fusing the surface defect boundary with boundary Wilson or 't Hooft lines, which results in shifts of $p$ by $n$ or $m \Psi$:
\begin{equation}
W_n \times S_p \sim S_{p+n} \qquad \qquad H_m \times S_p \sim S_{p+m \Psi} 
\end{equation} 
The general vertex operator reduces to the ``degenerate'' ones when we set $p = n \Psi + m$, the values at which the surface defect in the gauge theory description disappears. 

Under $\Psi \to \Psi^{-1}$ the parameter $p$ transforms as $p \to p \Psi^{-1}$, as expected from the duality properties of the surface defect. 

In order to realize analogous vertex operators in $Y_{1,0,0}[\frac{1}{1-\Psi}]$ we can combine 
an operator of momentum $\tilde p$ for $U(1)_{-\frac{1}{1-\Psi}}$ and $- \tilde p$ for $U(1)_{-\frac{\Psi}{\Psi-1}}$.
This gives a BRST closed operator of momentum $p = (\Psi - 1) \tilde p$ for $J_\Psi$. 
The 3d picture is that of a monodromy defect of parameter $\tilde p$ crossing the interface. 

\subsection{Two $U(1)$ corners}

This is a very instructive example. The final answer for the junction VOA is simple, but it is realized in a very non-trivial manner in 
all duality frames. For clarity, we will anticipate here the final answer and then detail the derivation in the three alternative duality frames. 

We claim that the junction VOA is the product 
\begin{equation}
Y_{0,1,1}[\Psi] = U(1)_{\Psi^{-1} -1} \times \mathrm{Sf}_0
\end{equation}
where $\mathrm{Sf}_0$ is a vertex algebra which can be compactly defined as the charge $0$ 
subalgebra of a vertex algebra $\mathrm{Sf}$ defined by two ``fermionic currents'' $x(z)$, $y(z)$ 
of dimension $1$, OPE
\begin{equation}
x(z) y(w) \sim \frac{1}{(z-w)^2}
\end{equation}
and charges $\pm 1$ under a global $U(1)_o$ symmetry. \footnote{Equivalently, $\mathrm{Sf}$ can be defined as a $PSU(1|1)$ current algebra.}

The vertex algebra $\mathrm{Sf}$ appears in a variety of bosonization constructions, including the bosonization of 
symplectic bosons and the bosonization of $U(1|1)$ WZW models. That is how it will appear in our construction. 
We refer the reader to Appendix \ref{app:conventions} for definitions and references. 

The VOA $\mathrm{Sf}_0$ has two natural classes of modules: 
\begin{itemize}
\item The other charge sectors $\mathrm{Sf}_n$ in $\mathrm{Sf}$. They have highest weight vectors of conformal dimension $\frac{n^2 + |n|}{2}$. 
\item The charge $0$ sector $V^{xy}_{\lambda}$ of the twisted modules for $\mathrm{Sf}$. 
They have highest weight vectors of conformal dimension $\frac{\lambda^2 - \lambda}{2}$ which induce 
singularities $z^{- \lambda}$ in $x(z)$ and $z^{\lambda-1}$ in $y(z)$. 
\end{itemize}
with fusion rules 
\begin{align}
\mathrm{Sf}_n \times \mathrm{Sf}_m &\sim \mathrm{Sf}_{n+m}  \cr
\mathrm{Sf}_n \times V^{xy}_\lambda &\sim V^{xy}_{\lambda+n} \cr
V^{xy}_{\lambda} \times V^{xy}_{\lambda'} &\sim V^{xy}_{\lambda+ \lambda'} + V^{xy}_{\lambda+ \lambda' -1}
\end{align}
We refer the reader to Appendix \ref{app:conventions} for a few more details and references. 

The $W_n$ and $D_n$ degenerate modules will result from dressing $\mathrm{Sf}_n$ respectively with a 
magnetic operator of charge $n$ or an electric operator of charge $-n$ for 
$U(1)_{\Psi^{-1} -1}$:
\begin{align}
W_n &= V^{U(1)_{\Psi^{-1} -1}}_{n (\Psi^{-1} -1)} O_n \cr
D_n &= V^{U(1)_{\Psi^{-1} -1}}_{-n} O_n
\end{align}
They fuse and braid as expected, with the symplectic fermion operators going along for the ride. 
They have dimensions 
\begin{align}
\Delta_{W_n} = \frac{n^2}{2}\frac{1}{\Psi} +\frac{|n|}{2} \cr
\Delta_{D_n} = \frac{n^2}{2}\frac{1}{1-\Psi} +\frac{|n|}{2}
\end{align}

The magnetic degenerate modules $H_{s,t}$ will instead involve $V^{xy}_{\lambda = \Psi s + t}$:
\begin{equation}
H_{s,t} = V^{U(1)_{\Psi^{-1} -1}}_{(1-\Psi) s} V^{xy}_{\Psi s + t}
\end{equation}
Locality with $W_n$ and $D_n$ is the result of a delicate cancellation 
between the two ingredients of the VOA: the $U(1)$ vertex operator induces singularities of 
order $z^{n s (1-\Psi)}$ in $W_n$ and $z^{- n s \Psi}$ in $D_n$ which cancel the non-integral part of the 
singularities induced by $V^{xy}_{\Psi s + t}$ on operators in $\mathrm{Sf}_n$. 

Furthermore, we can compare the fusion rules
\begin{equation}
H_{s,t} \times H_{s',t'} \sim H_{s+s',t+t'} + H_{s+s',t+t'-1}
\end{equation}
with the fusion rules of $\mathfrak{u}(1|1)$ irreps: typical finite-dimensional representations of $\mathfrak{u}(1|1)$
are labelled by two complex numbers $(e,n)$ \cite{Gotz:aa}, with non-zero $e$. Under tensor product, the $e$ label is additive. 
If $e+e'\neq 0$, the representations multiply as
\begin{equation}
(e,n) \otimes (e', n') = (e+e', n+n') \oplus (e+e', n+n'-1)
\end{equation}
If $e+e'$=0 the tensor product is a single decomposable representation. 
The $H_{s,t}$ thus fuse as $\mathfrak{u}(1|1)$ representations, with $t$ and $\tilde t = s + t$ being the weights of the irrep. 

The conformal dimension of $H_{s,t}$ is 
\begin{equation}
\Delta_{H_{s,t}} = \frac{(1-\Psi) \Psi s^2}{2} + \frac{(\Psi s + t)(\Psi s + t-1)}{2} = \frac{s (2t+s -1)}{2} \Psi + \frac{t(t-1)}{2}
\end{equation}
The braiding is controlled by the pairing $(s s' + s t' + s' t)\Psi$. It is perhaps more natural to label the operators by $t$ and $\tilde t = s +t$, so that the pairing is the natural pairing $\tilde t \tilde t' - t t'$ for weights of $\mathfrak{u}(1|1)$. 

Finally, we will build a general module as 
\begin{equation}
S_{p,p'} = V^{\Psi^{-1} -1}_{(\Psi^{-1} -1)p - p'} V^{xy}_{p + p'}
\end{equation}
with fusion rules
\begin{align}
W_n &\times S_{p,p'} \sim S_{p+n,p'} \cr
D_n &\times S_{p,p'} \sim S_{p,p'+n} \cr
H_{s,t} &\times S_{p,p'} \sim S_{p+ \Psi (s+t),p' + (1-\Psi)t} +S_{p+ \Psi (s+t-1),p' + (1-\Psi)(t-1)}
\end{align}
and dimension 
\begin{equation}
\Delta_{S_{p,p'}} = \frac{p^2}{2} \frac{1}{\Psi} +  \frac{(p')^2}{2} \frac{1}{1-\Psi} - \frac{ p + p'}{2}
\end{equation}

Now we can see that some of the ``degenerate'' modules are actually degenerate. Indeed, specializing $p = n$, $p'=0$ will gives a module 
which is not quite the same as $W_n$, as the limit $\lambda \to -n$ of $V^{xy}_\lambda$ is a non-trivial extension of two modules, one of which is  
$W_n$. Similar considerations apply to the $p=0$ and $p' = m$ specialization and $D_m$. On the other hand, 
$p = t \Psi$ and $p' = \Psi s + (1-\Psi) t$ gives directly $H_{s,t}$. 

This is as expected from gauge theory: these special values of $p$ and $p'$ are all such that the surface defect becomes transparent,
disappearing away from the interfaces and leaving behind some interface line defects.

For completeness, we can present some relevant characters.
The character of $\mathrm{Sf}$ admits a useful expansion:
\begin{equation}
\chi^{xy}  = \prod_{n=0}^\infty (1- q^{n+1}t )(1- q^{n+1}t^{-1}) = \frac{1}{\prod_{n=0}^\infty (1- q^{n+1})}\sum_{n=0}^\infty \sum_{m=-n}^n t^m (-1)^{n-m} q^{\frac{n(n+1)}{2}}
\end{equation}
where $t$ is the $U(1)_o$ fugacity. Thus we can write
\begin{equation}
\chi_{Y_{0,1,1}}(q)=\frac{1}{\prod_{n=0}^\infty (1- q^{n+1})^2}\sum_{n=0}^\infty (-1)^{n} q^{\frac{n(n+1)}{2}}
\end{equation}
and 
\begin{align}
\chi^{W_m}_{Y_{0,1,1}}(q)&=\frac{q^{\Delta_{W_m}}}{\prod_{n=0}^\infty (1- q^{n+1})^2}\sum_{n=|m|}^\infty (-1)^{n-m} q^{\frac{n(n+1)}{2}}\cr
\chi^{W_m}_{Y_{0,1,1}}(q)&=\frac{q^{\Delta_{D_m}}}{\prod_{n=0}^\infty (1- q^{n+1})^2}\sum_{n=|m|}^\infty (-1)^{n-m} q^{\frac{n(n+1)}{2}}
\end{align}

The character of $V^{xy}_\lambda$ for the full $xy$ VOA is even simpler
\begin{equation}
\chi^{xy}_\lambda  = \frac{1}{\prod_{n=0}^\infty (1- q^{n+1})}\sum_{n=- \infty}^\infty t^n (-1)^{n} q^{\frac{(n-\lambda)(n-\lambda+1)}{2}}
\end{equation}
so that
\begin{equation}
\chi_{Y_{0,1,1}}^{H_{s,t}}  = \frac{q^{\Delta_{H_{s,t}}}}{\prod_{n=0}^\infty (1- q^{n+1})^2}
\end{equation}
and similarly 
\begin{equation}
\chi_{Y_{0,1,1}}^{S_{p,p'}}  = \frac{q^{\Delta_{H_{s,t}}}}{\prod_{n=0}^\infty (1- q^{n+1})^2}
\end{equation}

Next, we can derive these facts from the various dual images of the junction. 
\begin{align} \label{eq:uoneoneduals}
&Y_{0,1,1}[\Psi] = Y_{1,0,1}[\Psi^{-1}] = Y_{1,0,1}[\frac{1}{1-\Psi}] = \cr &Y_{1,1,0}[\frac{\Psi}{\Psi-1}]=Y_{0,1,1}[1-\Psi] = Y_{1,1,0}[1-\Psi^{-1}]
\end{align}

\subsubsection{The $Y_{0,1,1}[\Psi]$ description}

\begin{wrapfigure}{l}{0.22\textwidth}
\vspace{-15pt}
  \begin{center}
      \includegraphics[width=0.175\textwidth]{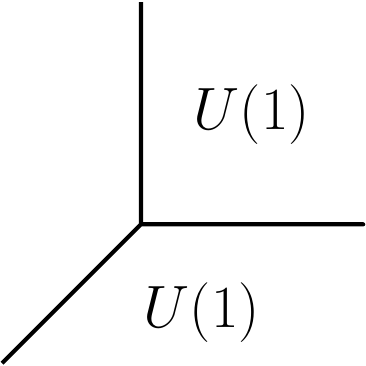}
\end{center}
\end{wrapfigure}
The first configuration involves a single set of symplectic bosons $(X,Y)$ with OPE 
\begin{equation}
X(z) Y(w) \sim \frac{1}{z-w}
\end{equation}
coupled at an interface a $U(1)$ Chern-Simons theory. The definition of $Y_{0,1,1}[\Psi]$ is a $\mathfrak{u}(1)$ BRST quotient 
of the product $U(1)_\Psi \times U(1)_{1-\Psi} \times \mathrm{Sb} \times \mathrm{bc}$. 

We can propose an explicit description of $Y_{0,1,1}[\Psi]$ with the help of the bosonization relation between the 
symplectic boson VOA and the symplectic fermion VOA.
The bosonization relation can be understood as follows: we bosonize the level $-1$ current $J_{XY} = :XY: = \partial \varphi_{XY}$
and write 
\begin{equation}
X(z) = e^{\varphi_{XY}(z)} x(z) \qquad Y(z) = e^{-\varphi_{XY}(z)} y(z)
\end{equation}
Then the symplectic bosons VOA decomposes as a sum of products of modules of charge $n$ for $J_{XY}$ and 
charge $n$ sectors $\mathrm{Sf}_n$ in the symplectic fermions VOA. 
\begin{equation}
\mathrm{Sb} = \oplus_{n \in \mathbb{Z}} V_n^{U(1)_{-1}} \otimes \mathrm{Sf}_n
\end{equation}
of the symplectic boson VOA into modules of a $U(1)_{-1} \times \mathrm{Sf}_0$ 
subalgebra. We refer the reader to Appendix \ref{app:conventions} for details and references. 

The BRST quotient only affects the $U(1)_{-1}$ current sub-algebra, reducing the product 
$U(1)_\Psi \times U(1)_{1-\Psi} \times U(1)_{-1} \times \mathrm{bc}$ to a single $U(1)$ current, 
which can be taken to be the BRST-closed representative 
\begin{equation}
U(1)_{\Psi^{-1}-1} = J _{1-\Psi}- \frac{1-\Psi}{\Psi} J_\Psi
\end{equation}
or, equivalently, 
\begin{equation}
U(1)_{\frac{\Psi}{1-\Psi}} = -J_\Psi + \frac{\Psi}{1-\Psi}J _{1-\Psi}.
\end{equation}

Thus we arrive to the anticipated claim 
\begin{equation}
Y_{0,1,1}[\Psi] = \mathrm{Sf}_0 \times U(1)_{\Psi^{-1}-1} 
\end{equation}

We can see the bosonization in action in the vacuum characters.
We begin from the following relation for the character of symplectic bosons: 
\begin{equation}
\chi^{XY}  = \frac{1}{\prod_{n=0}^\infty (1- q^{n+1})^2}\sum_{n=0}^\infty \sum_{m=-n}^n z^m (-1)^{n-m} q^{\frac{n(n+1)-m^2}{2}}
\end{equation}
Here $z$ is the fugacity for the $U(1)$ current $J_{XY}$. 
The $U(1)$ currents for $U(1)_\Psi$ and $U(1)_{1-\Psi}$ and the ghosts contributions (except the $c$ zeromode) cancel each other and the projection to charge $0$ 
leads to the expected character
\begin{equation}
\chi_{Y_{0,1,1}} = \frac{1}{\prod_{n=0}^\infty (1- q^{n+1})^2}\sum_{n=0}^\infty  (-1)^{n} q^{\frac{n(n+1)}{2}}
\end{equation}

\subsubsection{``Degenerate'' modules}
This description of the junction VOA makes it easy to identify 
$W_n$ and $D_n$. 

A line defect along the NS5 interface ending on the junction maps to a Chern-Simons Wilson loop ending at the interface from the direction of level $\Psi$.
At the interface it should be attached to a symplectic boson vertex operator of the correct gauge charge. 
That maps to a charge $-n$ vertex operator for $J_\Psi$ combined with a symplectic boson vertex operator 
of charge $n$ to give a BRST closed candidate for $W_n$.

With the help of bosonization, $W_n$ can be described as the product of a charge $-n$ magnetic vertex operator for $U(1)_{\Psi^{-1}-1}$ 
times a charge $n$ vertex operator in the symplectic fermions VOA, an element of $\mathrm{Sf}_n$, as anticipated. 
The characters can be readily matched as well. 

Similarly, a line defect along $\tilde B_{1,1}$ ending on the junction maps to a Chern-Simons Wilson loop 
ending at the interface from the direction of level $\Psi-1$.
This leads to a charge $m$ vertex operator for $J_{1-\Psi}$ combined with a symplectic boson vertex operator 
of charge $m$ to give a BRST closed candidate for $D_m$. After bosonization, this is an 
electric vertex operator of charge $m$ for $U(1)_{\Psi^{-1}-1}$, as anticipated. 
The characters can be readily matched as well. 

In order to produce $H_{s,t}$ as a BRST closed operator in the original $Y_{0,1,1}[\Psi]$ description 
we can employ a Ramond vertex operator $R_{s \Psi + t}$ for the symplectic bosons. See Appendix \ref{app:conventions} for
a definition. This has $U(1)_{-1}$ charge $(s +t + \frac12) \Psi + (1-\Psi)(t + \frac12)$. 
If we dress it with a $J_\Psi$ vertex operator of charge $-(s +t + \frac12) \Psi$ and a $J_{1-\Psi}$ 
vertex operator of charge $-(t + \frac12) (1-\Psi)$ we will get a BRST-closed representative for $H_{s,t}$.

The 4d gauge theory interpretation of these modules seems to be a generalized 't Hooft line defect along the D5 interface, which has magnetic
charges $\tilde t+\frac12$ and $t+\frac12$ in the two half-spaces and involves some non-trivial line defect for the 
interface hypermultiplets which somehow produces the $R_{s \Psi + t}$ module. 

\subsubsection{$Y_{1,0,1}[\Psi^{-1}]$ and relatives}
\begin{wrapfigure}{l}{0.22\textwidth}
\vspace{-15pt}
  \begin{center}
      \includegraphics[width=0.175\textwidth]{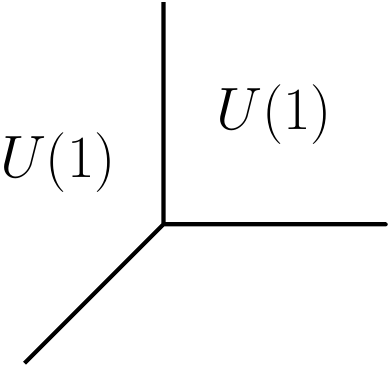}
\end{center}
\vspace{-15pt}
\end{wrapfigure}
The second, third, fourth and sixth descriptions in (\ref{eq:uoneoneduals}) involve an interface between an $U(1|1)$ and an $U(1)$ Chern-Simons theories.
In the second and sixth descriptions, we have some $U(1)$ BRST reduction of $U(1|1)_{\Psi^{-1}} \times U(1)_{\Psi^{-1}-1}$. 
In the third and fourth we have some $U(1)$ BRST reduction of  $U(1|1)_{(1-\Psi)^{-1}} \times U(1)_{\frac{\Psi}{1-\Psi} }$.
 
In our conventions, a $U(1|1)_{\kappa}$ VOA has currents
\begin{align}
J^1_1 = J + \frac{1}{2 \kappa} (I + J) \qquad J^2_1 = A \qquad J^1_2 = B \qquad J^2_2 = - I + \frac{1}{2 \kappa}  (I + J)
\end{align}
with OPE 
\begin{align}
J(z) J(w) &\sim \frac{\kappa}{(z-w)^2} \cr
J(z) A(w) &\sim \frac{A(w)}{z-w} \cr
J(z) B(w) &\sim - \frac{B(w)}{z-w} \cr
I(z) I(w) &\sim - \frac{\kappa}{(z-w)^2} \cr
I(z) A(w) &\sim -\frac{A(w)}{z-w} \cr
I(z) B(w) &\sim \frac{B(w)}{z-w} \cr
A(z) B(w) &\sim \frac{\kappa}{(z-w)^2} + \frac{J(w) + I(w)}{z-w}
\end{align}
The central charge is $0$. We refer to Appendix \ref{app:conventions} for more details. 

The BRST reduction employs the current $J^2_2$, whose level $1-\Psi^{-1}$ cancels the anomaly of $U(1)_{\Psi^{-1}-1}$.
The BRST close bosonic current surviving the coset can be taken to be 
\begin{equation}
J_{\Psi^{-1} -1}(z) = J(z) - \frac{\Psi}{2} \left(J(z) + I(z)\right)
\end{equation}
which is local with $J^2_2$, matching what we found in $Y_{0,1,1}[\Psi]$.

We can recover the anticipated form of the junction VOA by employing the bosonization of the $U(1|1)$ WZW model,
which decomposes it into a sum of products of modules for the $I$ and $J$ currents and $\mathrm{Sf}_0$: 
\begin{equation}
U(1|1)_{\Psi^{-1}} = \oplus_n V^{J,I}_{n,-n} \otimes \mathrm{Sf}_n
\end{equation} 
The $\mathfrak{u}(1)$-BRST quotient remodels the $U(1)$ into a single BRST closed currents $J_{\Psi^{-1} -1}(z)$ 
and leaves $\mathrm{Sf}_0$ unaffected, leading to 
\begin{equation}
Y_{1,0,1}[\Psi^{-1}] = U(1)_{\Psi^{-1} -1} \times \mathrm{Sf}_0
\end{equation}

\subsubsection{Degenerate Modules}

The gauge theory description suggests that $D_m$ should arise from a $U(1)$ Wilson loop ending on an operator of appropriate degree built from the boundary value of a fermionic $U(1|1)$ generator
and its derivatives. This means a charge $-m$ operator for $U(1)_{\Psi^{-1}-1}$ 
combined with an element of the $U(1|1)$ VOA of charge $m$ under $J^2_2(z)$. It has charge $m$ under the BRST closed
$J_{\Psi^{-1} -1}(z)$ as well and involves a charge $m$ vertex operator in the $(x,y)$ VOA. 
This agrees with the description of $D_m$ in the S-dual frame. 

Similarly, the Wilson loop of the $U(1|1)$ Chern-Simons theory  maps to a vertex operator 
$V^{U(1|1)}_{s,t}$ described in Appendix \ref{app:conventions} .
The corresponding module contains a descendant of the form $V^{J,I}_{s-\frac{\Psi}{2}s,\frac{\Psi}{2} s} V^{xy}_{s \Psi+t}$
which is BRST close and gives the anticipated form of $H_{s,t}$.

Finally, in the S-dual frame we have identified $W_n$ as a charge $n$ vertex operator for $U(1)_{\frac{\Psi}{1-\Psi}}$ 
times a charge $n$ vertex operator in the $(x,y)$ VOA.
That means it should have charge $(\Psi^{-1} -1)n$ under $J_{\Psi^{-1} -1}(z)$. 

We can engineer this from a nice $U(1|1)_{\Psi^{-1}}$ module, generated from a bosonized vertex operator 
of charge $n \Psi^{-1}-\frac{n}{2}$ under $J(z)$ and $\frac{n}{2}$ under $I(z)$, 
times a charge $n$ vertex operator in the $(x,y)$ VOA. This is a descendant of 
$V^{J,I}_{n \Psi^{-1}+\frac{n}{2},-\frac{n}{2}}$ and gives a simple BRST closed representative of $W_n$. 
In 3d, this must correspond to an interface vortex of some kind.

\subsubsection{General Modules}
We would like to identify in this context the general vertex operators $S_{p,p'}$. 
We need to recover the product of a vertex operator of momentum $\frac{1-\Psi}{\Psi} p' -  p$
for $J'(z)$ and $V^{xy}_{-p-p'}$. 

We can simply take $V^{U(1|1)_{\Psi^{-1}}}_{p'\Psi^{-1},-p \Psi^{-1}}$ 
and dress it with a $U(1)_{\Psi^{-1}-1}$ vertex operator of momentum $\Psi^{-1} p$ to get BRST invariance.
This is perfectly reasonable in the gauge theory.  

\subsection{Three $U(1)$ corners}

The most symmetric configuration comes from three $U(1)$ factors as in the figure on the left.
In all duality frames the construction of the algebra is the same, up to different choices of levels:
\begin{align}
&Y_{1,1,1}[\Psi] = Y_{1,1,1}[\Psi^{-1}] = Y_{1,1,1}[\frac{1}{1-\Psi}]= \cr
= &Y_{1,1,1}[\frac{\Psi}{\Psi-1}]=Y_{1,1,1}[1-\Psi] = Y_{1,1,1}[1-\Psi^{-1}]
\end{align}
In the first duality frame, we need to consider a $U(1|1)$ BRST quotient of 
\begin{equation}\label{eq:oneoneone}
U(1|1)_{\Psi} \times Sb^{1|1} \times U(1|1)_{- \Psi +1} 
\end{equation}

This setup is rather more intricate than the previous two examples. It is hard to describe BRST closed
vertex operators and even harder to make sure they are not BRST exact. The central charge of all ingredients 
vanishes independently of the value of $\Psi$ and so does the central charge of $Y_{1,1,1}[\Psi]$. 
\footnote{It is likely that some of these features are linked to the observation that the 
three D3 brane wedges in the brane setup can recombine to a single D3 brane 
and move away from the junction. This may mean that the BRST charge 
of $Y_{1,1,1}[\Psi]$ could be deformed to make the VOA trivial.}

One reason to believe that $Y_{1,1,1}[\Psi]$ itself should be non-trivial is that it should admit
three sets of degenerate modules $W_{s,t}$, $H_{s,t}$ and $D_{s,t}$ with non-trivial fusion and braiding
properties. 

The modules $W_{s,t}$ and $D_{s,t}$ should simply arise from operators built from the 
symplectic bosons and fermions, attached to Chern-Simons Wilson loops which carry the corresponding irreducible representations of 
$\mathfrak{u}(1|1)$. These should correspond to the modules $V^{U(1|1)_\Psi}_{s,t}$ or $V^{U(1|1)_{1-\Psi}}_{s,t}$ 
dressed by appropriate combinations of fermions and symplectic bosons. The overall conformal dimension 
of these vertex operators cannot vanish. 

Although a full analysis of the $U(1|1)$ BRST quotient goes beyond the scope of this paper, 
we can sketch a simpler procedure which we expect to be equivalent to it and to give a conjectural
free field realization of $Y_{1,1,1}[\Psi]$. Intuitively, we bosonize all the $U(1|1)$ WZWs 
and the fermions and symplectic bosons in $Sb^{1|1}$ and execute the coset of BRST reduction 
in two stages: we first deal with the bosonic currents in $U(1|1)$ and then with the leftover fermionic currents.

The bosonic reductions are identical to these considered for $Y_{0,1,1}[\Psi]$ and $Y_{1,0,0}[\Psi]$. 
They leave us with currents $I(z)$ and $J(z)$ of levels $1-\Psi^{-1}$ and $\Psi^{-1}-1$, together 
with the $xy$ currents from the symplectic boson and the $xy$ currents from the bosonized WZW models. 

From the perspective of a coset, the total fermionic currents 
\begin{equation}
A_{\mathrm{tot}}(z) = A_{U(1|1)_{\Psi}}(z) + \psi X(z) \qquad B_{\mathrm{tot}}(z) = B_{U(1|1)_{\Psi}}(z) + \chi Y(z)
\end{equation}
map after the bosonic coset to some combination of the rough form
\begin{eqnarray}\nonumber
x_{\mathrm{tot}}(z) &=& V_{-\Psi^{-1},\Psi^{-1}}^{IJ}(z) x_\Psi(z) + V_{1,-1}^{IJ}(z) x(z) 
\\ y_{\mathrm{tot}}(z) &=& V_{\Psi^{-1},-\Psi^{-1}}^{IJ}(z) y_\Psi(z) + V_{-1,1}^{IJ}(z) y(z)
\end{eqnarray}
Here $x_\Psi$, $y_\Psi$ denote the symplectic fermions which arise from bosonization of $U(1|1)_\Psi$, 
with OPE proportional to $\Psi$.

The composite fields $x_{\mathrm{tot}}(z)$ and $y_{\mathrm{tot}}(z)$ have the same OPE as free fermionic currents, 
thanks to a cancellation between the $I+J$ terms in the OPE. Thus we can consider the coset VOA 
generated by the currents in the bosonic coset which are local with $x_{\mathrm{tot}}$ and $y_{\mathrm{tot}}$.

From the perspective of BRST reduction, one would consider the combination of $I$, $J$ and three 
$xy$ systems, together with a BRST charge built with the help of two auxiliary $\beta \gamma$ ghost systems. 

\section{Examples with $U(2)$ gauge groups.}\label{sec:utwo}
\subsection{Virasoro $\times\ U(1)$}

\begin{wrapfigure}{l}{0.22\textwidth}
\vspace{-15pt}
  \begin{center}
      \includegraphics[width=0.175\textwidth]{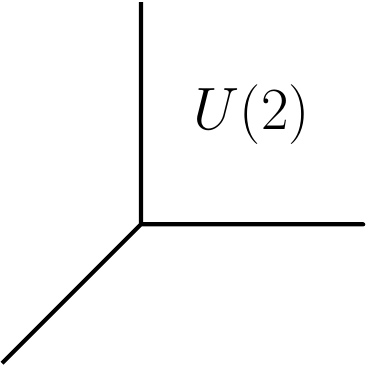}
\end{center}
\vspace{-15pt}
\end{wrapfigure}

\subsubsection{$Y_{0,0,2}[\Psi]$}

The simplest example involves a $U(2)$ gauge theory in the corner, i.e. $Y_{0,0,2}[\Psi]$. We already essentially analyzed 
this setup when looking at the three realizations of the Virasoro algebra, but it is instructive to 
add the $U(1)$ current algebra in order to get a full $U(2)$ gauge group.

Recall that according to our conventions, spelled out in Appendix \ref{app:conventions},
the diagonal current $J^1_1(z) + J^2_2(z)$ in $U(2)_\Psi$ has level $2 \Psi$ while the $SU(2)$ 
currents have level $\Psi - 2$. The OPE between $J^a_a$ and $J^a_a$ goes as $\frac{\Psi-1}{(z-w)^2}$. 
The OPEs between Cartan generators $J^1_1$ and $J^2_2$ take the form
\begin{align}
J^1_1(z) J^1_1(0) &\sim \frac{\Psi-1}{(z-w)^2} \cr
J^1_1(z) J^2_2(0) &\sim \frac{1}{(z-w)^2} \cr
J^2_2(z) J^2_2(0) &\sim \frac{\Psi-1}{(z-w)^2}.
\end{align} 

The definition of $Y_{0,0,2}[\Psi]$ involves the quantum DS reduction of $U(2)_{\Psi}$ by the regular $\mathfrak{su}(2)$ embedding. 
It produces the combination of the Virasoro VOA with $b^2 = - \Psi$ and 
a $U(1)_{2 \Psi}$ current. 

It is instructive to follow this at the level of vacuum characters. We begin with the $U(2)$ vacuum character
\begin{equation}
\chi^{U(2)_\Psi}(z_1,z_2;q) = \frac{1}{\prod_{n > 0}(1-q^n)^2 (1-\frac{z_1}{z_2} q^n)(1-\frac{z_2}{z_1} q^n)}.
\end{equation}
We add the ghost contribution and then adjust the Cartan fugacities $z_i$ to $z_2 = q z_1 = q^{\frac12} z$ in order to account for the shift of the stress tensor 
which gives dimension $0$ to $J^1_2$ and the symmetry breaking enforced by $J^1_2 = 1$:
\begin{equation}
\chi^{W_2[U(2)_\Psi]}(z;q) = \frac{1-\frac{z_2}{z_1}}{\prod_{n > 0}(1-q^n)^2} =  \frac{1-q}{\prod_{n > 0}(1-q^n)^2}
\end{equation}
which is the expected vacuum character for Virasoro times $U(1)$. Notice that the crucial factor of $1-q$ arises from the 
$c$ zeromode which does not cancel against the off-diagonal current contributions.

The central charge of $Y_{0,0,2}[\Psi]$ is, as expected,
\begin{equation}
c = 3 \frac{\Psi-2}{\Psi}+1-2- 6 (\Psi-2) = 14 - 6 \Psi - 6 \Psi^{-1} 
\end{equation}

\subsubsection{Modules}
A spin $j$ vertex operator for $SU(2)_{\Psi-2}$ can be combined with momentum $p$ vertex operators for 
$U(1)_{2\Psi}$ to give a vertex operator of dimension $\frac{j(j+1)+ p^2/4}{\Psi}$. It is natural to 
define $j = (\mu_1 - \mu_2)/2$ and $p = \mu_1 + \mu_2$ to get a vertex operator $V_{(\mu_1,\mu_2)}$
labelled by the $U(2)$ highest weight $(\mu_1, \mu_2)$ with $\mu_1 \geq \mu_2$. 
The conformal dimension of the highest weight vector is controlled by the $U(2)$ Casimir: 
\begin{equation}
\Delta_{\mu_1,\mu_2} = \frac{\mu_1^2 + \mu_1}{2\Psi}+\frac{\mu_2^2 - \mu_2}{2\Psi}.
\end{equation}

A Wilson loop ending at the boundary in the 3d Chern-Simons theory will give rise in 2d to the DS reduction of the 
module generated by $V_{(\mu_1,\mu_2)}$. This is our definition of $W_\mu$.
The resulting degenerate modules have conformal dimension   
\begin{equation}
\Delta_{W_\mu} =\frac{\mu_1^2 + \mu_1}{2\Psi} - \frac{\mu_1}{2} + \frac{\mu_2^2 - \mu_2}{2\Psi}+ \frac{\mu_2}{2} 
\end{equation}
At the level of characters, we begin with 
\begin{equation}
\chi^{U(2)_\Psi}_{\mu_1, \mu_2}(z_1,z_2;q) = q^{\frac{\mu_1^2 + \mu_1}{2\Psi}+\frac{\mu_2^2 - \mu_2}{2\Psi}}\frac{z_1^{\mu_1 + 1} z_2^{\mu_2} - z_2^{\mu_1 + 1} z_1^{\mu_2}}{(z_1 - z_2)\prod_{n > 0}(1-q^n)^2 (1-\frac{z_1}{z_2} q^n)(1-\frac{z_2}{z_1} q^n)}
\end{equation}
As before, we add the ghost contribution and adjust the fugacities to $z_2 = q z_1 = q^{\frac12} z$ in order to account for the shift of the stress tensor 
which gives dimension $0$ to $J^1_2$ and the symmetry breaking enforced by $J^1_2 = 1$:
\begin{equation}
\chi^{W_2[U(2)_\Psi]}_{\mu_1, \mu_2}(z;q) =q^{\frac{\mu_1^2 + \mu_1}{2\Psi}+\frac{\mu_2^2 - \mu_2}{2\Psi}} \frac{z_1^{\mu_1} z_2^{\mu_2} - z_2^{\mu_1 + 1} z_1^{\mu_2-1}}{\prod_{n > 0}(1-q^n)^2} = z^{\mu_1 + \mu_2}q^{\Delta_{W_\mu}} \frac{ 1 - q^{\mu_1-\mu_2 + 1}}{\prod_{n > 0}(1-q^n)^2}
\end{equation}
which is the expected degenerate character for Virasoro times $U(1)$, with a null vector at level $\mu_1-\mu_2 + 1$. 

The realization of the second family of degenerate modules $H_\nu$ is less obvious, but still straightforward. 
We propose to combine spectral flow images of the vacuum module of $SU(2)_{\Psi-2}$ 
and magnetic vertex operators (possibly with half-integral charge) for $U(1)_{2\Psi}$. The former are simply 
vertex operators for the bosonized Cartan current in $SU(2)_{\Psi-2}$, with momenta multiple of $\Psi - 2$. 
These have charges $-\nu_1 (\Psi-1)-\nu_2$ and $-\nu_2 (\Psi-1) -\nu_1$ under $J^1_1$ and $J^2_2$. 
At the level of characters, we have 
\begin{equation}
\tilde\chi^{U(2)_\Psi}_{\nu_1, \nu_2}(z_1,z_2;q) =  \frac{z_1^{-\nu_1 (\Psi-1)-\nu_2} z_2^{-\nu_2 (\Psi-1)-\nu_1}q^{\Psi\frac{\nu_1^2 + \nu_2^2}{2}-\frac{(\nu_1- \nu_2)^2}{2}}}{\prod_{n > 0}(1-q^n)^2 (1-\frac{z_1}{z_2} q^{n+\nu_2 - \nu_1})(1-\frac{z_2}{z_1} q^{n+ \nu_1 - \nu_2})}
\end{equation}

Next, we implement the DS reduction, which is well defined for $\nu_1 > \nu_2$. The result is  
\begin{equation}
\tilde\chi^{W_2[U(2)_\Psi]}_{\nu_1, \nu_2}(z;q) =  (-1)^{\nu_1 - \nu_2} z^{-\Psi(\nu_1 + \nu_2)}q^{\Delta_{H_\nu}} \frac{ 1 - q^{\nu_1-\nu_2 + 1}}{\prod_{n > 0}(1-q^n)^2}
\end{equation}
where we simplified $\nu_1 - \nu_2$ ratios of the form $\frac{1-\frac{z_2}{z_1} q^k}{1-\frac{z_1}{z_2} q^{-k}}$.

We recognize the expected character for the degenerate
operators $H_\mu$ of dimension 
\begin{equation}
\Delta_{H_\nu} =\frac{\nu_1^2 + \nu_1}{2}\Psi - \frac{\nu_1}{2} + \frac{\nu_2^2 - \nu_2}{2}\Psi + \frac{\nu_2}{2} 
\end{equation}

This construction has a simple interpretation in gauge theory. In 3d, we are defining a boundary vortex operator 
by a Hecke modification of the oper boundary condition. In 4d, we have a boundary 't Hooft line defect 
superimposed to the Nahm pole. 

The $W_\mu$ and $H_{\mu'}$ vertex operators are mutually local. 

We expect that general modules $S_p$ can be obtained from the DS reduction of highest weight modules for $U(2)_\Psi$
with generic, non-integral weights $p_1$ and $p_2$. In particular, they are associated to infinite highest 
weight representations of the zeromode algebra. At the level of characters, 
\begin{equation}
\hat \chi^{U(2)_\Psi}_{p_1, p_2}(z_1,z_2;q) = q^{\frac{p_1^2 + p_1}{2\Psi}+\frac{p_2^2 - p_2}{2\Psi}}\frac{z_1^{p_1} z_2^{p_2}}{(1-\frac{z_2}{z_1})\prod_{n > 0}(1-q^n)^2 (1-\frac{z_1}{z_2} q^n)(1-\frac{z_2}{z_1} q^n)}
\end{equation}
The usual manipulations lead to the obvious
\begin{equation}
\hat\chi^{W_2[U(2)_\Psi]}_{p_1, p_2}(z;q) =  z^{p_1 + p_2}q^{\frac{p_1^2 + p_1}{2\Psi}+\frac{p_2^2 - p_2}{2\Psi}+\frac{p_2-p_1}{2} } \frac{1}{\prod_{n > 0}(1-q^n)^2}
\end{equation}
of dimension 
\begin{equation}
\frac{p_1^2 + p_2^2}{2\Psi}+\frac{p_1 - p_2}{2}(\Psi^{-1} -1)
\end{equation}

\subsubsection{$Y_{2,0,0}[\Psi]$}
\begin{wrapfigure}{l}{0.2\textwidth}
\vspace{-15pt}
  \begin{center}
      \includegraphics[width=0.175\textwidth]{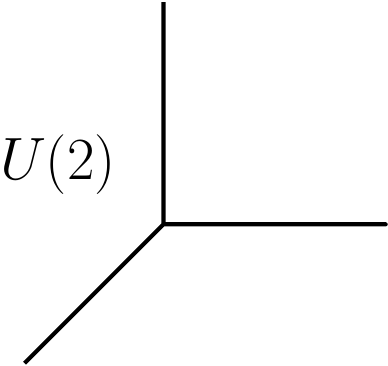}
\end{center}
\vspace{-15pt}
\end{wrapfigure}
In this duality frame we have a $U(2)$ BRST quotient of
\begin{equation}
U(2)_{\frac{1}{\Psi-1}} \times U(2)_{\frac{1}{\Psi^{-1}-1}} \times \mathrm{Ff}^{U(2)}
\end{equation}
where $\mathrm{Ff}^{U(2)}$ denotes a pair of complex fermions. The coset of the non-Abelian part is essentially the analytic continuation of the analytically continued 
Virasoro minimal model coset 
\begin{equation}
\frac{SU(2)_{\frac{1}{\Psi-1}-2} \times SU(2)_1}{SU(2)_{\frac{1}{\Psi-1}-1} }.
\end{equation}
Thus we expect the BRST coset to give again the product of Virasoro and an a $U(1)$ current with the correct total central charge 
\begin{equation}
c = 3 \left(\frac{1}{\Psi-1}-2\right)(\Psi-1) + 1 + 2 - 3 \left(\frac{1}{\Psi-1}-1 \right)(1-\Psi^{-1}) -1= 14 - 6 \Psi - 6 \Psi^{-1}
\end{equation}

The $U(1)$ current can be taken to be the combination of diagonal currents
\begin{equation}
J_{2\Psi}(z) = \Psi J_{\frac{2}{\Psi-1}} + J_{\frac{2}{\Psi^{-1}-1}} 
\end{equation}

It is instructive to follow this at the level of vacuum characters. 
We start from the product of characters 
\begin{equation}
\chi^{U(2)}(z_1,z_2;q)^2\chi^{\mathrm{Ff}^2}(z_1,z_2;q) = \frac{\prod_{n > 0}(1-q^{n+\frac12} z_1)(1-q^{n+\frac12}z_1^{-1})(1-q^{n+\frac12}z_2)(1-q^{n+\frac12}z_2^{-1})}{\prod_{n > 0}(1-q^n)^2 (1-\frac{z_1}{z_2} q^n)^2(1-\frac{z_2}{z_1} q^n)^2}
\end{equation}
The $\mathfrak{u}(2)$ ghosts cancel out the whole denominator. The $c$ ghost zeromodes 
contribute a Vandermonde determinant for the projection to gauge-invariant operators 
\begin{equation}
\chi^{Y_{2,0,0}}(q)= \oint \frac{dz_1 dz_2}{2 z_1 z_2} (1-\frac{z_2}{z_1})(1-\frac{z_1}{z_2})\prod_{n > 0}(1-q^{n+\frac12} z_1)(1-q^{n+\frac12}z_1^{-1})(1-q^{n+\frac12}z_2)(1-q^{n+\frac12}z_2^{-1})
\end{equation}
Expanding the product through a basic theta function identity gives the desired answer
\begin{equation}
\chi^{Y_{2,0,0}}(q)= \oint \frac{dz_1 dz_2}{2 z_1 z_2} (1-\frac{z_2}{z_1})(1-\frac{z_1}{z_2})\frac{\sum_{n_1,n_2} (-1)^{n_1 + n_2} z_1^{n_1} z_2^{n_2} q^{\frac{n_1^2 +n_2^2}{2}}}{\prod_{n > 0}(1-q^n)^2}
\end{equation}
i.e. 
\begin{equation}
\chi^{Y_{2,0,0}}(q)= \frac{1-q}{\prod_{n > 0}(1-q^n)^2}
\end{equation}

\subsubsection{Modules}

The $W_{\mu_1,\mu_2}$ and $H_{\nu_1,\nu_2}$ vertex operators descend from the corresponding electric vertex operators in 
either $U(2)$ VOA, dressed appropriately with the free fermions to make them gauge invariant. 

For example, 
\begin{equation}
\Delta_{W_\mu} = \frac{\mu_1^2 + \mu_1}{2}(\Psi^{-1}-1)+\frac{\mu_2^2 - \mu_2}{2}(\Psi^{-1}-1) + \frac{\mu_1^2 +\mu_2^2}{2}
\end{equation}
where the last term is the dimension of the free fermion operators of appropriate charge and the other terms
the dimension of the electric module for $U(2)_{\frac{1}{\Psi^{-1}-1}}$

At the level of characters,
\begin{equation}
\chi^{Y_{2,0,0}}_{W_\mu}(q)= q^{\Delta_\mu }\oint \frac{dz_1 dz_2}{2 z_1 z_2} (1-\frac{z_1}{z_2})(z_1^{\mu_1} z_2^{\mu_2} - z_2^{\mu_1 + 1} z_1^{\mu_2-1})\frac{\sum_{n_1,n_2} (-1)^{n_1 + n_2} z_1^{n_1} z_2^{n_2} q^{\frac{n_1^2 +n_2^2}{2}}}{\prod_{n > 0}(1-q^n)^2}
\end{equation}
i.e. 
\begin{equation}
\chi^{Y_{2,0,0}}_{W_\mu}(q)= q^{\Delta_\mu }\frac{q^{\frac{\mu_1^2 +\mu_2^2}{2}}-q^{\frac{(\mu_1+1)^2 +(\mu_2-1)^2}{2}}}{\prod_{n > 0}(1-q^n)^2}
=q^{\Delta_{W_\mu}}\frac{1-q^{1 + \mu_1 - \mu_2 }}{\prod_{n > 0}(1-q^n)^2}
\end{equation} 

In a similar manner, general modules $S_p$ arise from a combination of modules of general 
non-integral weights for both $U(2)$'s. In order to get a gauge-invariant combination, 
we need to combine Weyl modules induced from a highest weight representation of one 
$U(2)$ and a lowest weight representation of the other $U(2)$, with the same weight $(\tilde p_1, \tilde p_2)$. 

The resulting conformal dimension is 
\begin{equation}
\frac{\tilde p_1^2 + \tilde p_1}{2}(\Psi^{-1}-1)+\frac{\tilde p_2^2 - \tilde p_2}{2}(\Psi^{-1}-1)+\frac{\tilde p_1^2 + \tilde p_1}{2}(\Psi^{-1}-1)+\frac{\tilde p_2^2 - \tilde p_2}{2}(\Psi^{-1}-1)
\end{equation}
i.e. in terms of $p_i =(1-\Psi)\tilde p_i$
\begin{equation}
\frac{p_1^2 + p_2^2}{2\Psi}+\frac{p_1 - p_2}{2}(\Psi^{-1}-1)
\end{equation}

\subsection{Parafermions $\times\ U(1)$}
Next, we look at the VOA realized by 
\begin{align}
& Y_{0,1,2}[\Psi] = Y_{1,0,2}[\Psi^{-1}] = Y_{2,0,1}[\frac{1}{1-\Psi}]= \cr
&=Y_{2,1,0}[\frac{\Psi}{\Psi-1}]=Y_{0,2,1}[1-\Psi] = Y_{1,2,0}[1-\Psi^{-1}]
\end{align}

The result will be a combination of a $U(1)$ current at level $\Psi^{-1}(\Psi-1)(\Psi - 2)$ and 
the analytic continuation $\mathrm{Pf}_{\Psi-2}$ of a well known VOA $\mathrm{Pf}_{k}$
of $Z_k$ parafermions.

We will encounter two well-known coset constructions of parafermions and a less well-known one. 
In the process, we will define three families of modules which combine with $U(1)$ vertex operators into 
the degenerate modules $W_{\mu_1, \mu_2}$, $D_s$ and $H_{\nu_1, \nu_2, \nu_3}$. 

\subsubsection{$Y_{0,1,2}$ and parafermions.}
\begin{wrapfigure}{l}{0.20\textwidth}
\vspace{-10pt}
  \begin{center}
      \includegraphics[width=0.175\textwidth]{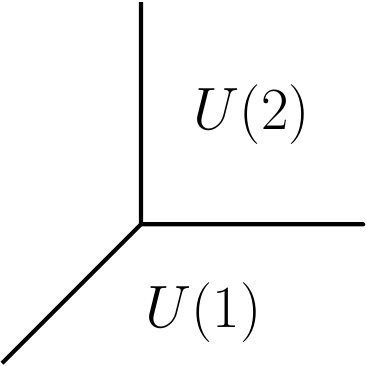}
\end{center}
\vspace{-10pt}
\end{wrapfigure}

The first realization involves a $U(1)$ BRST quotient of\footnote{Recall the conventions $U(2)_{\Psi}=U(1)_{2\Psi}\times SU(2)_{\Psi-2}$ and see appendix \ref{app:conventions}.}
\begin{equation}
U(2)_{\Psi} \times U(1)_{1 - \Psi} 
\end{equation}
by the sum of the $J_1^1$, which has level $\Psi - 1$, and the $U(1)_{1 - \Psi}$ current. 

The BRST cohomology of the vacuum module can only involve vertex operators in $U(2)_{\Psi}$ of $J_1^1$ charge $0$. 
That sub-algebra factorizes as
\begin{equation}
U(1)_{J^1_1} \times U(1)_{J^2_2} \times \mathrm{Pf}_{\Psi-2}
\end{equation}
where $\mathrm{Pf}_{\kappa}$ is by definition the coset VOA
\begin{equation}
\mathrm{Pf}_{\kappa} \equiv \frac{SU(2)_{\kappa}}{U(1)_{2\kappa}}
\end{equation}
For integral $\kappa$, this is known as the $Z_\kappa$-parafermion VOA.

The BRST quotient, as usual, reduces the tree $U(1)$ currents $U(1)_{J^1_1} \times U(1)_{J^2_2} \times U(1)_{1 - \Psi}$ 
to a single $U(1)$ current. A convenient choice 
\begin{equation}
J_c(z) = (1-\Psi^{-1})J_2^2(z) - \Psi^{-1} J_1^1(z)
\end{equation} 
has level $\Psi^{-1}(\Psi-1)(\Psi - 2)$ and gives integral charges to the off-diagonal WZW currents. 

The $\mathrm{Pf}_{\Psi-2}$ VOA goes along for the ride and thus we can write 
\begin{equation}
Y_{0,1,2}[\Psi] = U(1)_{\Psi^{-1}(\Psi-1)(\Psi - 2)} \times \mathrm{Pf}_{\Psi-2}
\end{equation}
\begin{equation}
c[Y_{0,1,2}[\Psi]] = 3 - 6 \Psi^{-1}
\end{equation}

Computing the vacuum character requires some judicious manipulations of the $U(2)$ character:
\begin{align}
\chi[Y_{0,1,2}] &= \oint \frac{dz}{z} \frac{1}{\prod_{n>0} (1-q^n)(1- z q^n)(1- z^{-1} q^n)} = \cr
 & =\oint \frac{dz}{z} \frac{1-z}{\prod_{n=0}^\infty (1- q^{n+1})^3}\sum_{n=0}^\infty \sum_{m=-n}^n z^m (-1)^{n-m} q^{\frac{n(n+1)-m(m+1)}{2}}  = \cr
 & =\frac{1}{\prod_{n=0}^\infty (1- q^{n+1})^3}\left( 1 + 2 \sum_{n=1}^\infty (-1)^{n} q^{\frac{n(n+1)}{2}} \right)
\end{align}

\subsubsection{Degenerate modules}
The $W_\mu$ modules will be the BRST reduction of the $V_\mu$ modules for $U(2)_{\Psi}$ 
times the vacuum module of $U(1)_{1 - \Psi}$.

The basic BRST closed representative will involve a vector of weight $(0, \mu_2 + \mu_1)$ in $V_{\mu_1,\mu_2}$.
It has charge $(1-\Psi^{-1})(\mu_2 + \mu_1)$ under $J_c$. The dimension of the highest weight vector 
is different depending on $(0, \mu_2 + \mu_1)$ being an element of $(\mu_1,\mu_2)$ irrep of the $\mathfrak{u}(2)$ current zeromodes
or not. Recall the dimension of the $U(2)_\Psi$ primaries
\begin{equation}
\Delta_{\mu_1,\mu_2} = \frac{\mu_1^2 + \mu_1}{2\Psi}+\frac{\mu_2^2 - \mu_2}{2\Psi}.
\end{equation}

We can compute the character as before. 
\begin{align}
\chi_{W_\mu}&[Y_{0,1,2}](y;q) = q^{\Delta_{\mu_1,\mu_2}} \oint \frac{dz}{z} \frac{z^{\mu_2}-z^{\mu_1+1} }{(1-z)\prod_{n>0} (1-q^n)(1- z q^n)(1- z^{-1} q^n)} = \cr
 & =q^{\Delta_{\mu_1,\mu_2}} \oint \frac{dz}{z} \frac{z^{\mu_2}-z^{\mu_1+1}}{\prod_{n=0}^\infty (1- q^{n+1})^3}\sum_{n=0}^\infty \sum_{m=-n}^n z^m (-1)^{n-m} q^{\frac{n(n+1)-m(m+1)}{2}}  = \cr
 & =q^{\Delta_{\mu_1,\mu_2}} \frac{\sum_{n=|\mu_1|}^\infty  (-1)^{n+\mu_1} q^{\frac{n(n+1)-\mu_1(\mu_1+1)}{2}}+ \sum_{n=|\mu_2|}^\infty  (-1)^{n+\mu_2} q^{\frac{n(n+1)-\mu_2(\mu_2-1)}{2}} }{\prod_{n=0}^\infty (1- q^{n+1})^3}
 \end{align}
 
The $SU(2)/U(1)$ parafermion VOA has modules $M_{j,m}$ which arise from 
vertex operators of weight $m$ in the $SU(2)$ module of spin $j$. Here we take such a module with $j = \frac{\mu_1 - \mu_2}{2}$ 
and $m = - \frac{\mu_1 + \mu_2}{2}$ and dress it with a $U(1)$ vertex operator of charge $(1-\Psi^{-1})(\mu_2 + \mu_1)$.

The $D_s$ modules will be the BRST reduction of the vacuum modules for $U(2)_{\Psi - 2}$ 
times the charge $s$ module of $U(1)_{1 - \Psi}$. It has charge $s$ under $J_c$.
The highest weight vector arises from $s$ powers of an off-diagonal current and thus 
\begin{equation}
\Delta_{D_s} = \frac{s^2}{2} \frac{1}{1-\Psi} + |s|
\end{equation}

We can compute the character as before:
\begin{align}
\chi_{D_s}&[Y_{0,1,2}] = q^{\frac{s^2}{2} \frac{1}{1-\Psi}}  \oint \frac{dz}{z} \frac{z^s}{\prod_{n>0} (1-q^n)(1- z q^n)(1- z^{-1} q^n)} = \cr
 & =q^{\frac{s^2}{2} \frac{1}{1-\Psi}}\oint \frac{dz}{z} \frac{z^s-z^{s+1}}{\prod_{n=0}^\infty (1- q^{n+1})^3}\sum_{n=0}^\infty \sum_{m=-n}^n z^m (-1)^{n-m} q^{\frac{n(n+1)-m(m+1)}{2}}  = \cr
 & =q^{\frac{s^2}{2} \frac{1}{1-\Psi}}\frac{\left( \sum_{n=s}^\infty (-1)^{n+s} q^{\frac{n(n+1)-s(s-1)}{2}}+\sum_{n=s+1}^\infty(-1)^{n+s} q^{\frac{n(n+1)-s(s+1)}{2}} \right)}{\prod_{n=0}^\infty (1- q^{n+1})^3}
\end{align}
Notice that the sums are economical for $s\geq 0$. For negative $s$ the first $2s$ terms in each sum cancel pairwise and 
the summations can start from $-s$ and $-s-1$ respectively. These alternative starting points are also valid for positive $s$. 
Changing the first sum in that manner, we can rewrite the sum as 
\begin{align}
\chi_{D_s}[Y_{0,1,2}] & =q^{\frac{s^2}{2} \frac{1}{1-\Psi}} \frac{\left( \sum_{n=s}^\infty (-1)^{n+s} q^{\frac{n(n+1)-s(s-1)}{2}}+\sum_{n=s+1}^\infty(-1)^{n+s} q^{\frac{n(n+1)-s(s+1)}{2}} \right)}{\prod_{n=0}^\infty (1- q^{n+1})^3} \cr
& =q^{\frac{s^2}{2} \frac{1}{1-\Psi}} \frac{\left( \sum_{n=0}^\infty (-1)^{n} q^{\frac{n(n- 2s +1)}{2}}+\sum_{n=1}^\infty(-1)^{n} q^{\frac{n(n+2 s +1)}{2}} \right)}{\prod_{n=0}^\infty (1- q^{n+1})^3}
\end{align}

The $SU(2)/U(1)$ parafermion VOA has modules $M_{s}$ which arise from 
vertex operators of weight $s$ in the $SU(2)$ vacuum module. These are essentially the parafermions themselves.
Here we take such a module and dress it with a $J_c$ vertex operator of electric charge $s$.

The mutual locality between $W_\mu$ and $D_s$ is obvious before the BRST coset. 
In terms of parafermion and $U(1)$ modules, it follows from a conspiracy 
between the braiding phases of individual factors.

We will not attempt a direct construction of the $H_{\nu}$ modules here. 
We will find a candidate S-dual description 
of $H_{\nu}$ in the next section. 

\subsubsection{$Y_{1,0,2}$ and parafermions.}
\begin{wrapfigure}{l}{0.20\textwidth}
\vspace{-10pt}
  \begin{center}
      \includegraphics[width=0.175\textwidth]{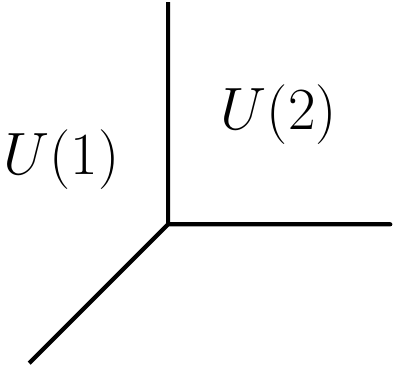}
\end{center}
\vspace{-10pt}
\end{wrapfigure}
The second realization gives a construction that combines DS-reduction with $U(1)$ coset
\begin{eqnarray}
Y_{1,0,2}[\Psi^{-1}]\equiv \frac{W_{2|1}\left [U(2|1)_{\Psi^{-1}}\right ]}{U(1)_{1-\Psi^{-1}}}:\quad c=3-\frac{6}{\Psi}
\end{eqnarray}
The DS reduction step is well-understood: it is known to give the product of a $U(1)$ current and a 
$\mathcal{N}=2$ super Virasoro algebra, generated by the stress-energy tensor $T$, two fermionic generators $G^{\pm}$ of conformal dimension $\frac{3}{2}$ and a $U(1)$ current \cite{Kac:qy}. The super Virasoro algebra has central charge $c=3-\frac{6}{\Psi}$. 

Thus we can write a simplified definition
\begin{eqnarray}
Y_{1,0,2}[\Psi^{-1}]=\frac{\mathrm{sVir}_{\mathcal{N}=2}[c=3-\frac{6}{\Psi}] \times U(1)_{\Psi^{-1}}}{U(1)_{1-\Psi^{-1}}}
\end{eqnarray}
In turn, the BRST cohomology of the vacuum module can only involve vertex operators in $\mathrm{sVir}_{\mathcal{N}=2}$ of $U(1)$ charge $0$. 
The parafermion algebra is known to arise as a coset \cite{Zamolodchikov:1986gh}
\begin{equation}
\mathrm{Pf}_{\kappa} = \frac{\mathrm{sVir}_{\mathcal{N}=2}[c=3-\frac{6}{\kappa+2}]}{U(1)_{1-\frac{2}{\kappa+2}}}
\end{equation}
and thus we recover the expected 
\begin{equation}
Y_{1,0,2}[\Psi^{-1}]= U(1)_{\Psi^{-1}(\Psi-1)(\Psi - 2)} \times \mathrm{Pf}_{\Psi-2}
\end{equation}

A direct calculation of the vacuum character of $Y_{1,0,2}$ is also reassuring.
The vacuum character for $U(2|1)$ is 
\begin{equation}
\chi[U(2|1)](z_1,z_2, w;q) = \frac{\prod_{n > 0}(1-q^n \frac{z_1}{w}) (1-q^n \frac{z_2}{w}) (1-q^n \frac{w}{z_1}) (1-q^n \frac{w}{z_2}) }
{\prod_{n > 0}(1-q^n)^3 (1-\frac{z_1}{z_2} q^n)(1-\frac{z_2}{z_1} q^n)}
\end{equation}
Adding the $bc$ ghosts for $J^1_2$ and the $\beta \gamma$ ghosts for $J^1_3$ we get a simpler product 
\begin{equation}
\chi[U(2|1)\times \mathrm{DS\,ghosts}](z_1,z_2, w;q) = \frac{(1-\frac{z_2}{z_1})\prod_{n > 0} (1-q^n \frac{z_2}{w}) (1-q^n \frac{w}{z_2}) }
{(1-\frac{w}{z_1})\prod_{n > 0}(1-q^n)^3}
\end{equation}
The shift of the stress tensor which makes $J^1_2$ of dimension $0$ and $J^1_3$ of dimension $\frac12$ and the reduction of symmetry 
enforced by $J^1_2=1$ are implemented in the character by the specialization of fugacities $z_2 = q z_1 = q^{\frac12} z$: 
\begin{equation}
\chi[W_{2|1} U(2|1)](z, w;q) = \frac{(1-q)\prod_{n > 0} (1-q^{n+\frac12} \frac{z}{w}) (1-q^{n+\frac12} \frac{w}{z}) }
{\prod_{n > 0}(1-q^n)^3}
\end{equation}
which is the vacuum character for ${\cal N}=2$ super-Virasoro times $U(1)$. 

Notice that the current $J^3_3$ has initial level $2-\Psi^{-1}$ in our conventions. The $\beta \gamma$ ghost system 
contributes an extra $-1$ to the level, giving a current $\tilde I$ after the DS reduction of level $1-\Psi^{-1}$. The current $J^1_1+J^2_2$ 
has initial level $2\Psi^{-1}+2$. The $\beta \gamma$ ghost system 
contributes an extra $-1$ to the level, giving a current $\tilde J$ after the DS reduction of level $2\Psi^{-1}+1$.
The OPE coefficient in $\tilde I \tilde J$ is $1$, shifted from $2$ by the $\beta \gamma$ contribution. 

The next step is the quotient by $U(1)_{1-\Psi^{-1}}$ given by $\tilde I$. We need to compute a contour integral 
\begin{align}
\chi[Y_{1,0,2}](z, w;q) &= \frac{(1-q)}{\prod_{n > 0}(1-q^n)^2}
\oint \frac{dz}{z}\prod_{n > 0} (1-q^{n+\frac12} z) (1-q^{n+\frac12} z^{-1}) \cr
&= \frac{(1-q)}{\prod_{n > 0}(1-q^n)^3}
\oint \frac{dz}{z}\frac{\sum_{n=-\infty}^\infty (-1)^n z^n q^{\frac{n^2}{2}}}{(1-q^{\frac12} z) (1-q^{\frac12} z^{-1})} \cr
&= \frac{(1-q)}{\prod_{n > 0}(1-q^n)^3}
\oint \frac{dz}{z}\sum_{a,b \geq 0 }\sum_{n=-\infty}^\infty q^{\frac{a+b}{2}} z^{a-b} (-1)^n z^n q^{\frac{n^2}{2}} \cr
&= \frac{(1-q)}{\prod_{n > 0}(1-q^n)^3}\sum_{a,b \geq 0 }q^{\frac{a+b}{2}}(-1)^{a+b} q^{\frac{(a-b)^2}{2}} \cr
 & =\frac{1}{\prod_{n=0}^\infty (1- q^{n+1})^3}\left( 1 + 2 \sum_{n=1}^\infty (-1)^{n} q^{\frac{n(n+1)}{2}} \right)
\end{align}
Adding up the contributions for fixed $a-b$ one gets a geometric series which cancels the $1-q$ prefactor, 
leaving the expected answer on the last row, the same as $\chi[Y_{0,1,2}]$

The current $(1-\Psi^{-1}) \tilde J - \tilde I$ is local with $\tilde I$ and survives the quotient. 
It has level $(1-\Psi^{-1})^2 (2\Psi^{-1}+1) - (1-\Psi^{-1})= \Psi^{-3}(\Psi-1)(\Psi-2)$.
We can pick the normalization 
\begin{equation}
J_c = (\Psi-1)\tilde J  - \Psi \tilde I
\end{equation}
to match with the current which appears in $Y_{0,1,2}$.

\subsubsection{Degenerate modules}
In this realization, $H_\rho$ and $D_s$ are simply given by the 
BRST reduction of $U(2|1)$ and $U(1)$ Weyl modules built from irreducible finite-dimensional 
representations of the corresponding Lie algebras. 

Notice that the charge $s$ module for $U(1)_{\Psi^{-1}-1}$ dressed by appropriate powers of the 
off-diagonal currents will have charge $s$ under $J_c$. The characters are computed as 
\begin{align}
\chi_{D_s}[Y_{1,0,2}](z, w;q) &= q^{\frac{s^2}{2}\frac{\Psi}{1-\Psi}}\frac{(1-q)}{\prod_{n > 0}(1-q^n)^2}
\oint \frac{dz}{z}z^s \prod_{n > 0} (1-q^{n+\frac12} z) (1-q^{n+\frac12} z^{-1}) \cr
&= q^{\frac{s^2}{2}\frac{\Psi}{1-\Psi}}\frac{(1-q)}{\prod_{n > 0}(1-q^n)^3}
\oint \frac{dz}{z}z^s\frac{\sum_{n=-\infty}^\infty (-1)^n z^n q^{\frac{n^2}{2}}}{(1-q^{\frac12} z) (1-q^{\frac12} z^{-1})} \cr
&= q^{\frac{s^2}{2}\frac{\Psi}{1-\Psi}}\frac{(1-q)}{\prod_{n > 0}(1-q^n)^3}
\oint \frac{dz}{z}z^s \sum_{a,b \geq 0 }\sum_{n=-\infty}^\infty q^{\frac{a+b}{2}} z^{a-b} (-1)^n z^n q^{\frac{n^2}{2}} \cr
&= q^{\frac{s^2}{2}\frac{\Psi}{1-\Psi}}\frac{(1-q)}{\prod_{n > 0}(1-q^n)^3}\sum_{a,b \geq 0 }q^{\frac{a+b}{2}}(-1)^{a+b+s} q^{\frac{(a-b+s)^2}{2}} \cr
 & =q^{\frac{s^2}{2}\frac{\Psi}{1-\Psi}}\frac{1}{\prod_{n=0}^\infty (1- q^{n+1})^3}\left(  \sum_{n=0}^\infty (-1)^{n+s} q^{\frac{(n+s)^2+n}{2}}+ \sum_{n=1}^\infty (-1)^{n+s} q^{\frac{(n-s)^2+n}{2}} \right)
\end{align}
and match the S-dual description. 

The finite-dimensional irreducible representation of $\mathfrak{u}(2|1)$ we will use to define the $H_\nu$ modules 
are Kac modules labelled by a weight $(\nu_1,\nu_2,\nu_3)$. They are familiar in physics: one splits the 
odd generators in two halves, pick an irrep of the bosonic subalgebra and declare it annihilated by 
half of the odd generators. The rest of the module is built by acting with the other half of the odd generators. 

The character of a Weyl module $V_\nu$ of this type should take the form 
\begin{align}
\chi_{V_\nu}&[U(2|1)](z_1,z_2, w;q) = q^{\Delta_\nu} w^{\nu_3}\frac{z_1^{\nu_1} z_2^{\nu_2} - z_2^{\nu_1+1} z_1^{\nu_2-1}}{1-\frac{z_2}{z_1}} \cdot \cr 
&\cdot \frac{(1-\frac{z_2}{w})(1-\frac{w}{z_1})\prod_{n > 0}(1-q^n \frac{z_1}{w}) (1-q^n \frac{z_2}{w}) (1-q^n \frac{w}{z_1}) (1-q^n \frac{w}{z_2}) }
{\prod_{n > 0}(1-q^n)^3 (1-\frac{z_1}{z_2} q^n)(1-\frac{z_2}{z_1} q^n)}
\end{align}
Adding the ghosts and specializing the fugacities for the DS reduction gives
\begin{equation}
\chi_{V_\nu}[W_{2|1} U(2|1)](z, w;q) =  q^{\Delta'_\nu}w^{\nu_3}z^{\nu_1 + \nu_2}
\frac{(1 - q^{\nu_1-\nu_2+1}) \prod_{n \geq 0} (1-q^{n+\frac12} \frac{z}{w}) (1-q^{n+\frac12} \frac{w}{z}) }
{\prod_{n > 0}(1-q^n)^3}
\end{equation}
This appears to be the character of a degenerate module for ${\cal N}=2$ super-Virasoro times $U(1)$, with a 
highest weight vector of generic $U(1)$ charges and a single null vector at level $\nu_1-\nu_2+1$. 

Taking next the $U(1)$ quotient we get a very simple character 
\begin{equation}
\chi[Y_{1,0,2}]_{H_\nu}(z;q) = (-1)^{\nu_3}  q^{\Delta'_\nu+\frac{\nu_3^2}{2}} z^{\nu_1 + \nu_2+\nu_3 }\frac{1 - q^{\nu_1-\nu_2+1}}{\prod_{n > 0}(1-q^n)^3} 
\end{equation}

This description of the VOA makes $W_\mu$ into a possibly intricate magnetic object. 
It would be interesting to reconstruct the ancestor $U(2|1)_{\Psi^{-1}}$ module
and give a gauge theory interpretation. 

\subsubsection{$Y_{2,0,1}$ and parafermions.}
\begin{wrapfigure}{l}{0.20\textwidth}
\vspace{-10pt}
  \begin{center}
      \includegraphics[width=0.175\textwidth]{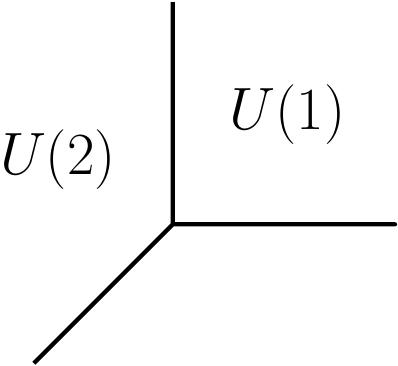}
\end{center}
\vspace{-10pt}
\end{wrapfigure}

The last realization of the VOA leads to an interface between $U(2|1)$ supergroup Chern-Simons theory and $U(2)$ Chern-Simons theory,
 with with $U(2)$ embedded inside $U(2|1)$ in the obvious block-diagonal way. This configuration leads to the BRST coset
\begin{eqnarray}
Y_{2,0,1}[\frac{1}{1-\Psi}]=\frac{U(2|1)_{-\frac{1}{1-\Psi}}}{U(2)_{-\frac{1}{1-\Psi}+1}}:\quad c=-\frac{6}{\Psi} +3
\end{eqnarray}
and we can see that central charge of the theory matches the previous two realizations.

The emergence of the parafermion VOA from such a coset is less familiar than the previous constructions, but it known \cite{Bowcock:kx}. 
Essentially, the para-fermions are used to dress spin $\frac12$ modules for $U(2)$ in order to assemble the odd currents of $U(2|1)$.

Notice that here $J^3_3$ has level $\frac{1}{1-\Psi}+2 =\frac{3-2\Psi}{1-\Psi}$ and OPE coefficient $2$ with $J^1_1 + J^2_2$, 
which has level $-\frac{2}{1-\Psi}+2 = -\frac{2 \Psi}{1-\Psi}$. The BRST-close combination $\frac{\Psi}{1-\Psi}J^3_3 + J^1_1 + J^2_2$
has level $\frac{\Psi^2 (2\Psi-3)}{(\Psi-1)^3} - 2 \frac{\Psi}{\Psi-1} = \frac{\Psi (\Psi-2)}{(\Psi-1)^3}$.
Hence we can identify tentatively 
\begin{equation}
J_c = (\Psi-1)J^3_3 + \frac{(\Psi-1)^2}{\Psi}(J^1_1 + J^2_2)
\end{equation}

The match of vacuum characters is striking. We begin with the familiar 
\begin{equation}
\chi[U(2|1)](z_1,z_2, w;q) = \frac{\prod_{n > 0}(1-q^n \frac{z_1}{w}) (1-q^n \frac{z_2}{w}) (1-q^n \frac{w}{z_1}) (1-q^n \frac{w}{z_2}) }
{\prod_{n > 0}(1-q^n)^3 (1-\frac{z_1}{z_2} q^n)(1-\frac{z_2}{z_1} q^n)}
\end{equation} 
In order to execute the (BRST) coset we multiply by the $U(2)$ character and the ghosts, take the contour integral
\begin{align}
\chi[V_{2,0,1}](w;q) &= \oint \frac{dz_1 dz_2}{2 z_1 z_2} (1-\frac{z_1}{z_2})(1-\frac{z_2}{z_1})\frac{\prod_{n > 0}(1-q^n \frac{z_1}{w}) (1-q^n \frac{z_2}{w}) (1-q^n \frac{w}{z_1}) (1-q^n \frac{w}{z_2}) }
{\prod_{n > 0}(1-q^n)} =\cr
&= \oint \frac{dz_1 dz_2}{2 z_1 z_2} (1-\frac{z_1}{z_2})(1-\frac{z_2}{z_1})\frac{\sum_{n,m\geq 0} \sum_{s=-n}^n \sum_{t=-m}^m(-1)^{n+m} q^{\frac{n(n+1)+m(m+1)}{2}}w^{s+t} z_1^{-s} z_2^{-t}}
{\prod_{n > 0}(1-q^n)^3} =\cr
&= \frac{\sum_{n,m\geq 0} (-1)^{n+m} q^{\frac{n(n+1)+m(m+1)}{2}}-\sum_{n,m\geq 1} (-1)^{n+m} q^{\frac{n(n+1)+m(m+1)}{2}} }
{\prod_{n > 0}(1-q^n)^3} =\cr
&= \frac{1+2 \sum_{n> 0} (-1)^{n} q^{\frac{n(n+1)}{2}}}
{\prod_{n > 0}(1-q^n)^3} 
\end{align} 
and obtain the expected vacuum character.

\subsubsection{Degenerate modules}
This realization of the VOA should gives a simple description of $H_\nu$ and $W_\mu$ 
in terms of standard representations of $U(2|1)$ and $U(2)$. 

For example, we can compute the character for $W_\mu$: 
\begin{align}
&\chi_{W_\mu}[V_{2,0,1}](w;q) = q^{\Delta_\mu}\oint \frac{dz_1 dz_2}{2 z_1 z_2} (1-\frac{z_1}{z_2})(z_1^{\mu_1} z_2^{\mu_2}-z_2^{\mu_1+1} z_1^{\mu_2-1}) \cdot  \cr
& \cdot\frac{\prod_{n > 0}(1-q^n \frac{z_1}{w}) (1-q^n \frac{z_2}{w}) (1-q^n \frac{w}{z_1}) (1-q^n \frac{w}{z_2}) }
{\prod_{n > 0}(1-q^n)} =\cr
&= q^{\Delta_\mu}\oint \frac{dz_1 dz_2}{2 z_1 z_2} (1-\frac{z_1}{z_2})(z_1^{\mu_1} z_2^{\mu_2}-z_2^{\mu_1+1} z_1^{\mu_2-1})\cdot \cr & \cdot \frac{\sum_{n,m\geq 0} \sum_{s=-n}^n \sum_{t=-m}^m(-1)^{n+m} q^{\frac{n(n+1)+m(m+1)}{2}}w^{s+t} z_1^{-s} z_2^{-t}}
{\prod_{n > 0}(1-q^n)^3} =\cr
&= q^{\Delta_\mu}w^{\mu_1 + \mu_2}\frac{\left(\sum_{n\geq |\mu_1|,m\geq |\mu_2|}-\sum_{n\geq |\mu_1+1|,m\geq |\mu_2-1|} \right) (-1)^{n+m} q^{\frac{n(n+1)+m(m+1)}{2}} }
{\prod_{n > 0}(1-q^n)^3}  
\end{align} 
It is not hard to match this with the dual calculation. 

We can also compute again the character for $H_\nu$: 
\begin{align}
\chi_{H_\nu}&[V_{2,0,1}](w;q) = q^{\Delta_\nu}w^{\nu_3} \oint \frac{dz_1 dz_2}{2 z_1 z_2} (1-\frac{z_1}{z_2})(z_1^{\nu_1} z_2^{\nu_2} - z_2^{\nu_1+1} z_1^{\nu_2-1})(1-\frac{z_2}{w})(1-\frac{w}{z_1})\cdot \cr & \cdot \frac{\prod_{n > 0}(1-q^n \frac{z_1}{w}) (1-q^n \frac{z_2}{w}) (1-q^n \frac{w}{z_1}) (1-q^n \frac{w}{z_2}) }
{\prod_{n > 0}(1-q^n)} =\cr
&= q^{\Delta_\nu}w^{\nu_3}\oint \frac{dz_1 dz_2}{2 z_1 z_2} (1-\frac{z_1}{z_2})(z_1^{\nu_1} z_2^{\nu_2} - z_2^{\nu_1+1} z_1^{\nu_2-1})\cdot \cr & \cdot \frac{\sum_{n,m=-\infty}^\infty (-1)^{n+m} q^{\frac{n(n+1)+m(m+1)}{2}}w^{n+m} z_1^{-n} z_2^{-m}}
{\prod_{n > 0}(1-q^n)^3} =\cr
&= (-1)^{\nu_1 + \nu_2} q^{\Delta_\nu}w^{\nu_1 + \nu_2+ \nu_3}\frac{ q^{\frac{\nu_1(\nu_1+1)+\nu_2(\nu_2+1)}{2}}-q^{\frac{(\nu_1+2)(\nu_1+1)+\nu_2(\nu_2-1)}{2}}}
{\prod_{n > 0}(1-q^n)^3}  \cr
&= (-1)^{\nu_1 + \nu_2} q^{\Delta_\nu+\frac{\nu_1(\nu_1+1)+\nu_2(\nu_2+1)}{2}}w^{\nu_1 + \nu_2+ \nu_3}\frac{ 1-q^{\nu_1-\nu_2+1}}
{\prod_{n > 0}(1-q^n)^3}  
\end{align} 
which reduces again to the dual result

\section{Central charges, characters and 3d partitions}\label{sec:central}
We are now ready for some preliminary investigation of the general $L$,$M$,$N$ setup. 
We begin by computing the central charge of the VOA and checking its duality invariance. 

\subsection{Central charges}
The definition of  $Y_{L,M,N}[\Psi]$ is somewhat different depending on the relative magnitude of $N$ and $M$. 
The final expression for the corresponding central charge $c_{L,M,N}[\Psi]$ will hold uniformly for all cases:
\begin{align}
c_{L,M,N}[\Psi]= &\frac12\frac{1}{\Psi} (L-N)\left((L-N)^2-1\right) +\frac12(1-\frac{1}{\Psi}) (N-L)\left((N-L)^2-1\right) + \cr
+&\frac12 \Psi(M-N)\left((M-N)^2-1\right) + \frac12 (1-\Psi)(N-M)\left((N-M)^2-1\right) \cr
+&\frac12\frac{1}{1-\Psi}(L-M)((L-M)^2-1)+\frac12 \frac{\Psi}{\Psi-1}(M-L)((M-L)^2-1)+ \cr
&+\frac12(2L-N-M)(2M-N-L)(2N-L-M)
\end{align}
which is manifestly $S_3$-symmetric. 

The calculation is straightforward, but the details are somewhat tedious. We present them in 
Appendix \ref{app:ccharge}. Notice that the answer only depends on the 
differences between $L$, $M$, $N$. Concretely, this happens because 
the central charge of $U(N|M)_\Psi$ only depends on $|N-M|$:
\begin{equation}
c_{U(N|M)_\Psi} = 1 + \frac{\Psi-N+M}{\Psi} \left((N-M)^2-1\right)
\end{equation}
More conceptually, it is likely a consequence of the fact that full D3 branes can be continuously added or removed from the system 
without breaking supersymmetry. 

\subsection{Characters}

Next, we can look at vacuum characters. For simplicity, we will focus at first on the situation where 
at least one of the three labels $L$, $M$, $N$ vanishes. This allows us to avoid dealing with the subtleties of
superghost zeromodes in the BRST reductions. At the end, we will give some conjectural statements about 
general $L$, $M$, $N$.

\subsubsection{The vacuum character of $Y_{0,M,N}$.}
At first, we can consider the $N=M$ subcase. 

In order to compute the character, we start with the product of the vacuum characters of two 
$U(N)$ WZWs and the $N$ symplectic bosons, as a function of fugacities $x_i$ for the 
Cartan generators. The $\mathfrak{u}(N)$-valued ghost non-zeromodes 
precisely cancel the contributions to the character of the two sets of WZW currents. 

We trade the $c$ zeromode contributions for a contour integral projecting on $\mathfrak{u}(N)$ invariants: 
\begin{equation} \label{eq:contourNN}
\chi[V_{0,N,N}](q) = \frac{1}{N!} \oint \prod_{i=1}^N \frac{dx_i}{x_i} \frac{\prod_{i<j} (1-\frac{x_i}{x_j})(1-\frac{x_j}{x_i})}{\prod_{i,n} (1-q^{n+\frac12} x_i)(1-q^{n+\frac12} x^{-1}_i)}
\end{equation}

Notice the integral identity 
\begin{align}
& \oint \prod_{i=1}^N \frac{dx_i}{x_i} x_i^{s_i} \frac{1}{\prod_{i,n} (1-q^{n+\frac12} x_i)(1-q^{n+\frac12} x^{-1}_i)}=\cr
&= \frac{1}{ \prod_{n> 0} (1-q^n)^{2N}}
\sum_{n_i=0}^\infty \prod_i  (-1)^{\sum_i n_i} q^{\sum_i \frac{n_i(n_i+1)}{2}}q^{(n_i + \frac12)s_i} 
\end{align}
demonstrated in appendix \ref{app:id}, showing that this type of contour integrals can be evaluated as a sum over residues at 
$x_i = q^{n_i + \frac12}$. 

Thus we have 
\begin{align}
\chi[V_{0,N,N}](q) &= \frac{
\sum_{n_i=0|n_1<n_2 <\cdots<n_N}^\infty \prod_i  (-1)^{\sum_i n_i } q^{\sum_i \frac{n_i(n_i+1)}{2}} \prod_{i<j} (1-q^{n_i - n_j})(1-q^{n_j - n_i})}{ \prod_{n> 0} (1-q^n)^{2N}}
 \end{align}

It should be possible to simplify this espression further. In particular, expanding the character explicitly 
one notices that $\chi[V_{0,N,N}](q)$ differs from $\chi[V_{0,N-1,N-1}](q)$ only from the order $q^{N^2}$ on. 
In particular, the character has a well-defined $N \to \infty$ limit, which coincides with the MacMahon function 
$\prod_n (1-q^n)^{-n}$. It is natural to conjecture a relation to some $W_{1+\infty}$ algebra. 

In a similar manner, we can consider the case $M = N-1$. Now the vacuum character becomes 
\begin{equation}
\chi[V_{0,M,M+1}](q) = \frac{1}{M!}  \frac{1}{ \prod_{n> 0} (1-q^n)} \oint \prod_{i=1}^M \frac{dx_i}{x_i} \frac{\prod_{i<j} (1-\frac{x_i}{x_j})(1-\frac{x_j}{x_i})}{\prod_{i,n} (1-q^{n+1} x_i)(1-q^{n+1} x^{-1}_i)}
\end{equation}
i.e.
\begin{align}
\chi&[V_{0,M,M+1}](q) =  \cr
&= \frac{
\sum_{n_i=0|n_1<n_2 <\cdots<n_M}^\infty \prod_i  (-1)^{\sum_i n_i} (1-q^{n_i + 1}) q^{\sum_i \frac{n_i(n_i+1)}{2}} \prod_{i<j} (1-q^{n_i - n_j})(1-q^{n_j - n_i})}{ \prod_{n> 0} (1-q^n)^{2M+1}}
\end{align}

This is easily generalized to any $N>M$. We refer the reader to Appendix \ref{app:dsindex} for the calculation of the character of the DS reduction:
\begin{align}
\chi_{W_{N-M,\cdots} U(N)}(x_i;q)= & \prod_{n=1}^{\infty}\Bigg [\prod_{j=1}^{M}\frac{1}{1-x_jq^{n+\frac{N-M-1}{2}}}\frac{1}{1-x_j^{-1}q^{n+\frac{N-M-1}{2}}}\Bigg ] \cdot \cr
&\prod_{n=1}^{\infty}\Bigg [\prod_{i=1}^{N-M}\frac{1}{1-q^{n+i-1}}\prod_{i,j=1}^{M}\frac{1}{1-x_ix_j^{-1}q^n}\Bigg ]
\end{align}
Then 
\begin{align}\label{eq:contourMN}
\chi[V_{0,M,N}](q) = &\frac{1}{M!}  \frac{1}{\prod_{n=0}^{\infty} \prod_{i=1}^{N-M}(1-q^{n+i})} \cdot \cr
&\oint \prod_{i=1}^M \frac{dx_i}{x_i} \frac{\prod_{i<j} (1-\frac{x_i}{x_j})(1-\frac{x_j}{x_i})}{\prod_{n=0}^{\infty} \prod_{j=1}^{M}(1-x_jq^{n+\frac{N-M+1}{2}})(1-x_j^{-1}q^{n+\frac{N-M+1}{2}})}
\end{align}

The contour integral can be computed as before, resulting in a sum over residues evaluated at $x_i = q^{n_i+\frac{N-M-1}{2}}$:
\begin{align}
&\chi[V_{0,M,N}](q) = \frac{1}
{ \prod_{n> 0} (1-q^n)^{2M} \prod_{j=1}^{N-M}(1-q^{n+j}) } \cdot \cr 
&\sum_{\substack{n_i=0    \\  n_1<\cdots<n_M}}^\infty \prod_i  (-1)^{\sum_i n_i}q^{\sum_i \frac{n_i(n_i+1)}{2}} \prod_{j=1}^{N-M}(1-q^{n_i + j})  \prod_{i<j} (1-q^{n_i - n_j})(1-q^{n_j - n_i})\end{align}

Alternatively, we can give a combinatorial description of the fields in $V_{0,M,N}$: they are labelled by $U(M)$-invariant words
built from the following letters: $\partial^n W_i$ singlets of $U(M)$ of weight $i+n$, with $1 \leq i \leq N-M$, $\partial^n U$ fundamentals of $U(M)$ of weight 
$n+\frac{N-M+1}{2}$ and $\partial^n V$ anti-fundamentals of $U(M)$ of weight $n+\frac{N-M+1}{2}$.

Equivalently, we can quotient the collection of words built from singlets $\partial^n W_i$ and bilinears $\partial^n U \cdot \partial^m V$ by the relations satisfied by 
products of bilinears. If we ignore these relations, the  $\partial^n U \cdot \partial^m V$ give 
$1$ generator of weight $N-M+1$, $2$ of weight $N-M+2$, etc. and combine with the $\partial^n W_i$
to give a $W_{1+\infty}$-like set of generators.   

The first non-trivial relation should be $\det_{(M+1) \times (M+1)}\left( \partial^n U \cdot \partial^m V \right) = 0$, 
occurring at level $N-M+1 + N-M+3 + \cdots N+M+1 = (M+1)(N+1)$. 

\subsubsection{The vacuum character of $Y_{L,0,N}$.}
Here we need to do a DS reduction of $U(N|L)$ and then quotient by $U(L)$. 
The calculation is almost identical as in the previous section, except that the 
off-diagonal blocks are fermionic. Thus we have
\begin{equation} \label{eq:contourLN}
\chi[V_{L,0,N}](q) = \frac{1}{L!}  \oint \prod_{i=1}^L \frac{dx_i}{x_i}\frac{ \prod_{i<j} (1-\frac{x_i}{x_j})(1-\frac{x_j}{x_i})\prod_{n=0}^{\infty} \prod_{j=1}^{L}(1-x_jq^{n+\frac{N+1}{2}})(1-x_j^{-1}q^{n+\frac{N+1}{2}})}{\prod_{n=0}^{\infty} \prod_{i=1}^{N}1-q^{n+i}} 
\end{equation}
We can give again a combinatorial description of the generators of $V_{L,0,N}$: they are labelled by $U(L)$-invariant words
built from the following letters: $\partial^n W_i$ singlets of $U(L)$ of weight $i+n$, with $1 \leq i \leq N$, $\partial^n A$ fermionic fundamentals of $U(L)$ of weight 
$n+\frac{N+1}{2}$ and $\partial^n B$ fermionic anti-fundamentals of $U(M)$ of weight $n+\frac{N+1}{2}$.

Equivalently, we can quotient the collection of words built from singlets $\partial^n W_i$ and bilinears $\partial^n A \cdot \partial^m B$ by the relations satisfied by 
products of bilinears.  If we ignore the relations which occur at finite $L$, the combinations of the form $\partial^n A \cdot \partial^m B$ give 
$1$ generator of weight $N+1$, $2$ of weight $N+2$, etc. and combine with the $\partial^n W_i$
to give a $W_{1+\infty}$-like set of generators.   

The first non-trivial relation should be $(A \cdot B)^{L+1}=0$, occurring at level $(L+1)(N+1)$. 

The symmetry under $N \leftrightarrow L$ of the character is far from obvious either from the character or from the combinatorial description. 
The full equality between the S-dual pairs of characters $\chi[V_{0,M,N}](q)$ and $\chi[V_{M,0,N}](q)$ is also far from obvious. 

We will soon derive these equalities by identifying the three contour integrals 
as different ways to count a certain class of 3d partitions by diagonal slicing. 

\subsubsection{A general conjecture}
The combinatorial description of the generators for $V_{0,M,N}$ and $V_{L,0,N}$
has an obvious generalization: they should be labelled by $U(M|L)$-invariant words
built from the following letters: $\partial^n W_i$ singlets of $U(M|L)$ of weight $i+n$, with $1 \leq i \leq N-M$, $\partial^n {\cal U}$ fundamentals of $U(M|L)$ (i.e. sets of 
$M$ bosons and $L$ fermions) of weight 
$n+\frac{N-M+1}{2}$ and $\partial^n {\cal V}$ anti-fundamentals of $U(M|L)$ (i.e. sets of 
$M$ bosons and $L$ fermions) of weight 
$n+\frac{N-M+1}{2}$

Equivalently, we can quotient the collection of words built from singlets $\partial^n W_i$ and bilinears $\partial^n {\cal U} \cdot \partial^m {\cal V}$ by the relations satisfied by 
products of bilinears.  If we ignore the relations which occur at finite $L$ and $M$, the combinations of the form $\partial^n {\cal U} \cdot \partial^m {\cal V}$ give 
$1$ generator of weight $N-M+1$, $2$ of weight $N-M+2$, etc. and combine with the $\partial^n W_i$
to give a $W_{1+\infty}$-like set of generators.   

The first non-trivial relation should involve a mixed symmetrization of the $\partial^n {\cal U}$ labels in a product of bilinears which vanishes 
for fundamentals of $U(M|L)$. The representations $R_{a,s}$ of $U(M|L)$ labelled by rectangular Young Tableaux obtained 
from mixed symmetrization of fundamentals of $U(M|L)$ are non-vanishing for $(a,s)$ inside the ``$(M,L)$-hook'', the 
difference between the positive quadrant and the shifted quadrant with $s=L+1$, $a=M+1$. 

The first non-trivial vanishing condition occurs for $R_{M+1,L+1}$. This is a modification of the $(L+1)$-th power of the determinant  
$\det_{(M+1) \times (M+1)}\left( \partial^n {\cal U} \cdot \partial^m {\cal V} \right)$ and should have weight 
$(L+1)(M+1)(N+1)$. 

The full triality of this set of generators is far from obvious. We will now make it obvious by identifying the three contour integrals 
as different ways to count a certain class of 3d partitions by diagonal slicing. 

\subsection{Crystal melting and vacuum characters}
We now formulate the following conjecture: the generators of $Y_{L,M,N}$ are in one-to-one correspondence 
with 3d partitions (as in the crystal melting story \cite{Okounkov:uq}) restricted to lie in the difference between 
the positive octant and the shifted positive octant 
with origin at $L$, $M$, $N$. Notice that unrestricted 3d partitions are counted by the McMahon function 
\begin{equation}
\chi_\infty(q) = \frac{1}{\prod_{n>0} (1-q^n)^n}
\end{equation}
which also counts generators of $W_{1+\infty}$. 

We will verify this conjecture in examples of increasing complexity.

\subsubsection{Melting crystals for $Y_{L,0,0}$}

Consider 3d partitions restricted to the slab $0 \leq z \leq L$, $0 \leq x$, $0 \leq y$. 
The diagonal slicing of such a 3d partition gives a sequence of 2d partitions 
$\mu_n$ restricted to lie in the horizontal 2d slab $0 \leq z \leq L$, $0 \leq w$. 
Consecutive 2d partitions must be interlaced: $\mu_{n} \prec \mu_{n+1}$ for negative $n$, 
$\mu_{n} \prec \mu_{n-1}$ for positive $n$. 

\begin{figure}[h]
  \centering
      \includegraphics[width=0.43\textwidth]{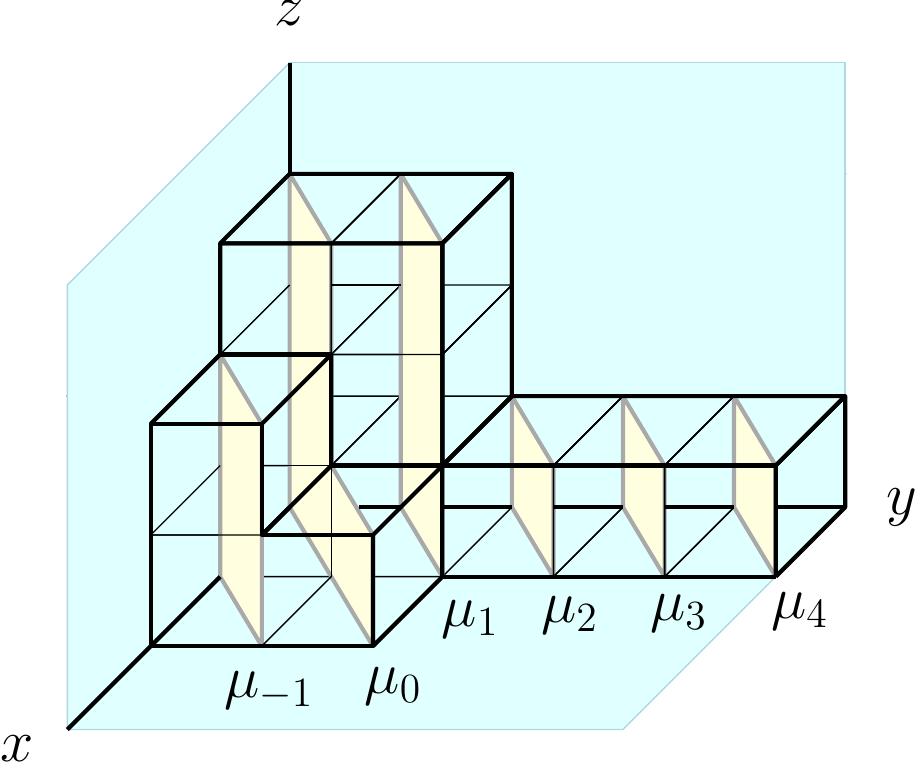}
\caption{Diagonal slicing of a 3d partition as described in the main text. Note that the two series of partitions are indeed interlaced $\mu_{-1}<\mu_0>\mu_1>\mu_2=\mu_3=\mu_4$.}
\end{figure}

Here interlaced means 
\begin{equation}
\mu \prec \nu := \nu_1 \geq \mu_1 \geq \nu_2 \geq \mu_2 \cdots
\end{equation}
and $\mu_i$ are the column heights of $\mu$. 

We can identify restricted 2d partitions with characters $\chi_{\lambda^t}(x_i)$ of $U(L)$ irreducible representations,
orthogonal under the Vandermonde measure. If we expand 
\begin{equation}
\prod_{n\geq 0} \prod_{i=1}^L (1+ x_i q^{n+\frac12}) = \sum_\lambda c_\lambda(q) \chi_{\lambda^t}(x_i)
\end{equation}
then the contour integral for the vacuum character for $Y_{L,0,0}$ can be written as 
the norm squared 
\begin{equation}
\chi[Y_{L,0,0}](q) = \sum_\lambda c_\lambda(q) c_\lambda(q)
\end{equation}

We would like to claim that $c_\lambda(q)$ counts the sequences of interlaced partitions $\mu_{n} \prec \mu_{n-1}$,
with $n \geq 0$ and $\mu_0 = \lambda$, weighed by $q^{\frac{|\lambda|}{2} + \sum_{n>0} |\mu_n|}$, 
with $|\mu|$ being the total number of boxes in $\mu$. This follows in a standard way from the basic identity
\begin{equation}
\chi_{\mu^t}(x_i) \prod_{j=1}^L (1+ x_j) = \sum_{\nu \succ \mu} \chi_{\nu^t}(x_i)
\end{equation}
Thus $\chi[Y_{L,0,0}](q)$ counts the 3d partitions restricted as above.
\subsubsection{Melting crystals for $Y_{0,0,N}$}
Next, we can try to count the same set of 3d partitions, but sliced along a different axis. 
We will thus consider 3d partitions restricted to lie in the slab $0 \leq z$, $0 \leq x$, $0 \leq y \leq N$.

As we slice such a partition, we get a sequence $\mu_n$ of interlaced 2d partitions with the following properties:
for $n\geq 0$ they are restricted to lie in the vertical 2d slab $0 \leq z$, $0 \leq w \leq N$, while for $n<0$
they lie in the vertical 2d slabs $0 \leq z$, $0 \leq w \leq N+n$.

We can identify a 2d partition $\lambda$ restricted to the vertical 2d slab $0 \leq z$, $0 \leq w \leq N$ with a character $\chi_{\lambda}$
for $U(N)$. We define
\begin{equation}
\frac{1}{\prod_{n\geq 0} \prod_{i=1}^N (1- x_i q^{n+\frac12})} = \sum_\lambda d_\lambda(q) \chi_{\lambda}(x_i)
\end{equation}
We claim that $d_\lambda(q)$ counts as before the sequences of interlaced 2d partitions $\mu_n$, $n \geq 0$, $\mu_0 = \lambda$, but
restricted to the vertical 2d slab $0 \leq z$, $0 \leq w \leq N$. This follows in a standard way from the basic identity
\begin{equation}
\frac{\chi_{\mu}(x_i)}{\prod_{j=1}^N (1- x_j)} = \sum_{\nu \succ \mu} \chi_{\nu}(x_i)
\end{equation}

Furthermore, we claim that the counting function for the sequences of interlaced 2d partitions $\mu_n$, $n \geq 0$, $\mu_0 = \lambda$
restricted to the vertical 2d slabs $0 \leq z$, $0 \leq w \leq N-n$ is 
\begin{equation}
\chi_{\lambda}(x_i = q^{i - \frac12})
\end{equation}
This follows in a standard way from the basic identity
\begin{equation}
\chi^{U(n)}_{\mu}(x_i = y_i, x_n = 1) = \sum_{\nu \prec \mu} \chi^{U(n-1)}_{\nu}(y_i)
\end{equation}

Thus we recover the character of $Y_{0,0,N}$ from this diagonal slicing of the restricted 3d partitions as
\begin{equation}
\chi[Y_{0,0,N}](q) = \sum_\lambda d_\lambda(q) \chi_{\lambda}(x_i = q^{N-i + \frac12}) = \frac{1}{\prod_{n\geq 0} \prod_{i=1}^N (1- q^{n+i})}
\end{equation}

\subsubsection{Melting crystals for $Y_{0,N,N}$}
The contour integral for the character of $Y_{0,N,N}$ \ref{eq:contourNN} gives immediately 
\begin{equation}
\chi[V_{0,N,N}](q) = \sum_\lambda d_\lambda d_\lambda
\end{equation}
The combinatorial interpretation is simple: this counts 3d partitions which lead to 
sequences $\mu_n$ of interlaced 2d partitions which are restricted to lie in the vertical 2d slab $0 \leq z$, $0 \leq w \leq N$
for all $n$. These are precisely 3d partitions restricted to lie in the difference between the positive octant and the shifted positive octant 
with origin at $z=0$, $x=N$, $y=N$!

\subsubsection{Melting crystals for $Y_{0,M,N}$, $M<N$}
Next, we want to count 3d partitions restricted to lie in the difference between the positive octant and the shifted positive octant 
with origin at $z=0$, $x=M$, $y=N$, $M<N$. 

As we take a diagonal slicing, we get a sequences $\mu_n$ of interlaced 2d partitions which are restricted to lie in the vertical 2d slab $0 \leq z$, $0 \leq w \leq N$ for $n\geq0$ and $0 \leq z$, $0 \leq w \leq N+n$ for $M-N \leq n \leq 0$ and $0 \leq z$, $0 \leq w \leq M$ for $n \leq M-N$. 

The counting function $e_\lambda(q)$ for the sequence $\mu_n$ for $n\geq M-N$ is 
\begin{align}
& \sum_\lambda e_\lambda(q) \chi^{U(M)}_{\lambda^t}(x_i)=\sum_\lambda d^{U(N)}_\mu(q) \chi^{U(N)}_{\mu^t}(y_i=q^{N-M} x_i, y_{M+j} = q^{N-M-j+\frac12}) =
\cr &= \frac{1}{\prod_{n\geq 0} \prod_{i=1}^M (1- x_i q^{n+N-M+\frac12})\prod_{i=1}^{N-M} (1- q^{n+i})}
\end{align}

The inner product with the counting function $d^{U(M)}_\lambda(q)$ for the sequence $\mu_n$ for $n\leq M-N$
gives precisely the contour integral representation \ref{eq:contourMN} of $\chi[V_{0,M,N}](q)$!

\subsubsection{Melting crystals for $Y_{L,0,N}$}
Now we are ready to slice the previous 3d partitions along a different axis. 
We want to count 3d partitions restricted to lie in the difference between the positive octant and the shifted positive octant 
with origin at $z=L$, $x=0$, $y=N$, $M<N$. 

The sequence of 2d partitions now includes partition restricted to the 2d L-hook $R_{n,L}$: 
the difference between the positive quadrant and the shifted positive quadrant 
with origin at $z=L$, $w = n$. These can be identified with characters $\chi_{\lambda}(x_i; y_a)$ of irreducible representations 
of $U(n|L)$. We will assume now some simple combinatorial 
relations which generalize the relations we used until now for $U(n)$ characters. 

In this particular setup, the slicing of a restricted 3d partition gives a sequence of interlaced 2d partitions 
which are restricted to lie in $R_{N,L}$ for $n\geq0$ and $R_{N+n,L}$ for $-N \leq n \leq 0$ and $0 \leq z \leq L$, $0 \leq w $ for $n \leq -N$.

We expect the following combinatorial statement to be true 
\begin{equation}
\chi_{\mu}(x_i;y_a)\frac{\prod_{a=1}^L (1+y_a)}{\prod_{j=1}^N (1- x_j)} = \sum_{\nu \succ \mu} \chi_{\nu}(x_i;y_a)
\end{equation}
We also expect 
\begin{equation}
\chi^{U(n|L)}_{\mu}(x_i = x'_i, x_n = 1;y_a) = \sum_{\nu \prec \mu} \chi^{U(n-1|L)}_{\nu}(x'_i;y_a)
\end{equation}

These statements then imply that the coefficients in 
\begin{equation}
\frac{\prod_{n\geq 0} \prod_{i=1}^N (1+ y_a q^{n+\frac12})}{\prod_{n\geq 0} \prod_{i=1}^N (1- x_i q^{n+\frac12})} = \sum_\lambda d^{U(n|L)}_\lambda(q) \chi_{\lambda}(x_i;y_a)
\end{equation}
counts as before the sequences of interlaced 2d partitions $\mu_n$, $n \geq 0$, $\mu_0 = \lambda$, but
restricted to $R_{N,L}$. 

They also imply that the coefficients in 
\begin{equation}
\frac{\prod_{n\geq 0} \prod_{i=1}^N (1+ y_a q^{N+n+\frac12})}{\prod_{n\geq 0} \prod_{i=1}^N (1- q^{n+i})} = \sum_\lambda f_\lambda(q) \chi^{U(L)}_{\lambda^t}(y_a)
\end{equation}
counts as before the sequences of interlaced 2d partitions $\mu_n$, $n \geq -N$, $\mu_{-N} = \lambda$, 
to lie in $R_{N,L}$ for $n\geq0$ and $R_{N+n,L}$ for $-N \leq n \leq 0$.

The contour integral \ref{eq:contourLN} for $Y_{L,0,N}$ then coincides with the inner product 
\begin{equation}
\chi[V_{L,0,N}](q) = \sum_\lambda c_\lambda(q) f_\lambda(q)
\end{equation}
i.e. the counting function of the restricted 3d partitions. 

\subsubsection{Melting crystals for $Y_{L,M,N}$}

We want to count 3d partitions restricted to lie in the region $R_{L,M,N}$, defined as the difference between the positive octant and the shifted positive octant with origin at $z=L$, $x=M$, $y=N$, $M\leq N$.

\begin{figure}[h]
  \centering
      \includegraphics[width=0.43\textwidth]{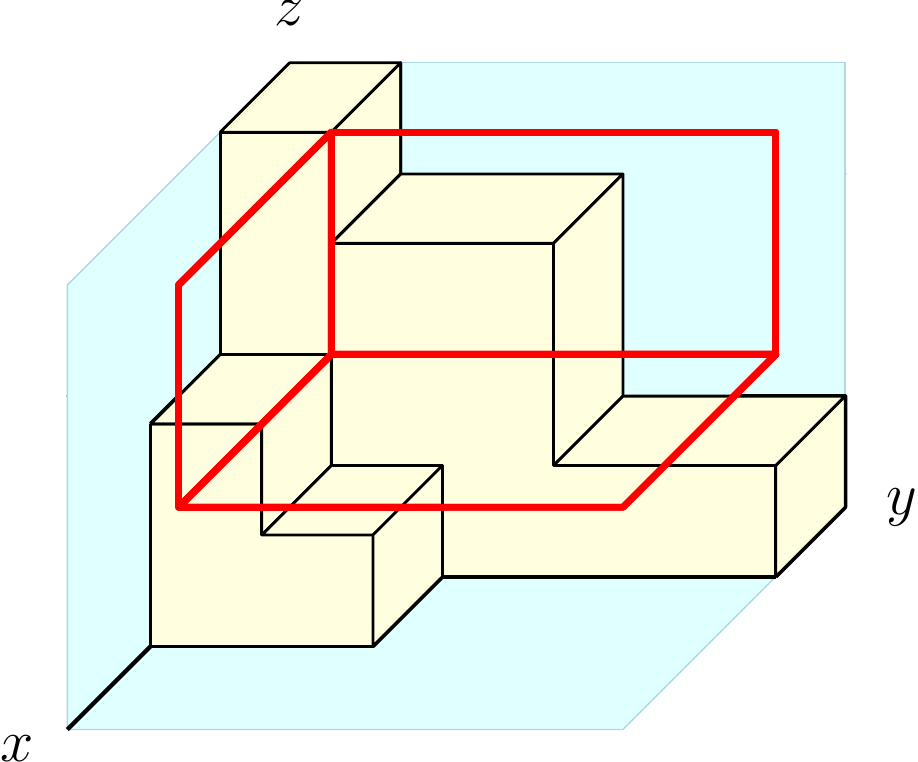}
\caption{Example of a 3d partition for algebra $Y_{2,1,1}[\Psi]$. All the boxes of all allowed partitions are constrained to lie between the corner with a vertex at the origin and shifted (red) corner with vertex at $(2,1,1)$.}
\end{figure}

We can proceed as before. The slicing of a restricted 3d partition gives a sequence of interlaced 2d partitions 
which are restricted to lie in $R_{N,L}$ for $n\geq0$ and $R_{N+n,L}$ for $M-N \leq n \leq 0$ and $R_{M,L}$ for $n \leq -N$.

The coefficients in 
\begin{equation}
\frac{\prod_{n\geq 0} \prod_{i=1}^N (1+ y_a q^{N-M+n+\frac12})}{\prod_{n\geq 0} \prod_{i=1}^{M} (1- x_i q^{N-M+n+\frac12}) \prod_{i=1}^{N-M} (1- q^{n+i})} = \sum_\lambda f^{U(M|L)}_\lambda(q) \chi^{U(M|L)}_{\lambda}(x_i;y_a)
\end{equation}
counts as before the sequences of interlaced 2d partitions $\mu_n$, $n \geq M-N$, $\mu_{M-N} = \lambda$, 
to lie in $R_{N,L}$ for $n\geq0$ and $R_{N+n,L}$ for $M-N \leq n \leq 0$.

The counting function of the restricted 3d partitions is the inner product 
\begin{equation}
\sum_\lambda d^{U(M|L)}_\lambda(q) f^{U(M|L)}_\lambda(q)
\end{equation}
is the natural projection to $U(M|L)$ invariants of the character for the ingredients of the BRST reduction 
defining $Y_{L,M,N}$. We expect it to be the correct vacuum character for $Y_{L,M,N}$.

\subsection{Characters of degenerate modules and melting crystals} \label{sec:dcharacters}
It is straightforward to modify the vacuum character calculations 
in order to compute the characters of degenerate modules of type $W$ or $D$:
essentially, one just inserts characters of finite-dimensional irreducible representations in the contour integrals,
with fugacities associated to DS-reduced directions specialized to the appropriate powers of $q$. 

There is an obvious extension of the 3d partition counting problem 
we associated to the vacuum characters of $Y_{L,M,N}$: one may consider 
3d partitions with semi-infinite cylindrical ends modelled on 2d partitions $\lambda$, $\mu$, $\nu$, 
as in the definition of the topological string vertex \cite{Okounkov:uq}.

The crucial observation is that the restriction for the 3d partition to lie in the region $R_{L,M,N}$
forces $\lambda$, $\mu$, $\nu$ to lie respectively in $R_{M,N}$, $R_{N,L}$ and $R_{L,M}$. 
Thus $\lambda$, $\mu$, $\nu$ have precisely the same form as the data labelling 
our degenerate modules $M_{\lambda, \mu, \nu} = W_\mu \times H_\lambda \times D_\nu$. 

It is nice to observe that the 3d counting for $\lambda = 0$ is particularly simple, 
in the same way as the computation of characters of $W_\mu \times D_\nu$ is 
particularly simple. 

Up to ``framing factors'', the computation simply inserts extra factors of $\chi^{U(N|L)}_\mu$ and 
$\chi^{U(M|L)}_\mu$ in order to implement the boundary conditions on the 2d partitions at large positive and negative $n$. 
This is precisely the same as what we would do to compute the characters of $W_\mu \times D_\nu$.

We are thus lead to the conjecture that the counting of 3d partitions with semi-infinite ends restricted to $R_{L,M,N}$ 
computes the character of $M_{\lambda, \mu, \nu}$ for $Y_{L,M,N}$.

\section{Ortho-symplectic $Y$-algebras} \label{sec:orthosymplectic}
\subsection{Branes and O3-planes}
In this section, we describe the generalization of the above construction to a $Y$-junction of defects in $\mathcal{N}=4$ SYM with orthogonal and symplectic gauge groups.
Theories with these gauge groups can be realized by $D3$-branes sitting on an O3-plane. 
The gauge theory perspective on boundary conditions and interfaces associated to fivebranes in the presence of 
O3-planes was developed in \cite{Gaiotto:2008aa}, building on a broad literature in string theory \cite{Gimon:aa,Kutasov:aa,Sen:aa} and gauge theory \cite{Hanany:aa,Feng:aa,Argyres:aa}.

There are four $O3$-planes in type IIB string theory. When superimposed to a stack of D3 brane, they give rise to four possible choices of gauge groups: 
$O3^-$ planes give an $SO(2n)$ gauge theory, $\tilde{O}3^-$ planes give an $SO(2n+1)$ gauge theory, $O3^+$ planes give an $Sp(2n)$ gauge theory
and $\tilde{O}3^+$ planes give a gauge theory denoted as $Sp(2n)'$, which is the same as $Sp(2n)$ but has a different convention for the $\theta$ angle, 
so that $\theta=0$ in $Sp(2n)'$ is the same as $\theta = \pi$ in $Sp(2n)$.

The $O3^-$ plane is unaffected by duality transformations. Correspondingly, $SO(2n)$ ${\cal N}=4$ SYM has a $PSL(2,Z)$ S-duality group. 
The remaining three types of $O3$ planes are exchanged by duality transformations. A $T$ transformation clearly maps $Sp(2n) \leftrightarrow Sp(2n)'$ and relates 
$O3^+$ and $\tilde{O}3^+$. It leaves $\tilde{O}3^-$ invariant. On the other hand, an $S$ transformation exchanges the $Sp(2n)$ and $SO(2n+1)$ gauge groups
and the $\tilde{O}3^-$ and $O3^+$ planes, while it maps $Sp(2n)'$ to itself and leaves $\tilde{O}3^+$ invariant. 

The story is further complicated by the fact that the elementary interfaces in the presence of O3 planes are associated to ``half-fivebranes'' that are $Z_2$ projections of ordinary fivebranes. 
The type of O3 planes jumps across these interfaces. As a consequence, half-NS5 interfaces must interpolate between $SO(2n)$ and $Sp(2m)$ 
or between $SO(2n+1)$ and $Sp(2m)'$:
\begin{figure}[h!]
  \centering
      \includegraphics[width=0.65\textwidth]{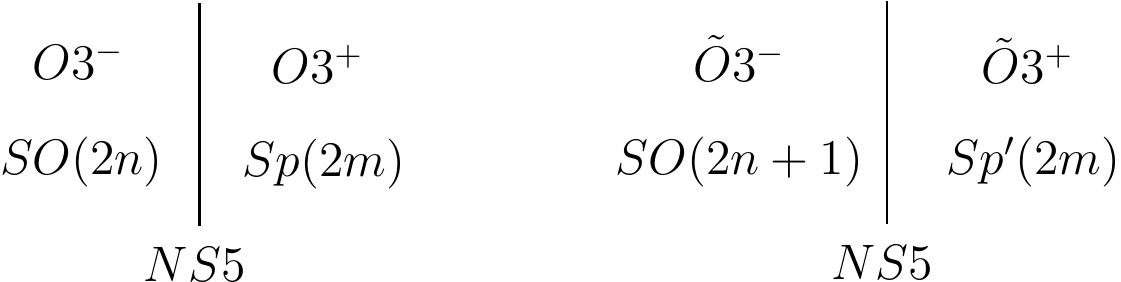}
\end{figure}\\
while half-D5 interfaces must interpolate between $SO(2n)$ and $SO(2m+1)$ or $Sp(2n)$ and $Sp(2m)'$: 
\begin{figure}[h!]
  \centering
      \includegraphics[width=0.65\textwidth]{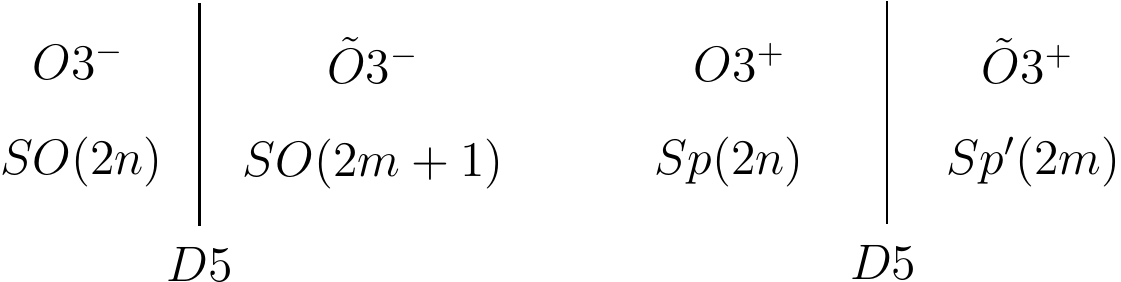}
\end{figure}

The gauge theory description of the interfaces is very similar to the unitary cases, except that the orbifold projection
cuts in half the interface degrees of freedom. Half-NS5 interfaces support ``half-hypermultiplets''
transforming as bi-fundamentals of $SO \times Sp$. \footnote{Notice that half-hypermultiplets must transform in a symplectic representation, 
precluding such elementary interfaces for $SO \times SO$ or $Sp \times Sp$. Furthermore, half-hypermultiplets have a potential anomaly which has to be cancelled by 
inflow from the bulk, constraining the choice of $Sp(2n)$ vs $Sp(2n)'$ as predicted by string theory.} 

Half-D5 interfaces between orthogonal groups involve a Nahm pole of odd rank \footnote{Notice that the rank of the Nahm pole must be odd for 
the $\mathfrak{su(2)}$ embedding to exist in an orthogonal group}. Half-D5 interfaces between symplectic groups 
involve a Nahm pole of even rank or a half-hypermultiplet in the fundamental representation of $Sp$. \footnote{Notice that the rank of the Nahm pole must be even for 
the $\mathfrak{su(2)}$ embedding to exist in an orthogonal group. Also, the type of $Sp$ theory must jump across the interface for the same anomaly inflow constraint mentioned in the previous footnote.}

The half-$(1,1)$-type interfaces work in a similar manner as half-NS5 interfaces, except that the role of $Sp'$ and $Sp$ is reversed because of the extra interface Chern-Simons terms. 

The relation between the four-dimensional gauge theory setup and analytically continued Chern-Simons theory works in the same manner 
as in the unitary case, up to matter of conventions for the levels of the corresponding Chern-Simons theories. 

We will employ $OSp(n|2m)_\kappa$ WZW models. We use conventions where $\kappa$ is the level of the $SO$ currents and $-\kappa/2$ the level of the Sp currents. 
The dual Coxeter number for $SO(n)$ is $n-2$ and for $Sp(2m)$ is $m+1$. The critical level for $OSp(n|2m)$ is $2-n+2m$. A half-NS5 interface in the presence of gauge theory parameter $\Psi$ 
will result in an $OSp(n|2m)_{\pm \Psi - n +2m+2}$ theory, depending on which side of the interface the $SO$ and $Sp$ or $Sp'$ groups lie. Notice that the level of the $Sp$ WZW currents 
differ by an half-integral amount from $\pm \Psi$ if $n$ is odd, which is when we have an $Sp(2m)'$ gauge group in four dimensions.

The relation between Nahm poles and DS reductions will be the same as before. Furthermore, half-hypermultiplets 
in the fundamental representation of $Sp(2m)$ will map to symplectic bosons which support $Sp(2m)_{-\frac12}$ WZW currents.
Adding $n$ Majorana chiral fermions will promote that to $OSp(n|2m)_{1}$ WZW currents. See Appendix \ref{app:conventions} for details. 

\subsection{Definition of ortho-symplectic $Y$-algebras}

Depending on the choice of O3 plane in the top right corner, the Y-junction setup for orthogonal and symplectic gauge groups 
gives rise to four classes of ortho-symplectic $Y$-algebras: $Y^\pm_{L,M,N}[\Psi]$ and $\tilde Y^\pm_{L,M,N}[\Psi]$. 

Because of the duality properties of O3 planes, $\tilde Y^+_{L,M,N}[\Psi]$ will have the same triality properties as $Y_{L,M,N}[\Psi]$. 
Instead, triality will map into each other $Y^\pm_{L,M,N}[\Psi]$ and $\tilde Y^-_{L,M,N}[\Psi]$, up to the usual $S_3$ action 
on labels and coupling. 

In particular, the definition of the algebras will imply 
\begin{equation}
Y^+_{L,M,N}[\Psi] = Y^+_{L,N,M}[1-\Psi] \qquad \qquad Y^-_{L,M,N}[\Psi] = \tilde Y^-_{L,N,M}[1-\Psi] 
\end{equation}
and the non-trivial S-duality conjecture is
\begin{equation}
Y^+_{L,M,N}[\Psi] = \tilde Y^-_{M,L,N}[\frac{1}{\Psi}] \qquad \qquad Y^-_{L,M,N}[\Psi] = Y^-_{M,L,N}[\frac{1}{\Psi}]
\label{orthoS} 
\end{equation}
etcetera. 

\begin{figure}[h!]
  \centering
      \includegraphics[width=0.6\textwidth]{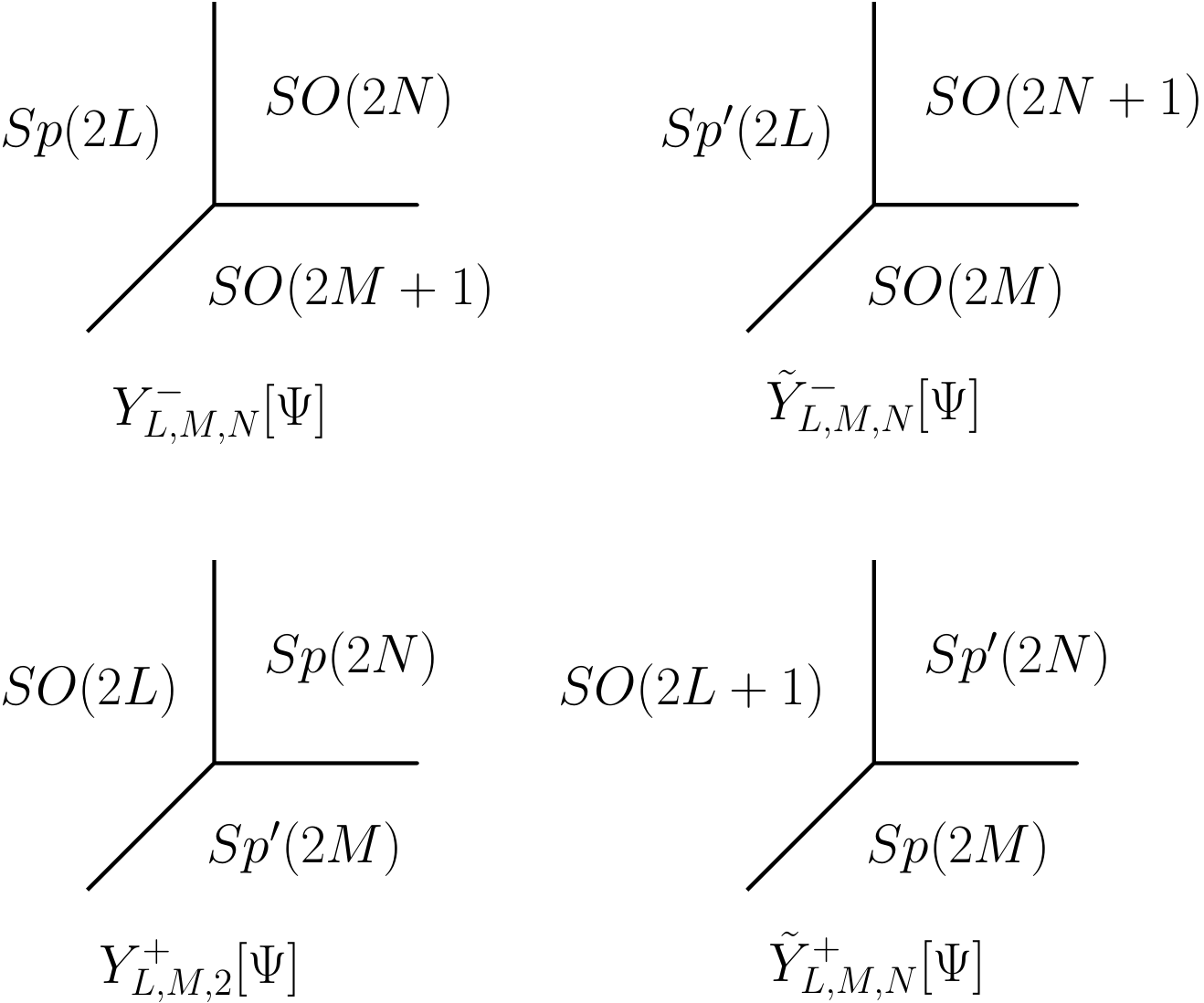}
\caption{Configurations defining ortho-symplectic $Y$-algebras.}
\label{fig:2}
\end{figure}

We will give now a brief definition of these vertex algebras. 

The VOAs $Y^-_{L,M,N}[\Psi]$ corresponding to the first figure in \ref{fig:2} are defined as follows. 

There are a super Chern-Simons theory with gauge groups $OSp(2N,2L)$ and $OSp(2M+1,2L)$ induced at the NS5 interfaces.
For $L=0$, $N=M$ or $N=M+1$, there is no Nahm-pole present and corresponding $Y$-algebra is a BRST reduction of
\begin{eqnarray}\nonumber
&SO(2M)_{\Psi-2M+2}\times SO(2M+1)_{-\Psi-2M+2}\\
&SO(2M+2)_{\Psi-2M}\times SO(2M+1)_{-\Psi-2M+2}
\end{eqnarray} 
that lead to cosets
\begin{eqnarray}\nonumber
Y^-_{0,M,M}[\Psi]&=&\frac{SO(2M+1)_{-\Psi-2M+2}}{SO(2M)_{-\Psi-2M+2}}\\
Y^-_{0,M,M+1}[\Psi]&=&\frac{SO(2M+2)_{\Psi-2M}}{SO(2M+1)_{\Psi-2M}}.
\end{eqnarray}

For $L=0$ and $N>M+1$ , the VOA is defined as a BRST reduction of the DS-reduction  by the $(2N-2M-1)\times (2N-2M-1)$ block
\begin{eqnarray}
W_{2N-2M-1}[SO(2N)_{\Psi-2N+2}]\times SO(2M+1)_{-\Psi-2M+2}
\end{eqnarray}
i.e. coset
\begin{eqnarray}
Y^-_{0,M,N}[\Psi]=\frac{W_{2N-2M-1}[SO(2N)_{\Psi-2N+2}]}{SO(2M+1)_{\Psi-2M}}.
\end{eqnarray}
and similary for $N<M$
\begin{eqnarray}
Y^-_{0,M,N}[\Psi]=\frac{W_{2M+1-2N}[SO(2M+1)_{\Psi-2M}]}{SO(2N)_{\Psi-2N+2}}.
\end{eqnarray}

For $L\neq 0$, levels of the super Chern-Simons theories are $\Psi-2N+2L+2$ and $-\Psi-2M+2L$ respectively. In the four cases described above, one gets BRST reductions of similar combinations of DS-reduced and not reduced theory leading to
\begin{eqnarray}\nonumber
Y^-_{L,M,M}[\Psi]&=&\frac{OSp(2M+1|2L)_{-\Psi-2M+2+2L}}{OSp(2M|2L)_{-\Psi-2M+2+2L}},\\ \nonumber 
Y^-_{L,M,M+1}[\Psi]&=&\frac{OSp(2M+2|2L)_{\Psi-2M+2L}}{OSp(2M+1|2L)_{\Psi-2M+2L}},\\ \nonumber
Y^-_{L,M,N}[\Psi]&=&\frac{W_{2N-2M-1}[OSp(2N|2L)_{\Psi-2N+2L+2}]}{OSp(2M+1|2L)_{\Psi-2M+2L}} \qquad  N>M+1,\\
Y^-_{L,M,N}[\Psi]&=&\frac{W_{2M+1-2N}[OSp(2M+1|2L)_{\Psi-2M+2L}]}{OSp(2N|2L)_{\Psi-2N+2+2L}} \qquad  N<M.
\end{eqnarray}

The VOA $\tilde Y^-_{L,M,N}[\Psi]$ corresponding to the second configuration in \ref{fig:2} are defined simply as 
\begin{eqnarray}
\tilde Y^-_{L,M,N}[\Psi] = Y^-_{L,N,M}[1-\Psi].
\end{eqnarray}

Let us now define the VOAs $Y^+_{L,M,N}[\Psi]$ corresponding to the bottom left diagram in \ref{fig:2}. 
Let $L=0$ and $N=M$. An $Sp(2N)$ Chern-Simons theory is induced at the vertical boundary with shift in the level by $\frac{1}{2}$. The anomaly mismatch compensated by a (half)-symplectic boson in a fundamental representation of $Sp(2N)$. The VOA is then identified with the BRST reduction of
\begin{eqnarray}
Sp(2N)_{\frac{\Psi}{2}-N-1} \times Sb^{Sp(2N)}\times Sp(2N)_{- \frac{\Psi}{2} -N-\frac12}
\end{eqnarray}
i.e. the coset
\begin{eqnarray}
Y^+_{0,N,N}[\Psi]=\frac{Sp(2N)_{\frac{\Psi}{2}-N-1}  \times Sb^{Sp(2N)}}{Sp(2N)_{\frac{\Psi}{2}-N-\frac32} }.
\end{eqnarray}

If $M\neq N$, there are no symplectic bosons present but Nahm-pole boundary conditions appears leading for $N>M$ to
\begin{eqnarray}
Y^+_{0,M,N}[\Psi]=\frac{W_{2N-2M}Sp(2N)_{\frac{\Psi}{2}-N-1}}{Sp(2M)_{\frac{\Psi}{2}-M-\frac32}}
\end{eqnarray}
and for $N<M$ to
\begin{eqnarray}
Y^+_{0,M,N}[\Psi]=\frac{W_{2M-2N}[Sp(2M)_{- \frac{\Psi}{2} -M-\frac12}]}{Sp(2N)_{-\frac{\Psi}{2}-N-1}}.
\end{eqnarray}

If $L\neq 0$, one gets analogous expression with super-groups and dual super-Coxeter numbers:
\begin{eqnarray}\nonumber
Y^+_{L,N,N}[\Psi]=\frac{OSp(2L|2N)_{-\Psi+2N-2L+2}\times Sb^{OSp(2L|2N)}}{OSp(2L|2N)_{-\Psi+2N-2L+3}},\\
N>M \qquad Y_{L,M,N}[\Psi]=\frac{W_{2N-2M}[OSp(2L|2N)_{-\Psi+2N-2L+2}]}{OSp(2L|2M)_{-\Psi+2M-2L+3}} \\
N<M \qquad Y_{L,M,N}[\Psi]=\frac{W_{2M-2N}[OSp(2L|2M)_{-\Psi+2M-2L+1}]}{OSp(2L|2N)_{-\Psi+2M-2L+2}}.
\end{eqnarray}

The last diagram of \ref{fig:2} gives rise to $\tilde Y^+_{L,M,N}[\Psi]$:
\begin{eqnarray}\nonumber
\tilde Y^+_{L,N,N}[\Psi]=\frac{OSp(2L+1|2N)_{-\Psi+2N-2L+1}\times Sb^{OSp(2L+1|2N)}}{OSp(2L+1|2N)_{-\Psi+2N-2L+2}},\\
N>M \qquad \tilde Y^+_{L,M,N}[\Psi]=\frac{W_{2N-2M}[OSp(2L+1|2N)_{-\Psi+2N-2L+1}]}{OSp(2L+1|2N)_{-\Psi+2M-2L+2}} \\
N<M \qquad \tilde Y^+_{L,M,N}[\Psi]=\frac{W_{2M-2N}[OSp(2L+1|2M)_{-\Psi+2M-2L}]}{OSp(2L+1|2N)_{-\Psi+2M-2L+1}}.
\end{eqnarray}
where $Sb^{OSp(n|2N)}$ denotes a combination of $N$ symplectic bosons and $n$ real fermions which 
supports bilinear $OSp(n|2N)$ currents. 

\subsection{Super-Virasoro}

\begin{wrapfigure}{l}{0.22\textwidth}
  \begin{center}
      \includegraphics[width=0.185\textwidth]{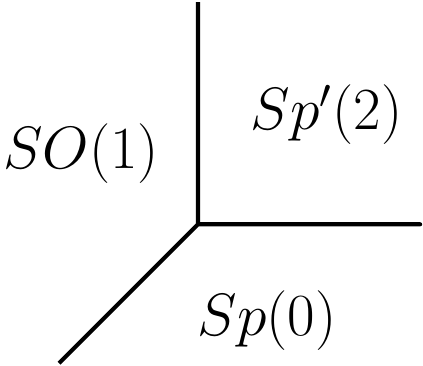}
\end{center}
\end{wrapfigure}

We will quickly look at the simplest example of $OY$ algebra: $\tilde Y^+_{1,0,2}[\Psi]$. This turns out to coincide with the 
$\mathcal{N}=1$ super Virasoro vertex algebra. The analysis is completely parallel to the 
Virasoro case and triality manifests itself in the same manner: the first two realizations lead to two descriptions related by Feigin-Frenkel duality and the third one to the coset model realization. 

The triality is then of the form
\begin{eqnarray}
W_{1|2}[OSp(1|2)_{-\Psi}]\quad \leftrightarrow\quad W_{1|2}[OSp(1|2)_{-\Psi^{-1}}]\quad \leftrightarrow \quad \frac{SO(3)_{\frac{1}{1-\Psi}-2}\times \mathrm{Ff}^{SO(3)}}{SO(3)_{\frac{1}{1-\Psi}-1}}.
\end{eqnarray}

The DS construction that produces $\mathcal{N}=1$ super Virasoro algebra from DS-reduction of $OSp(1|2)$ can be found in the appendix of \cite{Kac:qy}. 
The central charge of the VOA is in our conventions 
\begin{eqnarray}
c=\frac{15}{2}-3\Psi-\frac{3}{\Psi}.
\end{eqnarray}
We can see that this expression is indeed invariant under the Feigin-Frenkel duality $\Psi \leftrightarrow \frac{1}{\Psi}$

This coset realization is also known to lead to $\mathcal{N}=1$ super-Virasoro algebra: it is the analytic continuation of the well-known 
construction of $\mathcal{N}=1$ super Virasoro minimal models. It produces the $\mathcal{N}=1$ super Virasoro algebra with central charge indicated above.

\subsection{Central charge}

Central charge of orthosymplectic $Y$-algebras are given by (see appendix \ref{app:ccharge} for detailed calculation)
Central charge of $\tilde{Y}^-_{L,M,N}[\Psi]$ can then be identified as $\tilde{c}^-_{L,M,N}[\Psi]=c^-_{L,N,M}[1-\Psi]$. Recall that central charge of $OSp(2N|2L)_{\Psi-2N+2L+2}$ is
\begin{eqnarray}\nonumber
c^-_{L,M,N}[\Psi]&=&\tilde{c}^-_{L,N,M}[1-\Psi]\\ \nonumber
&=&-\frac{(2 (L-M)-1) (2 (L-M)+1) (L-M)}{\Psi -1}\\ \nonumber
&&+\frac{2 (2 (L-N)+1) (L-N+1) (L-N)}{\Psi }\\ \nonumber
&&+2 \Psi  (2 (M-N)+1) (M-N+1) (M-N)\\ \nonumber
&& - 2 L (1 + 6 M^2 + M (6 - 12 N) - 6 N + 6 N^2) \\
&&+4 M^3 - 3 M (1 - 2 N)^2+ N (5 - 12 N + 8 N^2)
\end{eqnarray}
and
\begin{eqnarray}\nonumber
c^+_{L,M,N}[\Psi]&=&\tilde{c}^+_{L-\frac{1}{2},M,N}[\Psi]\\ \nonumber
&=&-\frac{2 (M-L) (2 (M-L)+1) (M-L+1)}{1-\Psi }\\ \nonumber
&&-\frac{2 (N-L) (2 (N-L)+1) (N-L+1)}{\Psi }\\ \nonumber
&&+\Psi  (2 (M-N)-1) (2 (M-N)+1) (M-N)\\ 
&&+L (1 - 12 (M - N)^2) - N + 2 (M - N)^2 (3 + 2 M + 4 N).
\end{eqnarray}
One can check that the expressions above are indeed invariant under transformations \ref{orthoS}. Note also that $S_3$ action preserves $\tilde{Y}^+$-algebras and we can indeed write their central charge $\tilde{c}^+_{L,M,N}[\Psi]$ in $S_3$ invariant way
\begin{eqnarray}\nonumber
\tilde{c}^+_{L,M,N}[\Psi]&=&\frac{1}{2}\frac{1}{\Psi}(L-N)(4(L-N)^2-1)+\frac{1}{2}(1-\frac{1}{\Psi})(N-L)(4(N-L)^2-1)\\ \nonumber
&&\frac{1}{2}\Psi(M-N)(4(M-N)^2-1)+\frac{1}{2}(1-\Psi)(N-M)(4(N-M)^2-1)\\ \nonumber
&&\frac{1}{2}\frac{1}{1-\Psi}(L-M)(4(L-M)^2-1)+\frac{1}{2}\frac{\Psi}{\Psi-1}(M-L)(4(M-L)^2-1)\\
&&-2(L+M-2N)(L-2M+N)(-2L+M+N)+\frac{1}{2}.
\end{eqnarray}

\section{From Junctions to Webs} \label{sec:junctions}

It is natural to consider brane or gauge theory configurations involving a more intricate 
junction, perhaps involving several semi-infinite interfaces converging to a single two-plane. 

It is also natural to consider intricate webs, involving finite interface segments 
as well as semi-infinite ones. Web configurations would break scale invariance. 
In the IR, they would approach a single junction.

Conversely, one may consider webs with several simpler junction as a regularization 
of an intricate junction. If all junctions are dual to our basic Y-junctions, this may become a computational tool 
to determine the VOA's at generic junctions. 

There is a precedent to this: complicated half-BPS interfaces in ${\cal N}=4$ SYM 
can often be decomposed as a sequence of simpler interfaces, with a smooth limit
sending to zero the relative distances between the interfaces. This is an important computational tool,
as it allows one to apply S-duality transformations to well-understood individual pieces 
and then assemble them to the S-dual of the original, intricate interface. 

A concrete example could be a Nahm pole associated to a generic $\mathfrak{su}(2)$ embedding $\rho$, 
realized as a sequence of individual simple Nahm pole interfaces. This is a smooth resolution, 
as long as the individual interfaces are ordered in a specific way \cite{Gaiotto:2008ac}. The S-dual configuration 
is a sequence of bi-fundamental interfaces building up a complicated three-dimensional interface gauge theory
with a good IR limit \cite{Gaiotto:2008aa}.

One may want to follow that example for junctions, say to decompose a Y-junction of complicated interfaces 
into a web of simpler Y-junctions. 

This idea raises a variety of hard questions, starting from figuring out criteria for a smooth IR limit 
of an interface web. Furthermore, the same configuration may be the limit of many different 
inequivalent webs. 

Any of these questions would bring us far from the scope of this paper. In this section we will limit ourselves to a few 
judicious speculations. 

In general, local operators at the final junction may arise either from local operators at each 
elementary junction in the web or from extended operators, such as a finite line defect segment 
joining two consecutive junctions. Thus we may hope that the final VOA 
will be an extension of the product of the VOA's at the vertices of the web, 
including products of degenerate modules associate to the finite line defect segment.

This picture is supported by the observation that although the dimensions of 
degenerate modules are not integral, the sum of the dimensions of the local operators 
at the two ends of a finite line defect segment will be integral. For example, 
a finite Wilson line $W_\mu$ on a finite segment of NS5 interface supports two local operators 
at the endpoints which have dimensions which differ by integral amounts from 
$\Delta_\mu[\Psi]$ and $\Delta_\mu[-\Psi] = - \Delta_\mu[\Psi]$ respectively,
where $\Delta_\mu[\Psi]$ is the dimension of the $\mu$ vertex operator 
in the $U(N|L)_\Psi$ WZW model. 

This is not quite a full definition of the final interface VOA, but it strongly 
restricts its form. \footnote{It may be possible to formalize this procedure as a sort of tensor product 
of VOAs over a common braided monoidal category.} 

A striking observation is the formal resemblance between this idea and the way the topological vertex is used to 
assemble the topological string partition function of general toric Calabi-Yau, by summing up over a choice of 
partition $\mu$ for each internal leg of the toric diagram \cite{Okounkov:uq}. Perhaps one may use this analogy to 
determine which products of degenerate modules to included in the extended VOA. 

The simplest possible situation for us is a web which can be interpreted as a collection of 
D5 branes ending on a NS5 brane: a sequence of $(1,q_i)$ fivebrane segments with 
Y-junctions to semi-infinite D5 branes coming from the left or the right. Such a configuration can be 
lifted directly to a sequence of interfaces in 3d Chern-Simons theory. If the 3d interfaces have a good collision limit, 
one can derive directly the junction VOA. 

This situation also allows one to start probing questions about the extension structure of the 
final VOA and the equivalence between different web resolutions of the same interface. 

\subsection{Four-way intersection}
The simplest possibility we can discuss is that of an infinite D5 interface crossing an infinite 
NS5 interface. The four-way junction has two obvious resolutions, akin to the toric diagram of the conifold, 
involving either a $(1,1)$ or a $(1,-1)$ finite interface segment. 

\begin{figure}[h]
  \centering
      \includegraphics[width=0.7\textwidth]{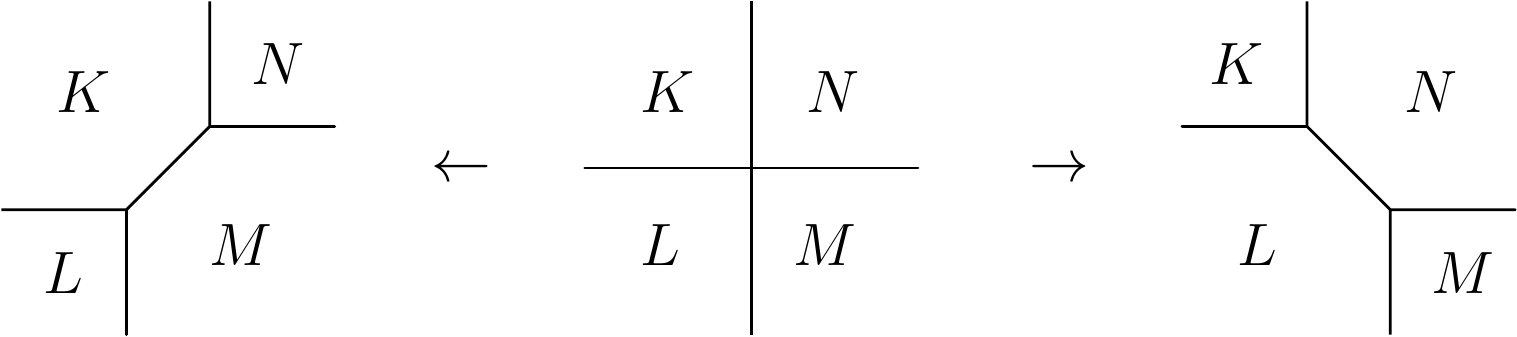}
\caption{Two possible resolutions of the configuration of D5-brane (horisontal) crossing NS5-brane (vertical). First resolution includes a finite segment of $(1,1)$-brane whereas the second resolution includes a $(1,-1)$-brane segment. $K,L,M,N$ D3-branes are attached to fivebranes leading to webs of interfaces between $U(K),U(L),U(M),U(N)$ theories.}
\label{fig:FourWay}
\end{figure}

We can denote the choices of gauge group in the four quadrants as $K,L,M,N$, counterclockwise 
from the top left quadrant. 

For some values of $K,L,M,N$, the two resolutions produce obviously the same 
3d interface in the scaling limit and then the same VOA. For example, if $K=L$ and $N=M$ 
then the CS theory interface results from the collision of interfaces which support some 2d matter 
coupled to the $U(N|K)$ CS gauge fields. The two resolutions give the same two interfaces 
in different order, and the collision/scaling limit is obviously the same: an interface which supports 
both 2d matter fields at the same location. 

On the other hand, in other configurations the two resolutions give clearly different answers
orproduce pairs of interfaces which do not have an obvious collision limit. 

In any case, the resolved web enjoys a non-trivial S-duality symmetry, 
exchanging, say, $K$ and $M$ while mapping $\Psi \to \Psi^{-1}$. 

We will consider a single entertaining example, a small variation of the parafermion example. 

\subsubsection{${\cal N}=2$ super-Virasoro}
\begin{wrapfigure}{l}{0.2\textwidth}
\vspace{-10pt}
  \begin{center}
      \includegraphics[width=0.175\textwidth]{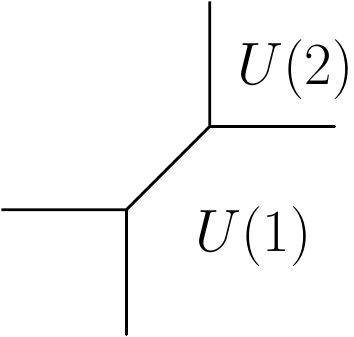}
\end{center}
\vspace{-10pt}
\end{wrapfigure}
Consider a four-way junction with $K=0$, $L=0$, $N=2$, $M=1$, 
resolved by a $(1,1)$ segment. 

In the 3d description, we have a $U(2)$ CS theory which is first 
reduced to a block-diagonal $U(1)$ and then coupled to a single complex 
fermion. There is no obstruction to bringing the interfaces together. 

The opposite resolution, involving a $(1,-1)$ segment, would have coupled a complex fermion 
doublet to $U(2)$ and then reduced the gauge symmetry to $U(1)$. In a scaling limit, this would 
differ from the original resolution by an extra spurious complex fermion decoupled from the 3d gauge theory. 

The system engineers a $\mathfrak{u}(1)$-BRST coset VOA
\begin{equation}
\frac{U(2)_\Psi \times \mathrm{Ff}^{U(1)}}{U(1)_{\Psi}}
\end{equation}
We can recast this as the product 
\begin{equation}
U(1) \times \mathrm{sVir}_{{\mathcal N}=2}
\end{equation}
using the Kazama-Suzuki coset description 
\begin{equation}
\mathrm{sVir}_{{\mathcal N}=2} = \frac{SU(2)_\kappa \times \mathrm{Ff}^{U(1)}}{U(1)_{2 \kappa+2}}
\end{equation}

\begin{wrapfigure}{l}{0.2\textwidth}\
\vspace{-5pt}
  \begin{center}
      \includegraphics[width=0.175\textwidth]{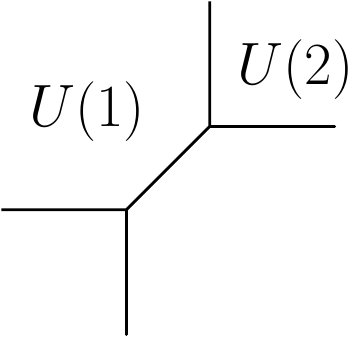}
\end{center}
\vspace{-5pt}
\end{wrapfigure}
The S-dual description involves a four-way junction with $K=1$, $L=0$, $N=2$, $M=0$, 
resolved by a $(1,1)$ segment. 

That gives simply the DS reduction of $U(2|1)$, which is know to coincide with 
\begin{equation}
U(1) \times \mathrm{sVir}_{{\mathcal N}=2}
\end{equation}
as well \cite{Kac:qy}.

\subsection{A trivalent junction with multiple D5 branes}
Another instructive example is a trivalent junction between a $(1,0)$, a $(1,k)$ and $k$ coincident $(0,1)$
fivebranes. We can resolve the stack of fivebranes into $k$ parallel D5 branes. 

If $N-M = k S$ for integer $S$, one may let the number of D3 branes drop by $S$ across each D5 brane. 
This configuration is expected to preserve a global $SU(k)$ symmetry on the limit of coincident fivebranes.

This symmetry manifests itself in the junction VOA: the vertex algebra involves a DS reduction, say, 
of $U(N|L)$ associated to an $\mathfrak{su}(2)$ embedding with $k$ blocks of size $S$ and $M$ blocks of size $1$. 
The result of that reduction has a $SU(k)$ WZW subalgebra which survives the coset by $U(M|L)$. 

If $N=M$, instead, we have $k$ copies of the symplectic bosons for $U(N|L)$, which also have an $SU(k)$ WZW subalgebra which survives the coset by $U(N|L)$.

It is natural to conjecture relations between this VOAs and some kind of $W_{k+\infty}$ algebra and the topological vertex and D-brane counting in $\C^3/{\mathbb Z}_k$ \cite{Young:2008aa}.

 \section*{Acknowledgements} 
We thank Kevin Costello, Thomas Creutzig, Toma\v{s} Proch\'{a}zka, and Edward Witten for interesting discussions.  The research of DG and MR was supported by the Perimeter Institute for Theoretical Physics.  Research at Perimeter Institute is supported by the Government of Canada through Industry Canada and by the Province of Ontario through the Ministry of Economic Development \& Innovation.

\appendix
\section{Conventions for current algebras}
\label{app:conventions}

\subsection{Free fermions}
\subsubsection{A single real fermion}
A single chiral Majorana free fermion $\psi(z)$ has OPE 
\begin{equation}
\psi(z) \psi(w) \sim \frac{1}{z-w}
\end{equation}
It has dimension $\frac12$ with stress tensor 
\begin{equation}
T^\psi = -\frac{1}{2} \psi \partial \psi
\end{equation}
of central charge $c^\mathrm{Ff} = \frac12$. 

We will denote the corresponding (spin-)VOA as $\mathrm{Ff}$. 

\subsubsection{$SO(n)$ fermions}
If we take $n$ chiral Majorana free fermions $\psi^i(z)$
we get a VOA $\mathrm{Ff}^{SO(n)}$. It includes $SO(n)_1$ WZW currents which we can sloppily normalize as 
\begin{equation}
J^{ij} = \psi^i \psi^j
\end{equation}
Indeed,
\begin{equation}
J^{ij}(z) J^{kt}(0) \sim\frac{\delta^{it}\delta^{jk}-\delta^{ik}\delta^{jt}}{z^2}+ \frac{\delta^{jk}J^{it}(0)-\delta^{jt}J^{ik}(0)-\delta^{ik}J^{jt}(0)+\delta^{it}J^{jk}(0)}{z}\end{equation}

This is a conformal embedding: the dimension of an $SO(n)$ WZW model at level $1$ is 
\begin{equation}
c^{SO(n)_1} = \frac{1 \times \frac{n(n-1)}{2}}{1+n-2} = \frac{n}{2} = n c^\mathrm{Ff}
\end{equation}
and 
\begin{equation}
:J^{ij}J^{kt}:= \delta^{jk}\partial\psi^i \psi^t -\delta^{jt}\partial\psi^i \psi^k-\delta^{ik}\partial\psi^j \psi^t+\delta^{it}\partial\psi^j \psi^k + \psi^i \psi^j\psi^k \psi^t
\end{equation}
hence the stress tensor coincides with the Sugawara stress tensor. 
\begin{equation}
T \equiv -\frac{1}{2} \psi^i \partial \psi^i = \frac{1}{1+n-2} \times \frac{1}{4}:J^{ij}J^{ji}
\end{equation}

\subsubsection{A single complex fermion and the $bc$ system}
Two Majorana free fermions can be combined into complex fermions 
 $\psi(z)$ and $\chi(z) \equiv (\psi)^\dagger(z)$ with OPE 
\begin{equation}
\psi(z) \chi(w) \sim \frac{1}{z-w}
\end{equation}
The fermions have dimension $\frac12$ with stress tensor 
\begin{equation}
T^{\psi \chi} = -\frac{1}{2} \psi \partial \chi-\frac{1}{2} \chi \partial \psi
\end{equation}
of central charge $c^\mathrm{Ff} = \frac12$. 

The VOA includes a $U(1)$ WZW current $J = \psi \chi$ of level $1$. 
Indeed
\begin{equation}
J(z) J(0) \sim \frac{1}{(z-w)^2}
\end{equation}
This is a conformal embedding:
\begin{equation}
T = \frac{1}{2}:JJ:
\end{equation}

We will denote the corresponding (spin-)VOA as $\mathrm{Ff}^{U(1)}$.

The $\mathrm{bc}$ ghost system is the same as a complex fermion, except that the 
stress tensor is shifted by $\frac12 \partial J$ to 
\begin{equation}
T^{bc} = -b \partial c
\end{equation}
so that the dimension of $c(z)$ is $0$ and of $b(z)$ is $1$. 
The central charge of the shifted stress tensor is $-2$ 

More generally, if we shift the stress tensor by $\delta \partial J$ 
we obtain a ghost system such that $c$ has dimension $\frac12-\delta$, 
$b$ has dimension $\frac12 + \delta$ and the central charge is $1 - 12\delta^2$. 

\subsubsection{$U(n)$ fermions}
If we take $2n$ chiral Majorana free fermions we can combine them into 
$n$ pairs of complex generators $\psi^a(z)$ and $\chi_a(z) \equiv (\psi^a)^\dagger(z)$
with OPE 
\begin{equation}
\psi^a(z) \chi_b(w) \sim \frac{\delta^a_b}{z-w}
\end{equation}
We denote this VOA as $\mathrm{Ff}^{U(n)}$. It includes $U(n)$ WZW currents 
\begin{equation}
J^a_b = \psi^a \chi_b
\end{equation}
with OPE 
\begin{equation}
J^a_b(z) J^c_d(w) \sim \frac{\delta^a_d \delta^c_b}{(z-w)^2} + \frac{\delta^c_b J^a_d(w) - \delta^a_d J^c_b(w)}{z-w}
\end{equation}
The level of the $SU(n)$ current sub-algebra $J^a_b- \frac{1}{n} \delta^a_b J^c_c$ is $1$. 
The diagonal $U(1)$ $J^c_c$ has level $n$. 

This is a conformal embedding: 
\begin{equation}
c^{SU(n)_1} + c^{U(1)} = \frac{1 \times (n^2-1)}{1+n} +1= n = 2 n c^\mathrm{Ff}
\end{equation}
\subsection{Symplectic fermions VOAs $\mathrm{Sf}$ and $\mathrm{Sf}_0$}
The vertex algebra $\mathrm{Sf}$ is generated by two ``symplectic fermions'' $x(z)$, $y(z)$,
which are fermionic currents of dimension $1$ and OPE\footnote{Equivalently, $\mathrm{Sf}$ can be defined as a $PSU(1|1)$ current algebra.}
\begin{equation}
x(z) y(w) \sim \frac{1}{(z-w)^2}
\end{equation}
The stress tensor is simply 
\begin{equation}
T = xy
\end{equation}
with central charge $-2$. The vertex algebra $\mathrm{Sf}$ has an $SU(2)_o$ global symmetry rotating the symplectic fermions as a doublet,
which is not promoted to an affine symmetry. It is useful to consider the Cartan subgroup $U(1)_o$ acting on the the two fermionic currents with charge $\pm1$.

There is a considerable body of work devoted to the study of $\mathrm{Sf}$, of the ``triplet algebra'' $\mathrm{Sf}_e$
consisting of bosonic vertex operators in $\mathrm{Sf}$ and of the ``singlet algebra'' $\mathrm{Sf}_0$
consisting of vertex operators in $\mathrm{Sf}$ of $U(1)_o$ charge $0$. We refer to \cite{Kausch:aa}
for the definition of $\mathrm{Sf}$ and to \cite{Creutzig:2013kq} and references therein for a detailed discussion of these subalgebras
and their modules.

The VOA $\mathrm{Sf}_0$ has two natural classes of modules: 
\begin{itemize}
\item The other charge sectors $\mathrm{Sf}_n$ in $\mathrm{Sf}$. They have highest weight vectors of conformal dimension $\frac{n^2 + |n|}{2}$. 
\item The charge $0$ sector $V^{xy}_{\lambda}$ of the twisted modules for $\mathrm{Sf}$. 
They have highest weight vectors of conformal dimension $\frac{\lambda^2 - \lambda}{2}$ which induce 
singularities $z^{- \lambda}$ in $x(z)$ and $z^{\lambda-1}$ in $y(z)$. 
\end{itemize}

In order to define $V^{xy}_{\lambda}$ we consider a vector $|\lambda;xy \rangle$
which satisfies 
\begin{align}
x_{\lambda +n} |\lambda;xy \rangle &= 0 \qquad n \geq 0 \cr
y_{-\lambda +n} |\lambda;xy \rangle &= 0 \qquad n \geq 1
\end{align}
and build the module by acting with the other generators.
Because of the fermionic nature of the generators, we can identify 
\begin{align}
x_{\lambda -1} |\lambda;xy \rangle &=|\lambda-1;xy \rangle \cr
y_{-\lambda} |\lambda;xy \rangle &=|\lambda+1;xy \rangle
\end{align}
so that $V^{xy}_\lambda$ generates a module for the $xy$ VOA which is defined modulo $\lambda \to \lambda + 1$ and includes the $\mathrm{Sf}_0$ modules $V^{xy}_{\lambda+n}$ for all $n$.

Notice that as we take $\lambda$ to $0$ we do not get the vacuum module, but rather a decomposable module. 
Similar considerations apply for other integer values of $\lambda$. In general we will assume $\lambda$ to be a non-integer 
complex number. 

Some of the fusion rules between these modules are obvious. For example, almost by definition 
we have 
\begin{align}
\mathrm{Sf}_n \times \mathrm{Sf}_m &\sim \mathrm{Sf}_{n+m}  \cr
\mathrm{Sf}_n \times V^{xy}_\lambda &\sim V^{xy}_{\lambda+n}
\end{align}
The OPE of two twisted modules is more subtle. As they are twist fields for the $xy$ currents, the tensor product 
is likely to contain modules of the form $V^{xy}_{\lambda+ \lambda' + k}$ with integer $k$. As 
$x(z)$ has a pole $z^\lambda$ and $y(z)$ a pole $z^{1-\lambda}$ near $V^{xy}_{\lambda}$, they can potentially 
have poles $z^{\lambda + \lambda'}$ and $z^{2-\lambda-\lambda'}$ near the result of the OPE. i
It is thus not unreasonable to expect 
\begin{equation}
V^{xy}_{\lambda} \times V^{xy}_{\lambda'} \sim V^{xy}_{\lambda+ \lambda'} + V^{xy}_{\lambda+ \lambda' -1}
\end{equation}
This is indeed the case, see e.g. \cite{Kausch:aa}.

\subsection{Symplectic bosons}
\subsubsection{Symplectic bosons, symplectic fermions and $\beta \gamma$ systems. }
The vertex algebra $\mathrm{Sb}$ of a single symplectic boson has two bosonic generators, $X(z)$ and $Y(z)$, with OPE
\begin{equation}
X(z) Y(w) \sim \frac{1}{z-w}
\end{equation}
and conformal dimension $1/2$. 
We can also denote the generators as a doublet $Z^\alpha$ with OPE 
\begin{equation}
Z^\alpha(z) Z^\beta(w) \sim \frac{\epsilon^{\alpha \beta}}{z-w}
\end{equation}
Several of the features we discuss below can be found discussed at length in \cite{Lesage:2002ch}.

The stress tensor can be written as 
\begin{equation}
T = \frac{1}{2} X \partial Y - \frac{1}{2} Y \partial X = \frac12 \epsilon_{\alpha \beta} Z^\alpha Z^\beta
\end{equation}
and gives a central charge of $c^{\mathrm{Sb}}=-1$.

The VOA has $SU(2)$ WZW currents 
\begin{equation}
J^{\alpha \beta} = Z^{\alpha} Z^{\beta}
\end{equation}
of level $-\frac12$:
\begin{equation}
J^{\alpha \beta}(z) J^{\gamma \delta}(0) \sim \frac{\epsilon^{\alpha \gamma}\epsilon^{\beta \delta}+\epsilon^{\alpha \delta}\epsilon^{\beta \gamma}}{z^2} + \frac{\epsilon^{\alpha \gamma}J^{\beta \delta}+\epsilon^{\alpha \delta}J^{\beta \gamma}+J^{\alpha \gamma}\epsilon^{\beta \delta}+\epsilon^{\beta \gamma}J^{\alpha \delta}}{z}
\end{equation}
This is likely to be a conformal embedding: 
\begin{equation}
c^{SU(2)_{-\frac12}} = \frac{3/2}{1/2-2} = -1
\end{equation}

The $U(1)_{-1}$ current subalgebra generated by the current $XY$ is also important. 
It plays an important role in the bosonization relation between symplectic bosons and symplectic fermions. 
\footnote{This is just the bosonization of $\beta \gamma$ system 
familiar in superstring perturbation theory, with currents $\eta$ and $\partial \xi$.}

In particular, $\mathrm{Sf}$ is the result of the $\mathfrak{u}(1)$-BRST quotient 
\begin{equation}
\mathrm{Sf}  = \left\{ \mathrm{Sb} \times \mathrm{Ff}^{U(1)} \times \mathrm{bc},\,Q_{BRST} \right\}
\end{equation}
with BRST charge $Q_{BRST} = \oint c (XY + \psi \chi)$. The fermionic currents 
are the BRST-closed generators 
\begin{equation}
x(z) = X(z) \psi(z) \qquad \qquad y(z) = Y(z) \chi(z)
\end{equation}

If we restrict ourselves to the $U(1)_1$ subalgebra in $\mathrm{Ff}^{U(1)}$ we get the 
VOA $\mathrm{Sf}_0$, the charge $0$ subalgebra of $\mathrm{Sf}$.
In other words, there is a coset relation
\begin{equation}
\mathrm{Sf}_0 = \frac{\mathrm{Sb}}{U(1)_{-1}}
\end{equation}

Conversely, one can reconstruct the symplectic boson VOA by dressing 
the fermionic currents with charge $\pm 1$ vertex operators for a $U(1)_{-1}$ current algebra.

The symplectic boson VOA has a variety of interesting modules, which have intricate relations with 
modules for $SU(2)_{-\frac12}$ and $\mathrm{Sf}^0$.

In particular, there is a family of Ramond modules $R_\lambda$, with non-integral $\lambda \in \mathbb{C}/\mathbb{Z}$ defined by relations 
\begin{align}
X_0 |n\rangle &= (\lambda + n)|n-1\rangle \cr
Y_0 |n\rangle &= |n+1\rangle \cr
X_k |n\rangle &=0 \qquad k>0 \cr
Y_k |n\rangle &=0 \qquad k>0
\end{align}
and $n \in \mathbb{Z}$. We denote as $R_\lambda$ the vertex operator associated to the vector $|0\rangle$ as well. It has conformal dimension 
$-\frac18$ and $U(1)$ charge $\lambda + \frac12$. 

This modules give Weyl modules for $SU(2)_{-\frac12}$ associated to infinite dimensional principal series representations
of the algebra of zeromodes. 

In the context of bosonization, $R_\lambda$ can be combined with a vertex operator of charge $\lambda$ for $U(1)_1$
to produce the vertex operators and modules $V^{xy}_\lambda$ for $\mathrm{Sf}^0$.



Finally, symplectic bosons are a special example of $\beta \gamma$ systems. Indeed, if 
we shift the stress tensor by appropriate multiples of $\partial (XY)$ we can get either the standard 
$\beta \gamma$ system of conformal dimensions $1$ and $0$ or more general ones 
of dimensions $\frac12\pm \delta$, with central charge is $-1 + 12\delta^2$. 

\subsubsection{$Sp(2n)$ and $U(n)$ symplectic bosons}
If we take $n$ pairs of symplectic bosons $Z^\alpha(z)$ with OPE 
\begin{equation}
Z^\alpha(z) Z^\beta(w) \sim \frac{\omega^{\alpha \beta}}{z-w}
\end{equation}
we get a VOA $\mathrm{Sb}^{Sp(2n)}$. It includes $Sp(2n)_{-\frac12}$ WZW currents 
\begin{equation}
J^{\alpha \beta} = Z^\alpha Z^\beta
\end{equation}
with OPE
\begin{equation}
J^{\alpha \beta}(z) J^{\gamma \delta}(0) \sim \frac{\omega^{\alpha \gamma}\omega^{\beta \delta}+\omega^{\alpha \delta}\omega^{\beta \gamma}}{z^2} + \frac{\omega^{\alpha \gamma}J^{\beta \delta}+\omega^{\alpha \delta}J^{\beta \gamma}+\omega^{\beta \delta}J^{\alpha \gamma}+\omega^{\beta \gamma}J^{\alpha \delta}}{z}
\end{equation}

This is likely to be a conformal embedding: the dimension of an $Sp(2n)$ WZW model at level $-\frac12$ is 
\begin{equation}
c^{Sp(2n)_{-\frac12}} = \frac{-\frac12 \times n(2n+1)}{-\frac12+n+1} = -n = n c^\mathrm{Sb}
\end{equation}

If we separate the symplectic bosons in two dual sets $X^a(z)$ and $Y_a(z)$
we get $U(n)$ WZW currents
\begin{equation}
J^a_b = X^a Y_b
\end{equation}
such that the level of the $SU(n)$ subalgebra is $-1$:
\begin{equation}
J^a_b(z) J^c_d(0) \sim -\frac{\delta^a_d \delta^c_b}{z^2}+ \frac{\delta^a_d J^c_b(0)-\delta^c_b J^a_d(0)}{z}
\end{equation}

For $n \neq 1$ this is likely to be a conformal embedding: 
\begin{equation}
c^{SU(n)_{-1}} +1 = \frac{-1 \times (n^2-1)}{-1+n} +1= -n = n c^\mathrm{Sb}
\end{equation}

\subsubsection{$OSp(n|2m)$ fermions}

If we combine $n$ Majorana fermions $\psi^i$ and $m$ pairs of symplectic bosons $Z^\alpha(z)$ 
we get a VOA $\mathrm{Ff}^{OSp(n|2m)}$. 
We can combine them into fields $u^a = (\psi^i, Z^\alpha)$ with OPE
\begin{equation}
u^A(z) u^B(w) \sim \frac{\eta^{AB}}{z-w}  
\end{equation}
where $\eta^{AB}$ is Koszul-antisymmetric: $\epsilon^{BA} = (-1)^{1+p(A)p(B)}\epsilon^{AB} $ with $p(i) = 1$ and $p(\alpha) = 0$. 

It includes $OSp(n|2m)_{1}$ WZW currents 
\begin{equation}
J= \begin{pmatrix}\psi^i \psi^j  & \psi^i Z^\beta \cr Z^\alpha \psi^j & Z^\alpha Z^\beta \end{pmatrix}
\end{equation}
i.e. $J^{AB} = u^A u^B$, with $J^{BA} = (-1)^{p(A)p(B)} J^{BA}$ and OPE
\begin{eqnarray}
J^{AB}(z) J^{CD}(0) &\sim&\frac{\eta^{BC}\eta^{AD} + (-1)^{p(A)p(B)}\eta^{AC}\eta^{BD}}{z^2}\\
&+& \frac{\eta^{BC}J^{AD}(0)+(-1)^{p(A)p(B)}\eta^{AC}J^{BD}(0))}{z}\\
&+& \frac{(-1)^{p(C)p(D)}\eta^{BD}J^{AC}(0)+(-1)^{p(A)p(B)+p(C)p(D)}\eta^{AD}J^{BC}(0)}{z}
\end{eqnarray}
This is likely to be a conformal embedding: the dimension of an $OSp(n|2m)$ WZW model at level $1$ is 
\begin{equation}
c^{OSp(n|2m)_1} = \frac{1 \times \frac{(n-2m)(n-2m-1)}{2}}{-1+n -2m -2} = \frac{n}{2}-m
\end{equation}

\subsubsection{$U(n|m)$ symplectic bosons}
If we combine $n$ pairs of symplectic bosons $X^a(z)$ and $Y_b(z)$ and $m$ complex fermions 
$\psi_i$, $\chi^j$ we get a VOA $\mathrm{Sb}^{U(n|m)}$.
we get $U(n)$ WZW currents
\begin{equation}
J= \begin{pmatrix}X^a Y_b & X^a \psi_i \cr \chi^j Y_b & \chi^j \psi_i \end{pmatrix}
\end{equation}
such that the level of the $SU(n|m)$ subalgebra is $-1$. 
For $n-m \neq 1$ this is likely to be a conformal embedding: 
\begin{equation}
c^{SU(n|m)_{-1}} +1 = \frac{-1 \times ((n-m)^2-1)}{-1+n-m} +1= -n+m
\end{equation}

We can collect the fermions and symplectic bosons in super-vectors $u^A = (X^a, \chi^i)$ and $v_A = (Y_a, \psi_i)$
with $p(i) = 1$ and $p(\alpha) = 0$, write the OPE as
\begin{equation}
u^A(z) v_B(w) \sim \frac{\delta^A_B}{z-w} \qquad \qquad v_B(z) u^A(w)  \sim -\frac{(-1)^{p(A)p(B)}\delta^A_B}{z-w}
\end{equation}
and the currents as 
\begin{equation}
J^A_B = u^A v_B 
\end{equation}
with OPE
\begin{align}
J^A_B(z) J^C_D(0) &\sim - \frac{(-1)^{p(B)p(C)}\delta^A_D \delta^C_B}{z^2}+ \cr
+& \frac{ (-1)^{p(A)p(B) + p(C) p(D)+p(C)p(B)} \delta^A_D J^C_B(0)-(-1)^{p(B)p(C)}\delta^C_B J^A_D}{z}
\end{align}

Notice that in these conventions the overall $U(1)$ is the super-trace $J = \sum_A (-1)^{p(A)} J^A_A$: 
\begin{equation}
J(z) J^C_D(0) \sim - \frac{\delta^C_D}{z^2}\qquad \qquad J(z) J(0) \sim -\frac{n-m}{z^2}
\end{equation}
We can obtain alternative normalizations of the overall $U(1)$ by defining 
$\hat J^A_B = J^A_B + c \delta^A_B J$ so that 
\begin{align}
\hat J^A_B(z) \hat J^C_D(0) &\sim - \frac{(-1)^{p(B)p(C)}\delta^A_D \delta^C_B+(2 c + c^2 (n-m) ) \delta^A_B \delta^C_D}{z^2}+ \cr
&+ \frac{ (-1)^{p(A)p(B) + p(C) p(D)+p(C)p(B)} \delta^A_D \hat J^C_B(0)-(-1)^{p(B)p(C)}\delta^C_B \hat J^A_D}{z}
\end{align}

Notice that we we can also exchange the role of fermions and symplectic bosons, 
with some $\tilde u^A = (\chi^a, X^i)$ and $\tilde v_A = (\psi_a, Y_i)$. If we use the same 
$p(i) = 1$ and $p(\alpha) = 0$ convention, we have OPE 
\begin{equation}
\tilde u^A(z) \tilde v_B(w) \sim \frac{\delta^A_B}{z-w} \qquad \qquad \tilde v_B(z) \tilde u^A(w)  \sim \frac{(-1)^{p(A)p(B)}\delta^A_B}{z-w}
\end{equation}
and currents  
\begin{equation}
\tilde J^A_B = -\tilde u^A \tilde v_B 
\end{equation}
with OPE
\begin{align}
\tilde J^A_B(z) \tilde J^C_D(0) &\sim \frac{(-1)^{p(B)p(C)}\delta^A_D \delta^C_B}{z^2}+\cr &+ \frac{ (-1)^{p(A)p(B) + p(C) p(D)+p(C)p(B)} \delta^A_D J^C_B(0)-(-1)^{p(B)p(C)}\delta^C_B J^A_D}{z}
\end{align}
of $SU(n|m)$ at level $1$

\subsection{$U(N)_\kappa$ currents}
Thorughout the paper, we use following notation for $U(N)_{\kappa}$ VOA
\begin{eqnarray}
U(N)_{\kappa}=U(1)_{N\kappa}\times SU(N)_{\kappa-N}.
\end{eqnarray}
The specific combination of levels of $SU(N)$ and $U(1)$ is natural \cite{Witten:1993xi} and corresponds to the 
Sugawara stress tensor and conformal dimensions being given by the standard $U(N)$ Casimir. The notation for the level 
is unusual, but very convenient for this paper. 

If $J^a_b$ are the $U(N)_{\kappa}$ currents, the coresponding OPE is given by
\begin{eqnarray}
J^a_b(z)J^{a'}_{b'}(w)\sim \frac{(\kappa-N)\delta^a_{b'}\delta^{a'}_b+\delta^{a}_{b}\delta^{a'}_{b'}}{(z-w)^2}+\frac{\delta_b^{a'}J^a_{b'}-\delta^a_{b'}J^{a'}_b}{z-w}
\end{eqnarray}
One can indeed check that $U(1)_{N\kappa}$ element $J_{N\kappa}$ given by
\begin{eqnarray}
J=\sum_{i=1}^{N}J^i_i
\end{eqnarray}
satisfy
\begin{eqnarray}
J(z)J(w)\sim \frac{N\kappa}{(z-w)^2}
\end{eqnarray}
and elements in the cartan of $ SU(N)_{\kappa-N}$ such as $H_{ij}=J^i_i-J^j_j$ satisfy
\begin{eqnarray}
H_{ij}(z)H_{ij}(w)\sim \frac{2(\kappa-N)}{(z-w)^2}
\end{eqnarray}
which is consistent with OPE of the off-diagonal components.

Notice that if we have WZW currents $\tilde J^a_b$ with OPE 
\begin{eqnarray}
\tilde J^a_b(z) \tilde J^{a'}_{b'}(w)\sim \frac{k \delta^a_{b'}\delta^{a'}_b}{(z-w)^2}+\frac{\delta_b^{a'}\tilde J^a_{b'}-\delta^a_{b'}\tilde J^{a'}_b}{z-w}
\end{eqnarray}
then $J^a_b + \tilde J^a_b$ are $U(N)_{\kappa+k}$ currents. 

This is the case, in particular, if $\tilde J^a_b$ are bilinears $\psi^a_i \psi^i_b$ or $X_i^a \cdot Y^i_b$ 
of $k N$ complex fermions (or $bc$ ghosts) or $-k N$ symplectic bosons (or $\beta \gamma$ ghosts) 
transforming in the fundamental representation of $U(N)$. 

Furthermore, notice that a block-diagonal $U(N-1)$ subalgebra of $U(N)_\kappa$ has the correct OPE 
to me identified with $U(N-1)_{\kappa-1}$. 

Finally, consider a $\mathfrak{u}(N)$-valued ghost system with currents
\begin{equation}
I^i_j = b^i_k c^k_j - b^k_j c^i_k
\end{equation}
then 
\begin{equation}
I^i_j(z) I^s_t(0)= \frac{2 N \delta^i_t \delta^s_j -2 \delta^s_t \delta^i_j}{z^2} + \frac{\delta^s_j I^i_t -\delta^i_t  I^s_j}{z}  
\end{equation}
which is precisely what is needed for the sum of $U(N)_\kappa$ currents, $U(N)_{-\kappa}$ currents and $I^i_j$ currents 
to have no $z^{-2}$ term in the OPE, as needed for a $\mathfrak{u}(N)$-BRST reduction. 

\subsection{$U(M|N)_\kappa$ currents}

The currents are labeled as components of supermatrix $J_a^b$ where $a,b=1,\dots,M$ are fermionic bosonic and $a,b=M+1,\dots, M+N$ are fermionic. The two diagonal blocks consist of bosonic generators and the two off-diagonal blocks are fermionic. By $U(M|N)_\kappa$, we really mean
\begin{eqnarray}
U(M|N)_{\kappa}=U(1)_{(M-N)\kappa} \times SU(M|N)_{\kappa-M+N}
\end{eqnarray}

As in the case of $U(N)$ currents, OPE of super-currents components is
\begin{align}
J^A_B(z) J^C_D(0) &\sim \frac{(-1)^{p(B)p(C)}(\kappa -M+N)\delta^A_D \delta^C_B+ \delta^A_B \delta^C_D}{z^2}+ \cr &+\frac{ (-1)^{p(A)p(B) + p(C) p(D)+p(C)p(B)} \delta^A_D J^C_B(0)-(-1)^{p(B)p(C)}\delta^C_B J^A_D}{z}
\end{align}
where $p(a)=0$ for $a=1,\dots, M$ and $p(a)=1$ otherwise.

Notice that if we have WZW currents $\tilde J^a_b$ with OPE 
\begin{align}
\tilde J^A_B(z) &\tilde J^C_D(0) \sim k \frac{(-1)^{p(B)p(C)}\delta^A_D \delta^C_B}{z^2}+\cr &+ \frac{ (-1)^{p(A)p(B) + p(C) p(D)+p(C)p(B)} \delta^A_D J^C_B(0)-(-1)^{p(B)p(C)}\delta^C_B J^A_D}{z}
\end{align}
then $J^A_B+ \tilde J^A_B$ are $U(N|M)_{\kappa+k}$ currents. 

This is the case, in particular, if $\tilde J^A_B$ are bilinears
of complex fermions (or $bc$ ghosts) and symplectic bosons (or $\beta \gamma$ ghosts) 
transforming in the fundamental representation of $U(N|M)$. 

Furthermore, notice that a block-diagonal $U(N-1|M)$ subalgebra of $U(N|M)_\kappa$ has the correct OPE 
to me identified with $U(N-1|M)_{\kappa-1}$. 

\subsection{The bosonization of $U(1|1)_\kappa$}
The typical convention for the $U(1|1)$ OPE's at level $\kappa$ is  
\begin{align}
J(z) J(w) &\sim \frac{\kappa}{(z-w)^2} \cr
J(z) A(w) &\sim \frac{A(w)}{z-w} \cr
J(z) B(w) &\sim - \frac{B(w)}{z-w} \cr
I(z) I(w) &\sim - \frac{\kappa}{(z-w)^2} \cr
I(z) A(w) &\sim -\frac{A(w)}{z-w} \cr
I(z) B(w) &\sim \frac{B(w)}{z-w} \cr
A(z) B(w) &\sim \frac{\kappa}{(z-w)^2} + \frac{J(w) + I(w)}{z-w}
\end{align}

In order to match our conventions for $U(1|1)_\kappa$, we can define 
\begin{align}
J^1_1 = J + \frac{1}{2 \kappa} (I + J) \qquad J^2_1 = A \qquad J^1_2 = B \qquad J^2_2 = - I + \frac{1}{2 \kappa}  (I + J)
\end{align}
We get the correct diagonal OPE's
\begin{align}
J^1_1(z) J^1_1(w) &\sim \frac{\kappa + 1}{(z-w)^2} \cr
J^1_1(z) J^2_2(w) &\sim \frac{1}{(z-w)^2} \cr
J^2_2(z) J^2_2(w) &\sim \frac{- \kappa + 1}{(z-w)^2} \cr
\end{align}

The bosonization of the $U(1|1)$ WZW model \cite{Creutzig:2008aa} is obtained by 
writing the odd currents 
\begin{equation}
A = V^{\kappa, - \kappa}_{1,-1} x \qquad \qquad B = \kappa V^{\kappa, - \kappa}_{-1,1} y
\end{equation}
as the product of vertex operators for the $U(1)_\kappa \times U(1)_{- \kappa}$ currents $J$ and $I$. 
Then the OPE of $x$ and $y$ is the free OPE of symplectic fermions. 

Accordingly, we can decompose the WZW vacuum module into products of modules for $U(1)_\kappa \times U(1)_{- \kappa} \times \mathrm{Sf}_0$
\begin{equation}
U(1|1)_\kappa = \oplus_n V^{\kappa, - \kappa}_{n,-n} \otimes \mathrm{Sf}_n
\end{equation} 

We can give a bosonized description of several other important modules for $U(1|1)_\kappa$.
A nice discussion of this VOA and its modules can be found in \cite{Creutzig:2011aa}
In particular, there are Weyl modules built from finite-dimensional irreducible representations of $\mathfrak{u}(1|1)$
(See e.g. \cite{Gotz:aa} for a review of the latter). 

Atypical modules associated to one-dimensional representations weights $(r,r)$ under $J^1_1$ and $J^2_2$ can be obtained 
from the highest weight vector $V^{\kappa, - \kappa}_{r,-r}$  of conformal dimension $0$ and decompose as
\begin{equation}
V_r^{U(1|1)_\kappa} = \oplus_n V^{\kappa, - \kappa}_{r+n,-r-n} \otimes \mathrm{Sf}_n
\end{equation} 
Typical modules associated to two-dimensional representations $(t+s,t)$ of weights $(t+s,t)$ under $J^1_1$ and $J^2_2$ can be obtained 
from $V^{\kappa, - \kappa}_{(1-\frac{1}{2 \kappa}) s + t,\frac{1}{2 \kappa} s - t} \otimes \mathrm{Sf}_{\kappa^{-1} s}$ of conformal dimension 
\begin{equation}
\frac{s^2}{2\kappa^2}-\frac{s}{2\kappa} + \frac{1}{2 \kappa}(s^2 + s (2 t - \frac{s}{\kappa} )) = \frac{1}{2 \kappa}s(s + 2 t -1) = \frac{1}{2 \kappa}(\tilde t^2 - \tilde t - t^2 +t)
\end{equation}
with $\tilde t = s + t$ and decompose as 
\begin{equation}
V_{s,t}^{U(1|1)_\kappa} = \oplus_n V^{\kappa, - \kappa}_{(1-\frac{1}{2 \kappa}) s + t-n,\frac{1}{2 \kappa} s - t+n} \otimes \mathrm{Sf}_{\kappa^{-1} s+n}
\end{equation} 

\section{General central charge}\label{app:ccharge}

\subsection{Unitary case}
Recall the central charge of $U(N|M)_\Psi$.
\begin{equation}
c_{U(N|M)_\Psi} = 1 + \frac{\Psi-N+M}{\Psi} \left((N-M)^2-1\right)
\end{equation}

That means that if $M=N$,
\begin{align}
c_{L,N,N}[\Psi]&= 1 + \frac{\Psi-N+L}{\Psi} \left((N-L)^2-1\right) + \cr
&-N+L -1- \frac{\Psi-N+L-1}{\Psi-1} \left((N-L)^2-1\right) = \cr
&=(N-L)((N-L)^2-1)\left(\frac{1}{\Psi-1}- \frac{1}{\Psi}\right) -N+L 
\end{align}

On the other hand, if $N=M+1$ we have 
\begin{align}
c_{L,N-1,N}[\Psi]&= 1 + \frac{\Psi-N+L}{\Psi} \left((N-L)^2-1\right) +\cr 
&-1- \frac{\Psi-N+L}{\Psi-1} \left((N-L-1)^2-1\right) = \cr
&=  (L-N)\left((N-L)^2-1\right)\frac{1}{\Psi} + \cr &+(N-L-1)\left((N-L-1)^2-1\right)\frac{1}{\Psi-1} + 2 N-2 L -1
\end{align}
and similarly if $N=M-1$ 
\begin{align}
c_{L,N-1,N}[\Psi]&= -1 - \frac{\Psi+N-L}{\Psi} \left((N-L)^2-1\right) +1+\cr +
& \frac{\Psi-1+N-L+1}{\Psi-1} \left((N-L+1)^2-1\right) = \cr
&=  (L-N)\left((N-L)^2-1\right)\frac{1}{\Psi} + \cr &+(N-L+1)\left((N-L+1)^2-1\right)\frac{1}{\Psi-1} +2N - 2L + 1
\end{align}

The general case requires a bit more work. A simplifying feature is that at all steps only the differences between $L$, $M$, $N$ 
will matter. Lets first set $N>M+1$. 

Let us first analyze the DS-reduction part. We need to both add the ghosts valued in $\mathfrak{n}$ and 
then shift the stress tensor by the derivative of the $t^3$ component of the total currents. 

The $\mathfrak{su}(2)$ embedding in $\mathfrak{u}(N|L)$ is given by decomposing the fundamental representation of 
$\mathfrak{u}(N|L)$ as the dimension $N-M$ irrep plus $M+L$ copies of the trivial representation.
Thus $t^3$ is the Cartan generator 
\begin{equation}
t^3 = (\frac{N-M-1}{2}, \cdots,-\frac{N-M-1}{2}, 0, \cdots,0| 0,\cdots 0)
\end{equation}
the level of $t^3 \cdot J_{U(N|L)_\Psi}$ is easily computed to be 
\begin{align}
\kappa_{t^3 \cdot J_{U(N|L)_\Psi}}&= (\Psi - N + L)\sum_{i=0}^{\frac{N-M-1}{2}}\frac{(N-M-1-2 i)^2}{2} = \cr
&=\frac{1}{12}(\Psi - N + L)(N-M)\left((N-M)^2-1\right)
\end{align}
Thus the central charge of the WZW part is shifted to 
\begin{equation}
\tilde c_{U(N|L)_\Psi} = 1 + \frac{\Psi-N+L}{\Psi} \left((N-L)^2-1\right) - (\Psi - N + L)(N-M)\left((N-M)^2-1\right)
\end{equation}

The $\frak{u}(N|L)$ generators can be decomposed into blocks accordingly as
\begin{eqnarray}
\begin{pmatrix}
                   D & C \\
                   B & A \\
\end{pmatrix}
\end{eqnarray}
Then $\mathfrak{n}$ consists of the upper triangular part of $D$, with $(N-M)(N-M-1)/2$ elements, 
together with $(N-M-1)M$ even and $(N-M-1)L$ odd elements in $C \oplus B$. 

After the shift of the stress tensor, the $c$ and $\gamma$ ghosts end up with dimension equal to the
$t^3$ charge $q_3$. The corresponding central charge is
\begin{eqnarray}
c_{bc}=-3(2q_3-1)^2+1.
\end{eqnarray}

Of the ghosts in $D$, $N-M-1$ have charge $1$, $N-M-2$ have charge $2$, etc. 
The corresponding central charge is 
\begin{eqnarray}\nonumber
c_{gh_D}&=&\sum _{n=1}^{N-M-1}(N-M-n)\left (-3(2n-1)^2+1\right )\\
&=&-(N-M) (N-M-1)\left((N-M) (N-M-1)-1 \right).
\end{eqnarray}
On the other hand, of the $bc$ ghosts in $C \oplus B$, for even $N-M$ we have $M$ of charge $1/2$, $2M$ of charge $3/2$, etc, while for odd 
$N-M$ we have $2M$ of charge $1$, etc. Combined with the $\beta \gamma$ ghosts we get for even $N-M$
\begin{eqnarray}\nonumber
c_{gh_{C \oplus B}}&=&2 M \sum _{n=2}^{(N-M)/2}(-12(n-1)^2+1)+M\\
&=&(L- M) (N-M-1) \left((N-M-1)^2-2 \right).
\end{eqnarray}
and the same expression for odd $N-M$

Thus the DS reduction has central charge 
\begin{align}
&c_{W_{N-M,\cdots}U(N|L)_\Psi}= 1 - \frac{N-L}{\Psi} \left((N-L)^2-1\right) - \Psi(N-M)\left((N-M)^2-1\right) + \cr
&- (N-M)(N-M-1)\left((N-M) (N-M-1)-1 \right) +\cr 
&+(L- M) (N-M-1) \left((N-M-1)^2-2 \right)+ \cr &+ \left((N-L)^2-1\right)+(N-L)(N-M)\left((N-M)^2-1\right)
\end{align}
i.e.
\begin{align}
&c_{W_{N-M,\cdots}U(N|L)_\Psi}= 1 - \frac{N-L}{\Psi} \left((N-L)^2-1\right) +\cr &- \Psi(N-M)\left((N-M)^2-1\right) + \cr &+
(2N+M-3L)(N-M-1)(N-M)+ \cr &+(N-L)(N-M-1)+ (N-L)^2-1
\end{align}

For the next step, we need to observe that the $U(M|L)$ currents in $W_{N-M,\cdots}U(N|L)_\Psi$ have 
indeed level $\Psi-1$, as we expected. In order to verify this fact, we need to take into account the contribution of the ghosts: 
before the DS reduction the currents in the block $A$ in $U(N|L)_\Psi$ form an $U(M|L)_{\Psi - N+M}$ 
current subalgebra. The ghosts which contribute to the level shift are these in the $C \oplus B$ blocks: 
$N-M-1$ sets of bosonic and fermionic ghosts, each being essentially an $\mathfrak{Sb}^{L|M}$ system. 
They shift the level of the $U(M|L)$ currents by precisely $N-M-1$ and thus the final level is $\Psi-1$, as it should. 

Doing the coset by $U(M|L)_{\Psi-1}$ we get
\begin{align}
&c_{L,M,N}[\Psi]= \frac{1}{\Psi} (L-N)\left((L-N)^2-1\right) + \Psi(M-N)\left((M-N)^2-1\right) + \cr
&+ \frac{1}{\Psi-1}(M-L)((M-L)^2-1)+(2N+M-3L)(N-M)^2+L-N
\end{align}
We can make the symmetries manifest by some simple manipulations:
\begin{align}
c_{L,M,N}[\Psi]= &\frac12\frac{1}{\Psi} (L-N)\left((L-N)^2-1\right) +\frac12(1-\frac{1}{\Psi}) (N-L)\left((N-L)^2-1\right) + \cr
&+\frac12 \Psi(M-N)\left((M-N)^2-1\right) + \frac12 (1-\Psi)(N-M)\left((N-M)^2-1\right) +\cr
&+\frac12\frac{1}{1-\Psi}(L-M)((L-M)^2-1)+\frac12 \frac{\Psi}{\Psi-1}(M-L)((M-L)^2-1)+ \cr
&+\frac12(2L-N-M)(2M-N-L)(2N-L-M)
\end{align}
We see a sum of all the $S_3$ images of the first term plus a symmetric function of $N$,$M$, $L$. 
The calculation for $M>N$ gives the same answer.

\subsection{Ortho-symplectic case}

This appendix gives some details of the calculation of central charges for ortho-symplectic algebras. 

Let us denote central charge of $Y^{-}_{L,M,N}[\Psi]$ as $c^-_{L,M,N}[\Psi]$. 
By definition, the central charge of $\tilde{Y}^-_{L,M,N}[\Psi]$ can be identified as $\tilde{c}^-_{L,M,N}[\Psi]=c^-_{L,N,M}[1-\Psi]$. 

Recall that the central charge of $OSp(2N|2L)_{\Psi-2N+2L+2}$ is
\begin{eqnarray}
c_{OSp(2N|2L)_{\Psi-2N+2L+2}}=\frac{(L-N)(2(L-N)+1)(\Psi+2(L-N)+2)}{\Psi}.
\end{eqnarray}
Assuming $N>M+1$, one needs to perform a DS-reduction in the $O(2(N-M)-1)$ diagonal glock of $OSp(2N|2L)$. The principal embedding of the $\mathfrak{su}(2)$ algebra inside $O(2(N-M)-1)$ can be identified with the $2(N-M)-1$ dimensional representation of $\mathfrak{su}(2)$. The stress-energy tensor modification term leads to a contribution to the central charge given by
\begin{eqnarray}\nonumber
c_{\partial H}&=&-6(\Psi+2(L-N)+2)\sum_{n=-N+M+1}^{N-M-1}n^2\\
&=&2(2 + 2(L - N) + \Psi) (1 + 2(M-N)) (M - N) (1 + M - N) .
\end{eqnarray}
There are againt two kinds of contributions coming from ghosts in different blocks. In the block where the DS-reduction is performed, different components of the current algebra decompose as
\begin{eqnarray}
SO(2(N-M)-1)\simeq 3\oplus7\oplus11\oplus\dots \oplus 4(N-M)-5
\end{eqnarray}
and one needs to fix all the components with positive weight under this decomposition by introducing appropriate ghosts. The contribution of these ghosts to the central charge is
\begin{eqnarray}\nonumber
c_{gh_D}&=&\sum_{n=1}^{N-M-1}\sum_{m=1}^{2n-1}(1-3(2m-1)^2)\\
&=&-2 (1 + M - N)^2 (1 + 4 M^2 - 8 M (-1 + N) + 4 (-2 + N) N).
\end{eqnarray}
Off-diagonal blocks contain $(2(M-N)-1)\times (2L+2M+1)$ components that are in fundamental representation of both $SO(2(N-M)-1)$ in the D-block and $OSp(2M+1|2L)$ in the A-block. To fix the fermionic components, one needs also to introduce bosonic ghosts of appropriate dimension. The central charge of such bosonic ghosts equals minus the central charge of fermionic ghosts and the contribution from this block can be identified with
\begin{eqnarray}
c_{B\oplus C}&=&(2(M-L)+1)\sum_{n=1}^{N-M-1}(1-3(2n-1)^2)\\ \nonumber
&=&-2(2(M-L)+1)(1+M-N)(1+2M^2-4M(N-1)+2(N-2)N).
\end{eqnarray}
Putting everythig together and subtracting the contribution coming from the coset part, one gets
\begin{eqnarray}\nonumber
c^-_{L,M,N}[\Psi]&=&-\frac{(2 (L-M)-1) (2 (L-M)+1) (L-M)}{\Psi -1}\\ \nonumber
&&+\frac{2 (2 (L-N)+1) (L-N+1) (L-N)}{\Psi }\\ \nonumber
&&+2 \Psi  (2 (M-N)+1) (M-N+1) (M-N)\\ \nonumber
&& - 2 L (1 + 6 M^2 + M (6 - 12 N) - 6 N + 6 N^2) \\
&&+4 M^3 - 3 M (1 - 2 N)^2+ N (5 - 12 N + 8 N^2)
\end{eqnarray}

The central charge for $Y^+_{L,M,N}[\Psi]$ will be denoted as $c^+_{L,M,N}[\Psi]$. It can be calculated in the same way as the one for $\tilde{Y}^+_{L,M,N}[\Psi]$ since the whole construction is independent of the value of $L$ and one can simply set $\tilde{c}^+_{L,M,N}[\Psi]=c^+_{L+\frac{1}{2},M,N}[\Psi]$.

The central charge of $OSp(2L|2N)_{-\Psi+2(N-L)+2}$ equals
\begin{eqnarray}
c_{OSp(2L|2N)_{-\Psi+2(N-L)+2}}=\frac{(N-L)(2(N-L)+1)(-\Psi+2(N-L)+2)}{-\Psi}
\end{eqnarray}
Now, we need to perform DS-reduction in the $Sp(2(N-M))$ block of $OSp(2L|2N)$. The principal embedding of $\mathfrak{su}(2)$ inside $Sp(2(N-M))$ can be identified with the $2(N-M)$ dimensional representation of $\mathfrak{su}(2)$ and modification term contributes to the central charge by
\begin{eqnarray}\nonumber
c_{\partial H}&=&12(-\Psi+2(N-L)+2)\sum_{n=1}^{N-M}\left ( \frac{2n-1}{2}\right )^2\\
&=&(\Psi+2(N-L)-2)(M-N)(2(M-N)+1)(2(M-N)-1)
\end{eqnarray}
The decomposition of the currents in D block is again of the form
\begin{eqnarray}
Sp(2(N-M))=3\oplus 7\oplus 11\oplus \dots \oplus 4(N-M)-1.
\end{eqnarray}
The corresponding ghosts contribute to the central charge as
\begin{eqnarray}
c_D=\sum_{n=1}^{N-M}\sum_{m=1}^{2n-1}(1-3(2m-1)^2)=-2(M-N)^2(-3+4(M-N)^2).
\end{eqnarray}
Finally, the currents in the off-diagonal block are in the product of fundamental representations of $Sp(2(N-M))$ and $OSp(2L|2M)$. 
Similar arguments as in the case of $Y^-$ applies here with only exception that fields of weight half are now present 
and we need to fix only half of the corresponding currents. One gets a contribution
\begin{eqnarray}\nonumber
c_{B\oplus C}&=& 2(M-L)\sum_{n=2}^{N-M}(1-12(n-1)^2)+M-L\\
&=&(M-L) (1 + 2(M-N)) (-1 - 4 N + 4 (M + M^2 - 2 M N + N^2)).
\end{eqnarray}
Putting everything together and subtracting the contribution coming from the coset part leads to
\begin{eqnarray}\nonumber
c^+_{L,M,N}[\Psi]&=&-\frac{2 (M-L) (2 (M-L)+1) (M-L+1)}{1-\Psi }\\ \nonumber
&&-\frac{2 (N-L) (2 (N-L)+1) (N-L+1)}{\Psi }\\ \nonumber
&&+\Psi  (2 (M-N)-1) (2 (M-N)+1) (M-N)\\ 
&&+L (1 - 12 (M - N)^2) - N + 2 (M - N)^2 (3 + 2 M + 4 N)
\end{eqnarray}

\section{Series, products and contour integrals}\label{app:id}

A useful contour integral identity with symplectic boson denominators
\begin{align} 
&\ \oint \prod_{i=1}^N \frac{dx_i}{x_i} x_i^{s_i} \frac{1}{\prod_{i,n} (1-q^{n+\frac12} x_i)(1-q^{n+\frac12} x^{-1}_i)}=\cr
&= \frac{1}{ \prod_{n> 0} (1-q^n)^{2N}}\oint \prod_{i=1}^N \frac{dx_i}{x_i} x_i^{s_i}
\sum_{n_i=0}^\infty \sum_{m_i = - n_i}^{n_i} \prod_i x_i^{m_i} (-1)^{\sum_i (n_i - m_i)} q^{\sum_i \frac{n_i(n_i+1)- m_i^2}{2}} = \cr
&=  \frac{1}{ \prod_{n> 0} (1-q^n)^{2N}}
\sum_{n_i=|s_i|}^\infty \prod_i  (-1)^{\sum_i (n_i +s_i)} q^{\sum_i \frac{n_i(n_i+1)- s_i^2}{2}} = \cr
&=   \frac{1}{ \prod_{n> 0} (1-q^n)^{2N}}
\sum_{n_i=0}^\infty \prod_i  (-1)^{\sum_i n_i} q^{\sum_i \frac{n_i(n_i+1)+ (2 n_i + 1)|s_i|}{2}} = \cr
&=  \frac{1}{ \prod_{n> 0} (1-q^n)^{2N}}
\sum_{n_i=0}^\infty \prod_i  (-1)^{\sum_i n_i} q^{\sum_i \frac{n_i(n_i+1)}{2}}q^{(n_i + \frac12)s_i} 
\end{align}
The only non-trivial step is the removal of the absolute value $|s_i| \to s_i$: if $s_i$ is negative the sum without the absolute value 
has the first $2 s_i$ terms cancelling out in pairs. The final result is that of a sum over residues 
of the contour integral at $x_i = q^{(n_i + \frac12)}$.

In a similar manner, a contour integral with current denominators
\begin{align}
& \oint \prod_{i=1}^N \frac{dx_i}{x_i} x_i^{s_i} \frac{1}{\prod_{i,n} (1-q^{n+1} x_i)(1-q^{n+1} x^{-1}_i)} =\cr
& \oint \prod_{i=1}^N \frac{dx_i}{x_i} q^{\frac{s_i}{2}} x_i^{s_i} \frac{\prod_i (1-q^{\frac12} x_i)}{\prod_{i,n} (1-q^{n+\frac12} x_i)(1-q^{n+\frac12} x^{-1}_i)} =\cr
&=  \frac{1}{ \prod_{n> 0} (1-q^n)^{2N}}
\sum_{n_i=0}^\infty \prod_i  (-1)^{\sum_i n_i} (1-q^{n_i + 1}) q^{\sum_i \frac{n_i(n_i+1)}{2}}q^{(n_i + 1)s_i} 
\end{align}

\section{Characters for $W_{N-M, 1, \cdots, 1} U(N)$}\label{app:dsindex}

The contribution to the index from the $U(N)$ currents can be split it to the contributions coming from different blocks as
\begin{eqnarray}
\chi _{U(N)}^{DS_M}=\chi_A\chi_B\chi_C\chi_D.
\end{eqnarray}
The fields in the $A$-sector do not have a modified conformal weight but they are graded with respect to the currents $J^{\{h\}}$ preserved by the DS-reduction. The corresponding character is then
\begin{eqnarray}
\chi_A=\prod_{n=1}^{\infty}\prod_{i,j=1}^{N-M}\frac{1}{1-x_ix_j^{-1}q^n}
\end{eqnarray}
where $x_i$ is a fugacity for the current $J_0^{\{h_i\}}$. In the $D$-block, there are no factors of $x_i$ but the conformal weights of the fields are non-trivially shifted. One gets the character
\begin{eqnarray}
\chi_{D}=\prod_{n=1}^{\infty}\prod_{j=1}^{N-M-1}\frac{1}{(1-q^{n+j})^{N-M-j}}\frac{1}{(1-q^n)^{N-M}}\frac{1}{(1-q^{n-j})^{N-M-j}}.
\end{eqnarray}
Both $B$- and $C$- blocks give the same contribution. The fields in these blocks are charged under $J^{\{h\}}$ (with opposite charges in the two blocks) 
and have dimensions shifted by the the stress-energy tensor modification. The characters are then
\begin{eqnarray}\nonumber
\chi_B&=&\prod_{n=1}^{\infty}\prod_{i=1}^{N-M}\prod_{j=1}^M\frac{1}{1-x_iq^{n+\frac{M+1}{2}-j}},\\
\chi_C&=&\prod_{n=1}^{\infty}\prod_{i=1}^{N-M}\prod_{j=1}^M\frac{1}{1-x_i^{-1}q^{n+\frac{M+1}{2}-j}}.
\end{eqnarray} 

The contributions to the character from the ghost sector can be again divided into contributions from different blocks. There are no ghosts associated to the $A$-block. 
The contribution from the $bc$-ghosts in the $D$-sector is 
\begin{eqnarray}
\chi_D^{bc}=\prod_{n=1}^{\infty}\prod_{i=1}^{N-M-1}(1-q^{n+i-1})^{N-M-i}(1-q^{n-i})^{N-M-i}.
\end{eqnarray}
The contributions from the $bc$-ghosts in the $B$- and $C$-sectors contains fugacities $x_i$ since they are charged under the total $U(M)$ currents, again with opposite 
charges in the two blocks, 
\begin{eqnarray}\nonumber
\chi_B^{bc}&=&\prod_{n=1}^{\infty}\prod_{i=1}^{\frac{N-M}{2}}\prod_{j=1}^{M}(1-x_jq^{n+i-\frac{3}{2}})(1-x_jq^{n-i+\frac{1}{2}}),\\
\chi_C^{bc}&=&\prod_{n=1}^{\infty}\prod_{i=1}^{\frac{N-M}{2}}\prod_{j=1}^{M}(1-x_j^{-1}q^{n+i-\frac{3}{2}})(1-x_j^{-1}q^{n-i+\frac{1}{2}}).
\end{eqnarray}
with some extra correction depending on $N-M$ being odd or even to account correctly for the ghosts with weight $1/2$. 

\section{Boundary conditions for hypermultiplets}\label{app:hypers}
The hypermultiplet SUSY transformations take the schematic form 
\begin{align}
\delta_{\alpha}^{A \dot A} q^B_a &=\epsilon^{AB} \rho^{\dot A}_{\alpha a} \cr
\delta_{\alpha}^{A \dot A}\rho^{\dot B}_{\beta a} &= \epsilon^{\dot A \dot B}\partial_{\alpha \beta} q^A_a
\end{align}
where $A,\cdots$, $\dot A, \cdots$ respectively denote indices for $SU(2)_H$ and $SU(2)_C$, $\alpha, \cdots$ 
spinor indices and $a$ is a flavor index. 


The supercurrents are 
\begin{equation}
S^{A \dot A}_{\alpha \beta \gamma} = \omega^{ab} \partial_{(\alpha \beta} q^A_a \rho^{\dot A}_{\gamma) b}
\end{equation}
 
We seek boundary conditions which preserve $(0,4)$ supersymmetry at the boundary.
Correspondingly, the normal components of the supercharges which are right-chiral 
on the boundary must vanish: 
\begin{equation}
S^{A \dot A}_{+--} =0
\end{equation}

There are two natural Lorentz-invariant boundary conditions for the fermions which preserve the 
$SU(2)_C$ R-symmetry and flavor groups: 
\begin{align}
\rho^{\dot A}_{+ a} &=0 \qquad (N) \cr
\rho^{\dot A}_{- a} &=0  \qquad (D)
\end{align}

These conditions then require respectively 
\begin{align}
\partial_{+-} q^A_a & =0 \qquad (N) \cr
\partial_{--} q^A_a & =0  \qquad (D)
\end{align}
which explain our monikers: the first possibility requires Neumann 
boundary conditions for all hypermultiplet scalars, while the 
second possibility (together with the CPT conjugate relation) requires 
Dirichlet boundary conditions for all hypermultiplet scalars:
\begin{equation}
q^A_a  =0  \qquad (D)
\end{equation}

Next, we need to consider some deformations of these boundary conditions which break 
Lorentz symmetry, but preserve a twisted Lorentz group which is defined either 
with the help of the Cartan of $SU(2)_H$ or the Cartan of $SU(2)_C$.

If we twist by $SU(2)_H$ then we have scalar supercharges $Q^{+ \dot A}_+$ and $Q^{- \dot A}_-$.
We may seek a boundary condition which preserves $Q^{- \dot A}_- + \zeta Q^{+ \dot A}_+$.
That means 
\begin{equation}
S^{- \dot A}_{+--} + \zeta S^{+ \dot A}_{++-}=0
\end{equation}

The natural boundary conditions on the fermions are unchanged
\begin{align}
\rho^{\dot A}_{+ a} &=0 \qquad (N) \cr
\rho^{\dot A}_{- a} &=0  \qquad (D)
\end{align}
which imply 
\begin{align}
\partial_{+-} q^-_a + \zeta \partial_{++} q^+_a & =0 \qquad (N) \cr
\partial_{--} q^-_a + \zeta \partial_{+-} q^+_a& =0  \qquad (D)
\end{align}
The first choice is an interesting deformation of the standard Neumann boundary conditions. 
It will be important for us. The second choice is {\it not} a deformation of Dirichlet boundary conditions.
Rather, it gives the parity conjugate of the deformed Neumann boundary conditions. Thus Dirichlet boundary conditions
do not admit a deformation of this type.  

There is a useful way to think about the deformation of boundary conditions induced by a deformation of the preserved supersymmetry. 
If we add some extra term to the boundary action which breaks $Q^{- \dot A}_-$, the variation of the term under $Q^{- \dot A}_-$ will appear 
as the boundary value $S^{- \dot A}_{+--}$ of the corresponding supercurrent. Thus we can find a deformation which preserves the deformed SUSY  
if we can write the boundary value $S^{+ \dot A}_{++-}$ of the other supercurrent as a $Q^{- \dot A}_-$ variation of some boundary action 
$O^{++}_{++}$. 

For example, for Neumann b.c. we have 
\begin{equation}
S^{+ \dot A}_{++-} = \omega^{ab} \partial_{++} q^+_a \rho^{\dot A}_{- b} = \delta_{-}^{- \dot A} \left( \omega^{ab} \partial_{++} q^+_a q^{+}_{b} \right)
\end{equation}
which is the variation of a natural boundary action which is equal to the action for symplectic bosons. 
On the other hand, for Dirichlet b.c. we have 
\begin{equation}
S^{+ \dot A}_{++-} = \omega^{ab} \partial_{+-} q^+_a \rho^{\dot A}_{+ b} 
\end{equation}
which is not a $\delta_{-}^{- \dot A}$ variation. 

If we twist by $SU(2)_C$ then we have scalar supercharges $Q^{A \dot +}_+$ and $Q^{A \dot -}_-$.
We may seek a boundary condition which preserves $Q^{A \dot -}_- + \dot \zeta Q^{A \dot +}_+$.
That means 
\begin{equation}
S^{A \dot -}_{+--} + \dot \zeta S^{A \dot +}_{++-}=0
\end{equation}

The boundary conditions on the fermions can now be twisted as well
\begin{align}
\rho^{\dot +}_{+ a}  &=\eta \rho^{\dot -}_{- a} \qquad (N) \cr
\rho^{\dot -}_{+ a} &=0 \qquad (N) \cr
\rho^{\dot +}_{- a} &=0  \qquad (D) \cr
\rho^{\dot -}_{- a} &=\eta \rho^{\dot +}_{+ a}  \qquad (D) 
\end{align}
which imply 
\begin{align}
(1 + 2 \eta \dot \zeta) \omega^{ab} \partial_{+-} q^A_a \rho^{\dot -}_{- b} + \dot \zeta \omega^{ab} \partial_{++} q^A_a \rho^{\dot +}_{- b}& =0 \qquad (N) \cr
2 \eta \omega^{ab} \partial_{+-} q^A_a \rho^{\dot +}_{+ b}  + \omega^{ab} \partial_{--} q^A_a \rho^{\dot -}_{+ b}  + 2\dot \zeta \omega^{ab} \partial_{+-} q^A_a \rho^{\dot +}_{+ b} & =0  \qquad (D)
\end{align}
The first choice is inconsistent. There is no linear boundary condition on the scalar fields which can 
satisfy this constraint. Thus Neumann boundary conditions do not admit this type of deformation. 

On the other hand, standard Dirichlet b.c. for the hypermultiplet scalars, together with $\eta = - \dot \zeta$,
give a useful deformation of Dirichlet boundary conditions. It will be important for us. 

Again, the deformability or lack thereof is related to the observation that for Dirichlet b.c. we have 
\begin{equation}
S^{A \dot +}_{++-} = \omega^{ab} \partial_{+-} q^A_a \rho^{\dot +}_{+ b} = \delta^{A \dot -}_- ( \omega^{ab} \rho^{\dot +}_{+ a} \rho^{\dot +}_{+ b})
\end{equation}
while for Neumann 
\begin{equation}
S^{A \dot +}_{++-} = \omega^{ab} \partial_{++} q^A_a \rho^{\dot +}_{- b} 
\end{equation}
cannot be written as an $\delta^{A \dot -}_- $ variation.

\subsection{Neumann boundary VOA}
Deformed Neumann boundary conditions support supersymmetric boundary local operators: the bulk SUSY transformations 
\begin{align}
\left( \delta_{-}^{- \dot A} + \zeta \delta_{+}^{+ \dot A}\right)q^B_a &=\epsilon^{-B} \rho^{\dot A}_{- a}+\zeta \epsilon^{+B} \rho^{\dot A}_{+ a} \cr
\left( \delta_{-}^{- \dot A} + \zeta \delta_{+}^{+ \dot A}\right)\rho^{\dot B}_{\beta a} &= \epsilon^{\dot A \dot B}\partial_{- \beta} q^-_a+\zeta  \epsilon^{\dot A \dot B}\partial_{+ \beta} q^+_a
\end{align}
restricted to the boundary give 
\begin{align}
\left( \delta_{-}^{- \dot A} + \zeta \delta_{+}^{+ \dot A}\right)q^B_a &=\epsilon^{-B} \rho^{\dot A}_{- a}\cr
\left( \delta_{-}^{- \dot A} + \zeta \delta_{+}^{+ \dot A}\right)\rho^{\dot B}_{- a} &= (1+\zeta \bar \zeta) \epsilon^{\dot A \dot B}\partial_{- -} q^-_a
\end{align}
showing that the $q^-_a$ are supersymmetric and holomorphic modulo operators in the image of the supercharges. 

Notice that for non-zero $\zeta$ the $q^-_a$ are supersymmetric only at the boundary. 
This fact allows them to have non-trivial holomorphic OPE. A simple calculation of 
boundary-to-boundary propagators recovers the symplectic boson OPE with coefficient 
proportional to $\zeta$. 

Let us introduce real fields $q_i$ such that
\begin{eqnarray} \nonumber
q^-&=&q_1+iq_2,\\
q^+&=&q_3-iq_4.
\end{eqnarray}
where we supressed the flavor indices $s$. in terms of the real fields, we can write above boundary conditions as
\begin{eqnarray}\nonumber
\partial q_1+\zeta(\partial_0q_3-\partial_1q_4)&=&0,\\ \nonumber
\partial q_2-\zeta(\partial_0q_4+\partial_1q_3)&=&0,\\ \nonumber
\partial q_3-\zeta(\partial_0q_1-\partial_1q_2)&=&0,\\
\partial q_4+\zeta(\partial_0q_2+\partial_1q_1)&=&0.
\end{eqnarray}
All the componens further satisfy bulk equations of motion $\Delta q_i=0$. Going to momentum space and introducing boundary source, and substituting $k_2^2\rightarrow k^2_0+k^2_1$, we can express boundary-to-boundary propagator as
\begin{eqnarray}
\langle q_i(k)q_j(0)\rangle =
\frac{1}{1-\zeta^2}\frac{1}{k_0^2+k_1^2} 
\begin{pmatrix}k_2 & 0 &\zeta k_0 & -\zeta k_1\\ \nonumber
                        0          & k_2&-\zeta k_1  &-\zeta k_0\\ \nonumber
\zeta k_0 &-\zeta k_1& k_2  & 0 \\ \nonumber
-\zeta k_1 & -\zeta k_0& 0 & k_2  \nonumber\end{pmatrix}.
\end{eqnarray}
The only non-vanishing boundary to boundary propagator of $Q$-closed operators is then
\begin{eqnarray}
\langle \bar{q}^+(x)q^-(0)\rangle =\frac{2\zeta}{1-\zeta^2}\int dk_0 dk_1 e^{ik_0x_0+ik_1x_1}\frac{k_0-ik_1}{k_0^2+k_1^2}.
\end{eqnarray}
Note that this correlation function can be expressed as
\begin{eqnarray}
=\frac{2\zeta}{\zeta^2+1}\bar{\partial} \int dk_0 dk_1 \frac{e^{i (x_0 k_0+x_1k_1)}}{(k_0^2+k_1^2)}=\frac{1}{2\pi}\frac{2\zeta}{\zeta^2+1}\bar{\partial} \ln |z|^2=\frac{1}{\pi}\frac{2\zeta}{\zeta^2+1}\frac{1}{z}.
\end{eqnarray}
This is the propagator of the symplectic boson.

\subsection{Dirichlet boundary VOA}
Deformed Dirichlet boundary conditions support supersymmetric boundary local operators: the bulk SUSY transformations are
\begin{align}
\left( \delta_{-}^{A \dot -} + \dot \zeta \delta_{+}^{A \dot +}\right)q^B_a &=\epsilon^{AB} \rho^{\dot -}_{- a} + \dot \zeta \epsilon^{AB} \rho^{\dot +}_{+ a} \cr
\left( \delta_{-}^{A \dot -} + \dot \zeta \delta_{+}^{A \dot +}\right)\rho^{\dot B}_{\beta a} &= \epsilon^{\dot - \dot B}\partial_{- \beta} q^A_a
+\dot \zeta \epsilon^{\dot + \dot B}\partial_{+ \beta} q^A_a
\end{align}
At the boundary, they simplify to 
\begin{align}
\left( \delta_{-}^{A \dot -} + \dot \zeta \delta_{+}^{A \dot +}\right)\rho^{\dot +}_{+ a} &= -\partial_{+-} q^A_a \cr
\left( \delta_{-}^{A \dot -} + \dot \zeta \delta_{+}^{A \dot +}\right)\rho^{\dot -}_{+ a} &=0
\end{align}
Thus $\rho^{\dot -}_{+ a}$ are supersymmetric at the boundary. This fact allows them to have non-trivial holomorphic OPE.

\section{Boundary conditions for gauge theory}\label{app:bound}

$\mathcal{N}=4$ super Yang-Mills admits many half-BPS boundary conditions and interfaces \cite{Gaiotto:2008ac}. 
These boundary conditions preserve a set of supercharges which form a 3d $\mathcal{N}=4$ superalgebra. 
These 3d $\mathcal{N}=4$ sub-algebras may be embedded in a variety of different ways in the four-dimensional super-algebra, depending on the choice of boundary condition. 

In particular, if we look at boundary conditions and interfaces which descend from IIB string theory configurations 
associated to $(p,q)$ fivebranes aligned along three specific directions in spacetime (say $456$, rotated by $SU(2)_H$)
we will find a corresponding two-parameter family of 
3d $\mathcal{N}=4$ sub-algebras $A^{3d}_{(p,q)}$ which are permuted by S-duality transformations \cite{Gaiotto:2008aa}. 

The corner configurations considered in this paper preserve four chiral supercharges in 2d, which are analogous to the supersymmetries 
preserved by the boundary conditions in Appendix \ref{app:hypers}. This 2d $(0,4)$ subalgebra will be a common sub-algebra 
to all the 3d $\mathcal{N}=4$ sub-algebras associated to $(p,q)$ boundary conditions of appropriate slope in the plane of the junction. 

We can denote the $(0,4)$ supercharges as $Q^{A \dot A}_-$. The 3d subalgebras will consist of $Q^{A \dot A}_-$
together with appropriate linear combinations of two sets of anti-chiral supercharges $Q^{A \dot A}_+$ and $\tilde Q^{A \dot A}_+$.
These have the properties that they anti-commute with $Q^{A \dot A}_-$ to translations in the plane of the junction:
\begin{equation}
\{ Q^{A \dot A}_-, Q^{B \dot B}_+ \} = \epsilon^{AB} \epsilon^{\dot A \dot B} P_2 \qquad \qquad \{ Q^{A \dot A}_-, \tilde Q^{B \dot B}_+ \} = \epsilon^{AB} \epsilon^{\dot A \dot B} P_3 
\end{equation}
We can organize the remaining supercharges into a set $\tilde Q^{A \dot A}_-$ with 
\begin{equation}
\{ \tilde Q^{A \dot A}_-, Q^{B \dot B}_+ \} = \epsilon^{AB} \epsilon^{\dot A \dot B} P_3 \qquad \qquad \{ \tilde Q^{A \dot A}_-, \tilde Q^{B \dot B}_+ \} = - \epsilon^{AB} \epsilon^{\dot A \dot B} P_2 
\end{equation}

Consider now an $SU(2)_H$ twist of the Lorentz generator in the $01$ plane, as in Appendix \ref{app:hypers}.
We have the scalar supercharges $Q^{- \dot A}_-$ and we may look for a way to deform them. 
The common deformed subalgebra should contain, in particular, the GL-twisted supercharge with parameter $\Psi$. 

It is clear that we cannot find a deformation analogous to the one Appendix \ref{app:hypers}
which belongs to all the  3d subalgebras: each 3d sub-algebra would require 
us to deform $Q^{- \dot A}_-$ by different linear combinations of $Q^{+ \dot A}_+$ and $\tilde Q^{+ \dot A}_+$.
Instead, we will need to deform simultaneously both the (twisted Lorentz scalar part of the) 3d subalgebras and the common 2d subalgebra. 

The deformation of Neumann b.c. compatible with a GL twist is well understood \cite{Witten:2010aa}. It is proportional to $\Psi$ and thus vanishes when $\Psi = 0$. 
In a similar manner, the $B_{(p,q)}$ boundary conditions will be undeformed when $\Psi$ is the appropriate rational number. 
Thus for $\Psi = 0$ we expect to be able to reach the GL twist by deforming $Q^{- \dot A}_-$ by $Q^{+ \dot A}_+$,
for $\Psi = \infty$ we expect to be able to reach the GL twist by deforming $Q^{- \dot A}_-$ by $\tilde Q^{+ \dot A}_+$, etc. 
In general, the deformed supercharge should look roughly as 
\begin{equation}
Q^{- \dot A}_- + \zeta \left(Q^{+ \dot A}_+ + \Psi  \tilde Q^{+ \dot A}_+ \right) 
\end{equation}

From the point of view of the 3d sub-algebra $(Q^{- \dot A}_-, Q^{+ \dot A}_+)$ preserved by Neumann boundary conditions, we are 
first deforming the 3d superalgebra to something like $(Q^{- \dot A}_- + \zeta \Psi \tilde Q^{+ \dot A}_+, Q^{+ \dot A}_+)$ 
and then the 2d subalgebra as in Appendix \ref{app:hypers}. Similar considerations apply to the other 3d subalgebras. 

Again, the deformability of 3d boundary conditions is encoded in the requirement that for every supercharge $Q$ we want to deform by another supercharge $Q'$, 
the boundary value of the supercurrent for $Q'$ should be the $Q$ variation of a local operator we are deforming the boundary condition by. 

We expect all boundary conditions and interfaces which arise from configurations of $(p,q)$ fivebranes with appropriate slope to admit deformations of this type.
We also expect deformability of 3d half-BPS boundary conditions to be rare. Analogous constraints were found in \cite{Gaiotto:2008ab}: Neumann boundary conditions 
coupled to generic 3d $\mathcal{N}=4$ matter are compatible with a bulk $\theta$ angle only if the matter theory moment maps satisfy certain quadratic identities. 
Turning on non-zero $\Psi$ is analogous to turning on a bulk $\theta$ angle.

\bibliography{Corner}
\bibliographystyle{JHEP}
\end{document}